\documentclass{aa}  

\usepackage{graphicx}
\usepackage{txfonts}
\begin{document}

   \title{Dense Gas in a Giant Molecular Filament}

   \author{Y. Wang
          \inst{1}
          \and
          H. Beuther\inst{1}       
          \and
         N. Schneider\inst{2}
          \and
         S. E. Meidt\inst{1}
         \and
         H. Linz\inst{1}
         \and
          S. Ragan \inst{3}
         \and
      	 C. Zucker\inst{4}
         \and      	 
         C. Battersby\inst{5}
      	 \and      	 
      	 J. D. Soler\inst{1}
      	 \and
         E. Schinnerer\inst{1}
         \and
         F. Bigiel\inst{6}
         \and
         D. Colombo\inst{7}
         \and
         Th. Henning\inst{1}
          }

   \institute{Max-Planck-Insitute for Astronomy,
              K\"onigstuhl 17, D-69117 Heidelberg\\
              \email{wang@mpia.de}
              \and 
              I. Physik. Institut, University of Cologne, Z\"ulpicher Str.77, 50937 Cologne, Germany
             \and
             School of Physics and Astronomy, Cardiff University, Queen’s Buildings, The Parade, Cardiff, CF24 3AA, UK
             \and
              Harvard Astronomy, Harvard-Smithsonian Center for Astrophysics, 60 Garden Street, Cambridge, MA 02138, USA
              \and
              Department of Physics, University of Connecticut, Storrs, CT 06269, USA
               \and
                Argelander Institut f\"ur Astronomie, Auf dem H\"ugel 71, 53121 Bonn, Germany
               \and
               Max-Planck-Institut f\"ur Radioastronomie, Auf dem H\"ugel 69, 53121 Bonn, Germany
               }

   \date{Received xxx; accepted xxx}

% \abstract{}{}{}{}{} 
% 5 {} token are mandatory
 
  \abstract
  % context heading (optional)
  % {} leave it empty if necessary  
   {Recent surveys of the Galactic plane in the dust continuum and CO emission lines reveal that large ($\gtrsim 50$~pc) and massive ($\gtrsim 10^5$~$M_\odot$) filaments, know as giant molecular filaments (GMFs), may be linked to galactic dynamics and trace the mid-plane of the gravitational potential in the Milky Way. Yet our physical understanding of GMFs is still poor.} 
  % aims heading (mandatory)
   {We investigate the dense gas properties of one GMF, with the ultimate goal of connecting these dense gas tracers with star formation processes in the GMF.}
  % methods heading (mandatory)
   {We have imaged one entire GMF located at $l\sim$52--54$^\circ$ longitude, GMF54 ($\sim$68~pc long), in the empirical dense gas tracers using the HCN(1--0), HNC(1--0), HCO$^+$(1--0) lines, and their $^{13}$C isotopologue transitions, as well as the N$_2$H$^+$(1--0) line. We study the dense gas distribution, the column density probability density functions (N-PDFs) and the line ratios within the GMF.}
  % results heading (mandatory)
   {The dense gas molecular transitions follow the extended structure of the filament with area filling factors between 0.06 and 0.28 with respect to $^{13}$CO(1--0). We constructed the N-PDFs of H$_2$ for each of the dense gas tracers based on their column densities and assumed uniform abundance. The N-PDFs of the dense gas tracers appear curved in log-log representation, and the HCO$^+$ N-PDF has the largest log-normal width and flattest power-law slope index. Studying the N-PDFs for sub-regions of GMF54, we found an evolutionary trend in the N-PDFs that high-mass star forming and Photon Dominated Regions (PDRs) have flatter power-law indices. The integrated intensity ratios of the molecular lines in GMF54 are comparable to those in nearby galaxies. In particular, the N$_2$H$^+$/$^{13}$CO ratio, which traces the dense gas fraction, has similar values in GMF54 and all nearby galaxies except ULIRGs.}
  % conclusions heading (optional), leave it empty if necessary 
   {The largest coherent cold gaseous structure in our Milky Way, GMFs, are outstanding candidates for connecting studies of star formation on Galactic and extragalactic scales. By analyzing a complete map of the dense gas in a GMF we have found that: 1) the dense gas N-PDFs appear flatter in more evolved regions and steeper in younger regions, and 2) its integrated dense gas intensity ratios are similar to those of nearby galaxies.}


   \maketitle
%
%-------------------------------------------------------------------

\section{Introduction}
\label{sect_intro}
Studies towards nearby molecular clouds (MCs) show that the dense regions of MCs are permeated with filaments that contain sites of star formation \citep[e.g,][]{Andre2014}. Recent observations identified a class of large ($\gtrsim$50~pc) and massive ($\gtrsim10^5$~$M_\odot$) filaments, known as giant molecular filaments (GMFs) \citep{Jackson2010, Goodman2014, Ragan2014, Zucker2015, Wang2015, li2016, Wang2016, Abreu-Vicente2016, wang2020}. Many of these giant filaments lie along, or extremely close to spiral arms in the position-position-velocity space, suggesting they may trace the dense ``spine'' of the spiral arms and the mid-plane of the gravitational potential in the Milky Way (MW) \citep{Goodman2014, Zucker2015}. Detailed inter-comparison of a sample of long filaments from the literature suggests that there may be different classes of filaments in both their physical properties and their association with Galactic structure \citep{Zucker2018}. GMFs are the largest coherent cold gas structures in our Milky Way yet our physical understanding of GMFs is still poor, limited to estimates of their occurrence, gas masses and lengths \citep{Ragan2014, Abreu-Vicente2016, Zucker2018}. 

Observations of nearby galaxies have resolved giant MCs with masses and sizes similar to those of GMFs (10 to 200 pc, e.g., \citealt{Hughes2013, Schinnerer2013, Leroy2016, li2016, Faesi2018, Herrera2020}), suggesting that GMFs could be analogous to these extragalactic giant MCs and could connect studies of star formation on Galactic and extragalactic scales. 

So far, GMF studies have relied on Galactic plane surveys, either in dust tracers like (sub)mm emission (e.g., Hi-GAL\footnote{The Herschel infrared Galactic Plane Survey.}, \citealt{Molinari2010}; ATLASGAL\footnote{The APEX Telescope Large Area Survey of the Galaxy.}, \citealt{Schuller2009}) or in low- to intermediate-spatial resolution surveys of CO and its isotopologues (e.g., GRS\footnote{The Galactic Ring Survey.}, \citealt{Jackson2006}; COHRS\footnote{The CO High-Resolution Survey}, \citealt{Dempsey2013}; SEDIGISM\footnote{Structure, excitation, and dynamics of the inner Galactic interstellar medium}, \citealt{Schuller2017}). However, there exists little knowledge about the high-volume-density gas, which will eventually form stars.

\begin{figure*}
   \centering
   \includegraphics[width=0.8\hsize]{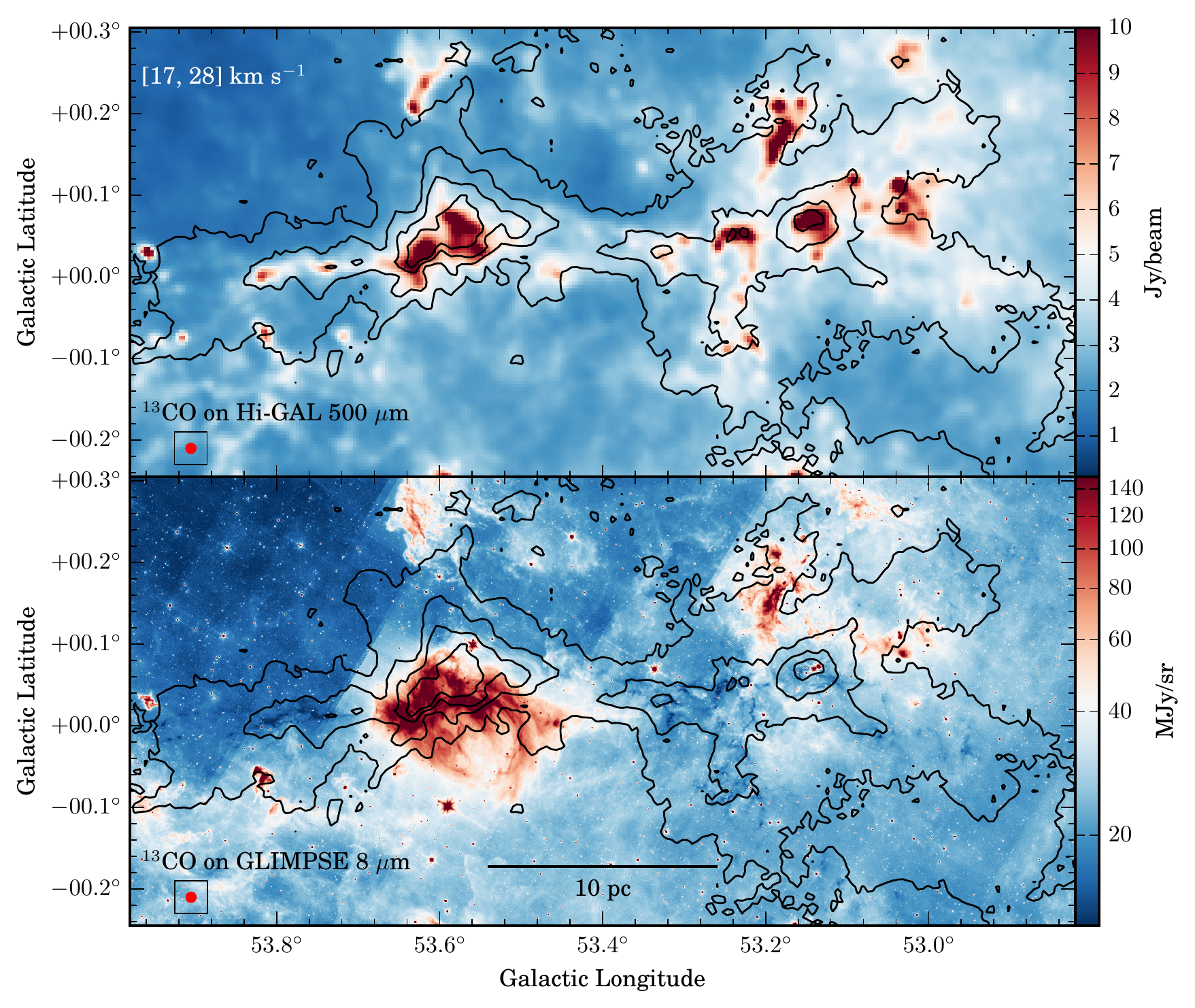}
      \caption{Integrated intensity contours of the GRS $^{13}$CO(1--0) emission (17--28~km~s$^{-1}$, \citealt{Jackson2006}) overlaid on the Hi-GAL 500~$\mu$m continuum map at an angular resolution of 36\arcsec\ (\citealt{Molinari2010}, {\it top panel}) and GLIMPSE 8~$\mu$m map (\citealt{Churchwell2009}, {\it bottom panel}) towards the giant molecular filament GMF54. Contour levels are 5, 15, 25, and 35$\sigma$ ($\sigma$=0.75~K~km~s$^{-1}$). The beam of the $^{13}$CO(1--0) emission is marked in the bottom left of each panel.}
         \label{fig_co_infrared}
\end{figure*}

To study this, we selected the giant molecular filament located at $l\sim52-54\deg$ longitude (GMF54, Fig.~\ref{fig_co_infrared}), that was identified by \citet{Ragan2014} in infrared extinction and confirmed using the GRS $^{13}$CO(1--0) data \citep{Jackson2006} as a velocity-coherent filament at a distance of $\sim2$~kpc. With the $^{13}$CO data, \citet{Ragan2014} estimated the total length and mass of GMF54 to be $\sim68$~pc and 6.8$\times10^4$~M$_\odot$, respectively.

As a common tool to study the physical properties of molecular clouds, column density probability density functions (N-PDFs) are widely used both in observational \citep[e.g.,][]{Lombardi2008, Kainulainen2009, AlvesdeOliveira2014, Sadavoy2014, Abreu-Vicente2015, Stutz2015, Schneider2015, Lin2017} and theoretical studies \citep[e.g.,][]{Ostriker2001, Federrath2010, Federrath2013, Burkhart2017, Chen2018, Koertgen2019}. The shape of the N-PDF depends on the physical processes dominating the cloud and can thus be used to study the evolution of the molecular clouds. From simulations, the early evolutionary stage of a molecular cloud is dominated by turbulence and the N-PDF shows a log-normal shape. The width of the log-normal N-PDF is determined by the turbulent motions \citep[see e.g.,][]{Federrath2010, Ballesteros-Paredes2011, Kritsuk2011, Federrath2013, Burkhart2015B, Bialy2017, Koertgen2019}. As the cloud evolves to a gravity dominated system, the N-PDF develops a high column density power-law tail with a slope of around --2 \citep{Klessen2000, Girichidis2014}. Observations indicate that star-forming clouds show such tails, providing support to this scenario \citep[e.g.,][]{Kainulainen2009, Schneider2013}. Furthermore, the slope of the power-law N-PDF can be related to evolutionary stages of the clouds as steeper slopes possibly indicating earlier quiescent stages \citep[e.g.,][]{Kritsuk2011, Federrath2013, Ward2014}, while flatter slopes or a second flatter power-law tail, often found in high-mass star-forming regions may indicate feedback effects \citep{Tremblin2014, Schneider2015b}.

N-PDFs are often constructed using dust extinction or continuum emission observations \citep[e.g,][]{Lombardi2008, Kainulainen2009, Sadavoy2014, Stutz2015, Schneider2015b}. However, since GMFs lie close to the Galactic mid-plane, heavy contamination from foreground and background emission makes it very difficult to study N-PDFs with continuum emission as it traces the entire line of sight through the Milky Way disk. With the velocity information of molecular line observations, we can easily distinguish different clouds along the line of sight. So far, a few studies investigated the molecular cloud properties with N-PDFs obtained from low J transitions of $^{12}$CO or $^{13}$CO observations \citep{Goldsmith2008, Wong2008, Goodman2009, Carlhoff2013, Schneider2015b, Sun2020}, which can not trace the column density of dense cores due to high optical depth. \citet{Schneider2016} found that compared to dust N-PDF, N-PDFs of the dense gas tracers N$_2$H$^+$(1--0) and CS(2--1) can recover the power-law tail well independent of the exact column-density.

%--------------------------------------------------------------------
\section{Observation and data reduction}
\subsection{Dense gas tracers with IRAM 30~m}
The giant filament GMF54 was observed with the IRAM~30~m telescope in September 2017 and November 2017 with the EMIR receiver and the Fourier Transform Spectrometer (FTS) backends at 3~mm. The receiver was tuned to center at 91.4~GHz with dual-polarisation to cover the dense gas tracers HCO$^+$(1--0), HCN(1--0), HNC(1--0) and their $^{13}$C isotopologues, including the cold and dense gas tracer N$_2$H$^+$(1-0). The FTS backends were used to cover the 16~GHz bandwidth of the receiver with a uniform spectral resolution of 200~kHz, which results in a velocity resolution of $\sim0.7$~km~s$^{-1}$ at 3~mm. The observations were carried out in the on-the-fly (OTF) mode employing position switching to an OFF position sufficiently far away. The filament was scanned at least once along the Galactic longitude and latitude directions, respectively, in order to reduce scanning effects. The system temperature and other calibration parameters were measured every 10 to 15~min. Jupiter and standard mm calibrators were observed regularly to calibrate the pointing and focus. During the observation, the weather condition in general was good and the radiometer opacity $\tau$ at 225 was measured around 0.5, except for two OTF maps observed on 9th September which have a precipitable water vapor (PWV) larger than 40~mm, the data was therefore dropped and the area was re-observed later. The spectra were calibrated with CLASS, which is part of the GILDAS software package\footnote{\url{http://www.iram.fr/IRAMFR/GILDAS}}. We converted the data from antenna temperature ($T_{\rm A}$) to main beam brightness temperature ($T_{\rm mb}$) using a forward efficiency ($F_{\rm eff}$ = 95\% at 3~mm) and main beam efficiency ($B_{\rm eff}$ = 81\% at 3~mm) following equation $T_{\rm mb}=T_{\rm A}\times F_{\rm eff}/B_{\rm eff}$. A linear function was fitted to the line free channels of the spectra to subtract the baseline. The final 3~mm data were all smoothed to a common spectral resolution of 0.7~km~s$^{-1}$ and a beam size of 32\arcsec ($\sim$0.32~pc at 2~kpc distance), and the rms noise level of $T_{\rm mb}$ measured at line free channels is around 0.05~K.

\subsection{$^{13}$CO(1--0)}
The $^{13}$CO(1--0) data are from the Galactic Ring Survey (GRS, \citealt{Jackson2006}), with a spectral resolution of 0.21~km~s$^{-1}$ and a FWHM beam size of 46\arcsec. After converting the antenna temperature ($T_{\rm A}$) to main beam brightness temperature ($T_{\rm mb}$) using a main beam efficiency of 0.48 \citep{Jackson2006}, the line free channel rms noise level of $T_{\rm mb}$ is around 0.22~K.

\subsection{Dust continuum data}
The dust temperature and column density maps were produced by \citet{Zucker2018} by performing pixel-by-pixel modified blackbody fits to the Hi-GAL 160, 250, 350, and 500~$\mu$m continuum map using the HiGal\_SEDfitter\footnote{\url{https://github.com/keflavich/HiGal_SEDfitter}} code \citep{Wang2015} with a fixed dust opacity of $\beta=1.75$ and a gas-to-dust ratio of 100. The 70~$\mu$m band was excluded from the fitting due to its high optical depth towards very dense region and to better tracing cold gas. We subtract a constant column density value (3.46$\times10^{21}$~cm$^{-2}$) measured from a 1\arcmin radius circular area centered at $l=53.69\degr$ and $b=-0.30\degr$ to remove the line-of-sight contamination \citep{Schneider2015b, Ossenkopf2016}.

\section{Results}
\subsection{Distribution of the dense gas}
\label{sect_gas}

\begin{table}
\caption[]{Observed tracers and their corresponding critical and effective densities.}
\label{tab_lines}
\centering 
\begin{tabular}{l c c c}
\hline\hline 
Tracer & $E_u/{\rm k}$\tablefootmark{a} & $n_{\rm crit}$\tablefootmark{b} & $n_{\rm eff}$\tablefootmark{c}\\
         & (K) & (cm$^{-3}$) & (cm$^{-3}$)\\
\hline                  
$^{13}$CO(1--0) & 5.3 &720. &--\\
 HCO$^+$(1--0) & 4.3 & 6.8$\times10^4$& 9.5$\times10^2$\\
 H$^{13}$CO$^+$(1--0) & 4.2 & 6.2$\times10^4$& 3.9$\times10^4$\\
 HCN(1--0) &4.3& 4.7$\times10^5$ &8.4$\times10^3$\\
 %H$^{13}$CN(1--0) &4.1& 5.3$\times10^5$ &3.5$\times10^5$\\
 HNC(1--0) &4.4& 1.4$\times10^5$ & 3.7$\times10^3$\\
 HN$^{13}$C(1--0) &4.2& 9.6$\times10^4$ & --\\
 N$_2$H$^+$(1--0) &4.5 &6.1$\times10^4$ &1.0$\times10^4$\\
\hline                                  
\end{tabular}
\tablefoot{
\tablefoottext{a}{$E_u/{\rm k}$ are from the Cologne Database for Molecular Spectroscopy (CDMS, \citealt{Mueller2001, Mueller2005, Endres2016}).}
\tablefoottext{b}{$n_{\rm crit}$ are from \citet{Shirley2015} assuming $T_{\rm k}$=10~K, except for $^{13}$CO(1--0) and HN$^{13}$C(1--0), which were calculated following the method described by \citet{Shirley2015}. We also assume HN$^{13}$C(1--0) shares the same collision rate coefficients as HNC. The relevant data used for the calculation are from \citet{Yang2010} and \citet{Dumouchel2010}, and the {\it Python} script can be found at \url{https://github.com/ZhiyuZhang/critical_densities}.}
\tablefoottext{c}{$n_{\rm eff}$ assuming $T_{\rm k}$=10~K are from \citet{Shirley2015}.}
}
\end{table}

\begin{figure*}
\centering
\includegraphics[height=14cm]{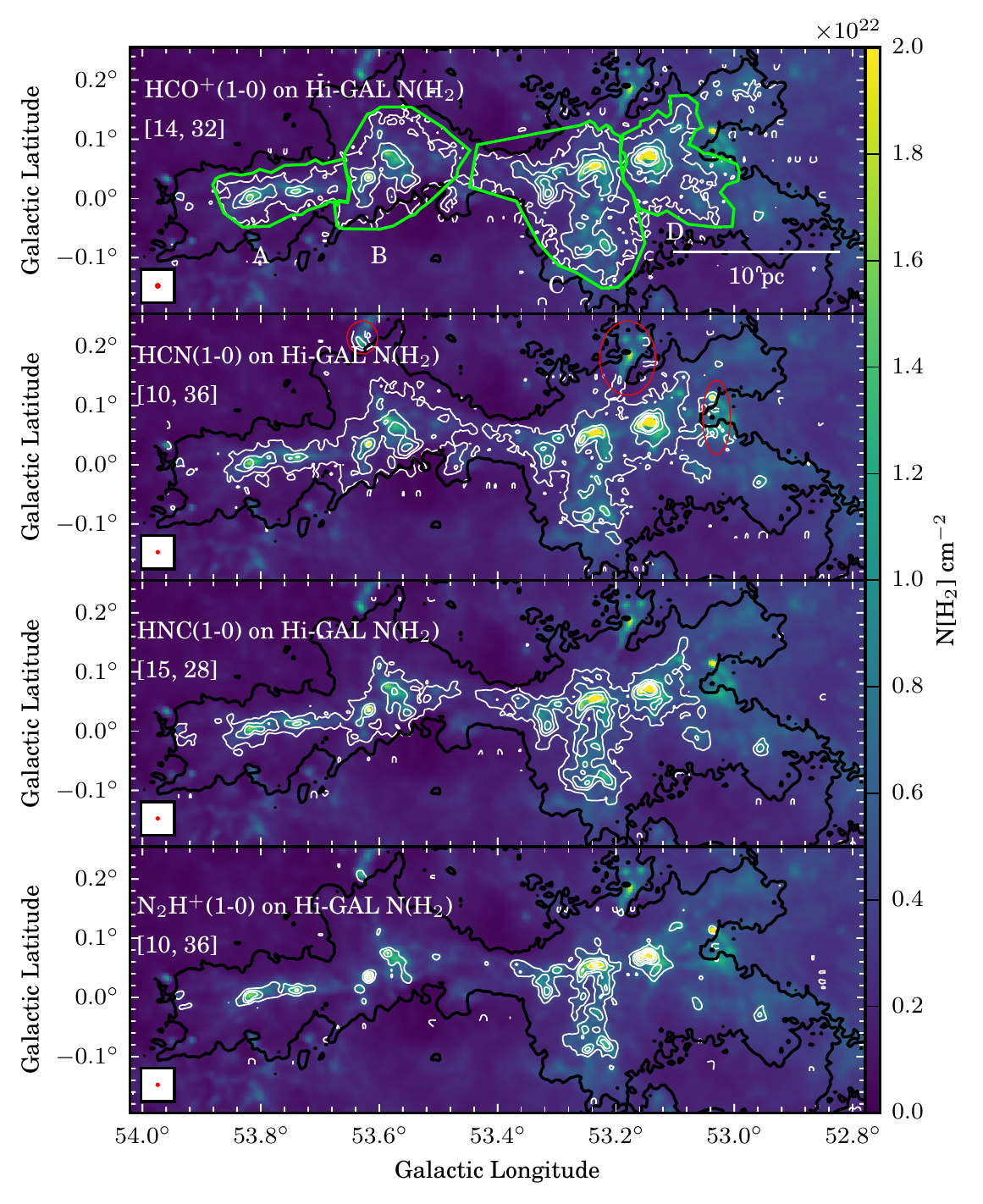} %for double col mode
\includegraphics[height=13.8cm]{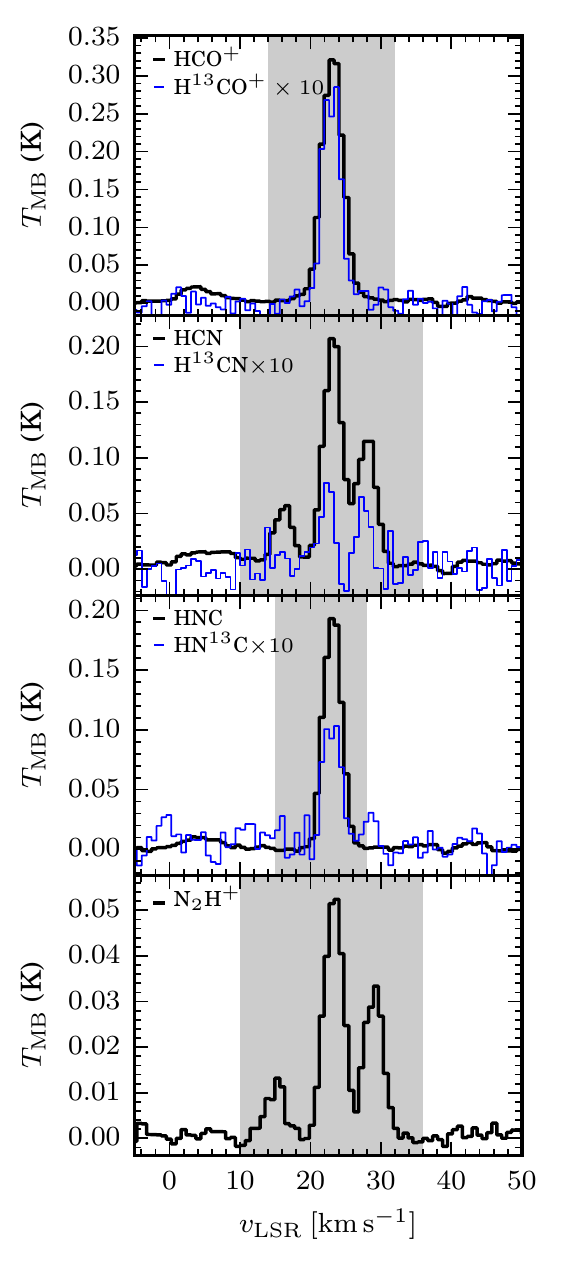}
\caption{{Left panels:} Integrated intensity contours of the HCO$^+$(1--0), HCN(1--0), HNC(1--0) and N$_2$H$^+$(1--0) overlaid on the molecular hydrogen column density map derived from fitting Hi-GAL 160, 250, 350, and 500~$\mu m$ continuum map \citep{Zucker2018}. Contour levels are 3, 9, 15, 25, and 35 times of the respective $\sigma$ of each integrated intensity map. The noise level of each panel from top to bottom is 0.28, 0.40, 0.22 and 0.29~K~km~s$^{-1}$, respectively. The integral velocity range for each line is labeled in each panel in the unit of km~s$^{-1}$. The polygons in the HCO$^+$(1--0) panel outline the four sub-regions, A, B, C, and D, that we studied in Sect.~\ref{sect_npdf}. The thick black contour in each panel traces the 5$\sigma$ level of the GRS $^{13}$CO(1--0) integrated intensity. The red ellipses mark the emissions that are at a different distance and not associated with the filament. The beam of the molecular line emission (32\arcsec) is marked in the bottom left of each panel. {Right panels:} The average spectra of the dense gas tracers across the filament. The thick black spectra are HCO$^+$(1--0), HCN(1--0), HNC(1--0), and N$_2$H$^{+}$(1--0) line, respectively. The thin blue spectra are the $^{13}$C isotopologues that are scaled up by a factor of 10. The shades mark the integral velocity range for each line.}
\label{fig_higal_lines}
\end{figure*}

While the $^{13}$CO(1--0) map (Fig.~\ref{fig_co_infrared}) shows the extended nature of the giant filament GMF54, the integrated intensity maps of HCO$^+$, HCN, HNC, and N$_2$H$^+$ in Fig.~\ref{fig_higal_lines} show that these dense gas spectral lines trace the extended structure of the filament although a much smaller area is detected compared to the $^{13}$CO map. The integral velocity range for each tracer was determined to include all emissions associated with the filament. Since we want to include the emission of the all hyper-fine structures, a larger integrated velocity range (as shown in Fig.~\ref{fig_higal_lines}) is adopted for HCN(1--0) and N$_2$H$^+$(1--0), which also includes some emission at velocities $\gtrsim$35~km~s$^{-1}$ or $\lesssim$10~km~s$^{-1}$. These emissions that lie in regions marked with red ellipses in Fig.~\ref{fig_higal_lines} are at different distance and unrelated to the filament, so we exclude them from the further discussion. As the spectra in Fig.~\ref{fig_higal_lines} show, the filament is dominated by one component centered at 20~km~s$^{-1}$. The contamination from the other two components (peaked at $\sim5$~km~s$^{-1}$ and $\sim36$~km~s$^{-1}$, respectively) is negligible.

To quantify how much extended structure is traced by different molecular lines, we calculate the area filling factor of the dense gas integrated intensity map relative to the $^{13}$CO(1--0) map. We measured the area of each molecular line map where S/N$>$3, and divided it by the $^{13}$CO map area to get the filling factor, and the results are listed in Table~\ref{tab_filling}. Comparing to the $^{13}$CO emission, these typical dense gas tracers cover between 6\% and 28\% of the area of the $^{13}$CO emission, depending on the tracer (Table~\ref{tab_filling}). If we smooth and regrid our 3~mm integrated intensity maps into the same angular resolution of 46\arcsec and pixel size of 22\arcsec as the $^{13}$CO data, the filling factors of these dense gas tracers increases by $\sim70$\% (see Table~\ref{tab_filling}). The filling factors should be taken in to account for extragalactic studies. 

We list the upper energy level, the critical density ($n_{\rm crit}$) under the simplifying assumptions of two-levels systems and optically thin emission, and the effective excitation density ($n_{\rm eff}$) of the lines we used in Table~\ref{tab_lines}.  Recent studies show that these molecular lines trace much lower densities than their critical densities \citep[e.g.,][]{Kauffmann2017, Pety2017}. Compared to $n_{\rm crit}$, $n_{\rm eff}$ gives a better estimate of the approximate density at which a modest (1~K~km~s$^{-1}$) molecular line can be observed \citep{Shirley2015}. Among these four dense gas tracers (HCO$^+$, HCN, HNC, and N$_2$H$^+$), we observe that HCO$^+$(1--0) has the smallest $n_{\rm eff}$, and also traces the most extended structure, while N$_2$H$^+$(1--0) has the highest $n_{\rm eff}$ and traces as expected the most compact structure. On the other hand, HCN(1--0) has both a higher $n_{\rm eff}$ and a higher $n_{\rm crit}$ than HNC(1--0), yet traces more extended structure than HNC(1--0). \citet{Kauffmann2017} also found that HCN(1--0) typically traces gas at a density of $\sim870$~cm$^{-3}$, while N$_2$H$^+$(1--0) traces a much higher density of $\sim4\times10^3$~cm$^{-3}$.

\begin{table}
\caption[]{Emission area and filling factor.}
\label{tab_filling}
\centering 
\begin{tabular}{l r r}
\hline\hline 
Molecule & Area\tablefootmark{*} & Filling factor\tablefootmark{*}\\
         & (arcmin$^2$) & relative to $^{13}$CO \\
\hline                  
$^{13}$CO & 1186.1&1 \\
 HCO$^+$ & 330.8 (534.7) & 0.28 (0.45) \\
 HCN &251.4 (421.3)& 0.21 (0.36)\\
 HNC &207.1 (304.5)& 0.17 (0.26)\\
 N$_2$H$^+$ &75.6 (130.8)& 0.06 (0.11)\\
\hline                                  
\end{tabular}
\tablefoot{
\tablefoottext{*}{The area and filling factors in brackets are measured with the data that are smoothed and regridded to the same angular resolution of 46\arcsec and pixel size 22\arcsec as the $^{13}$CO data.}
}
\end{table}

\begin{figure*}
\centering
\includegraphics[width=0.49\hsize]{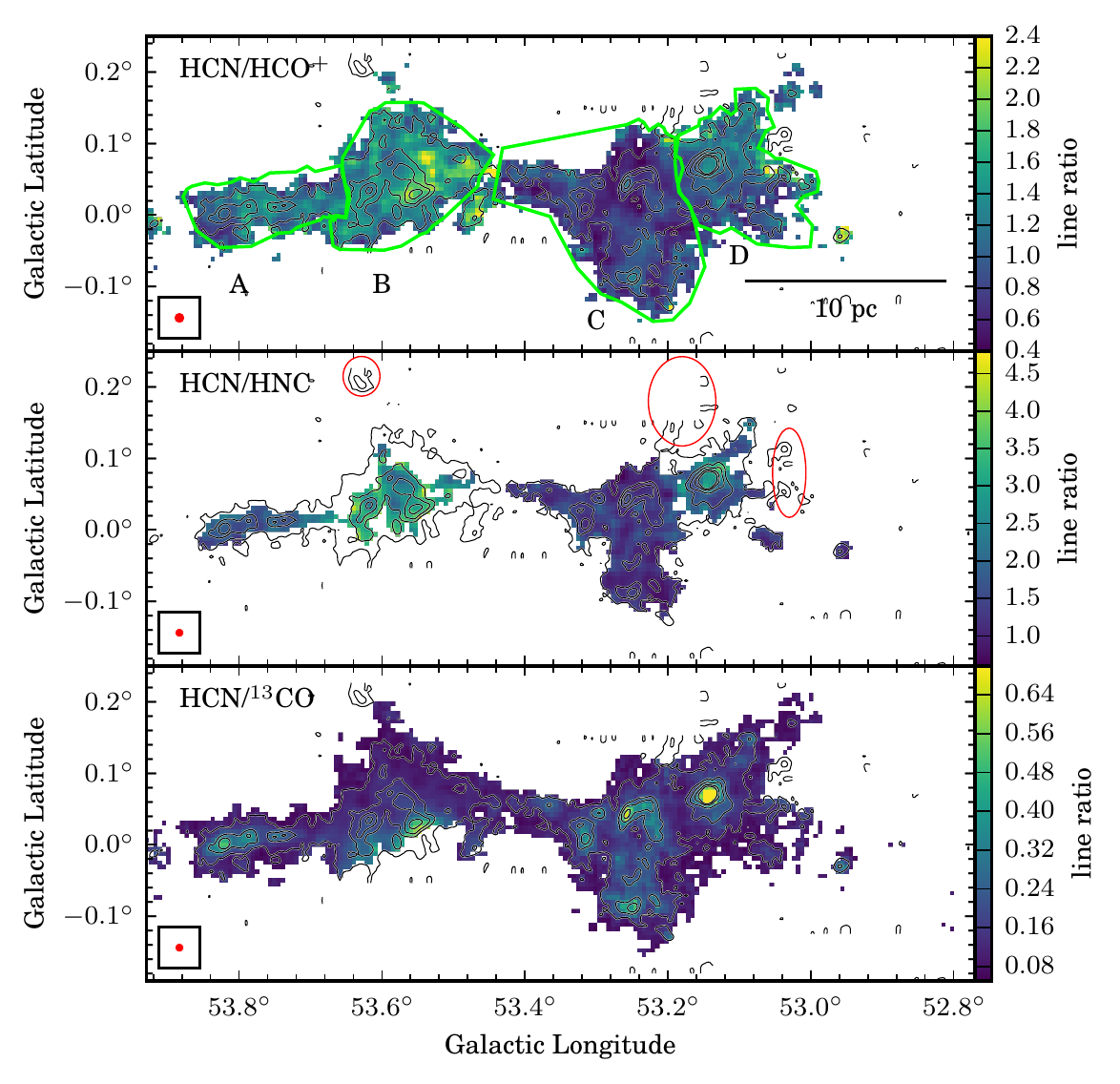}
\includegraphics[width=0.49\hsize]{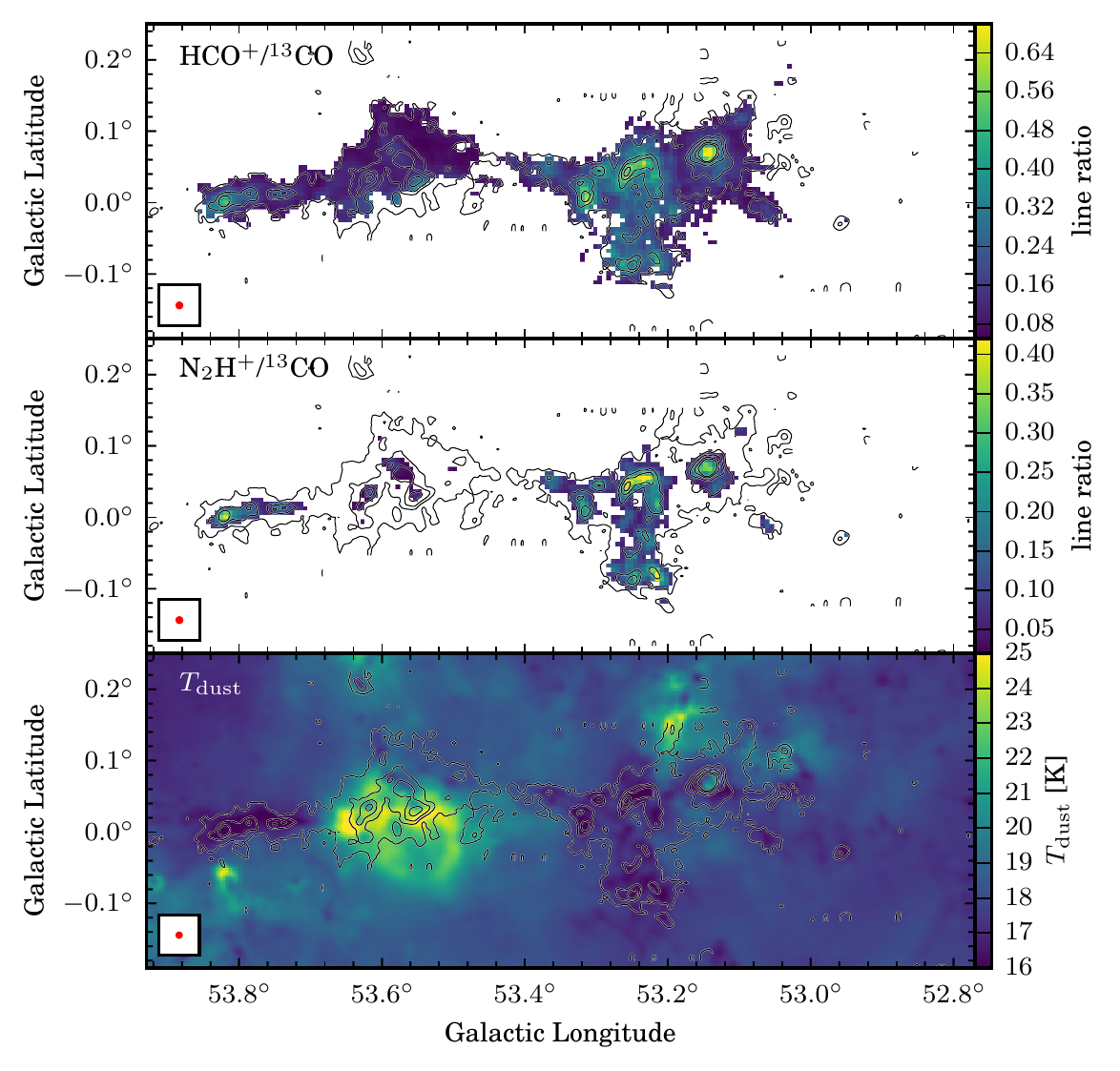}
\caption{Spatial distribution of the ratio of the line integrated intensity over [10, 36]~km~s$^{-1}$ and the dust temperature (bottom right panel). The contours in all panels trace the HCN(1--0) integrated intensity, and the contour levels are 3, 9, 15, 25, and 35 times of the noise level (0.4~K~km~s$^{-1}$). The beam of the line ratio maps (46\arcsec) is marked in the bottom left of each panel. The polygons and ellipses are the same as in Fig.~\ref{fig_higal_lines}.}
\label{fig_ratio_map}
\end{figure*}

The ratios of the molecular line integrated intensities are often used to trace different physical properties of the gas. The HCN/CO integrated intensity ratio is commonly used to trace the dense gas fraction in galactic and extragalactic studies \citep[e.g.,][]{Gao2004, Lada2012}, and HCN/HNC ratio is considered to be able to trace the evolutionary stages of the molecular cloud \citep{Schilke1992, Graninger2014, Hacar2019}. To compare the intensity of different lines properly, we first integrated all four dense gas tracers and the $^{13}$CO line in the velocity range $10<v_{\rm LSR}<36$~km~s$^{-1}$. We then convolved and regridded all dense gas maps to the same resolution as the $^{13}$CO one. We divided the HCN integrated intensity map by the HCO$^+$, HNC, N$_2$H$^+$ and $^{13}$CO map to obtain the line ratio maps shown in Fig.~\ref{fig_ratio_map}.

To study the dense gas properties in different evolutionary stages, we divided the filament into four regions based on the molecular line and dust emission maps. The border between Regions A and B is defined by where the narrow filament connects to the extended PDR (the narrowest part in HNC and Hi-GAL maps). Regions B and C are separated by the 3$\sigma$ closed contours in the HCO$^+$ and HCN maps. Regions C and D are separated by the narrowest part between the two strongest peaks in this region in the HNC map. All polygons are further extended to include all emission above 3$\sigma$.

\begin{figure*}
\centering
\includegraphics[height=10.2cm]{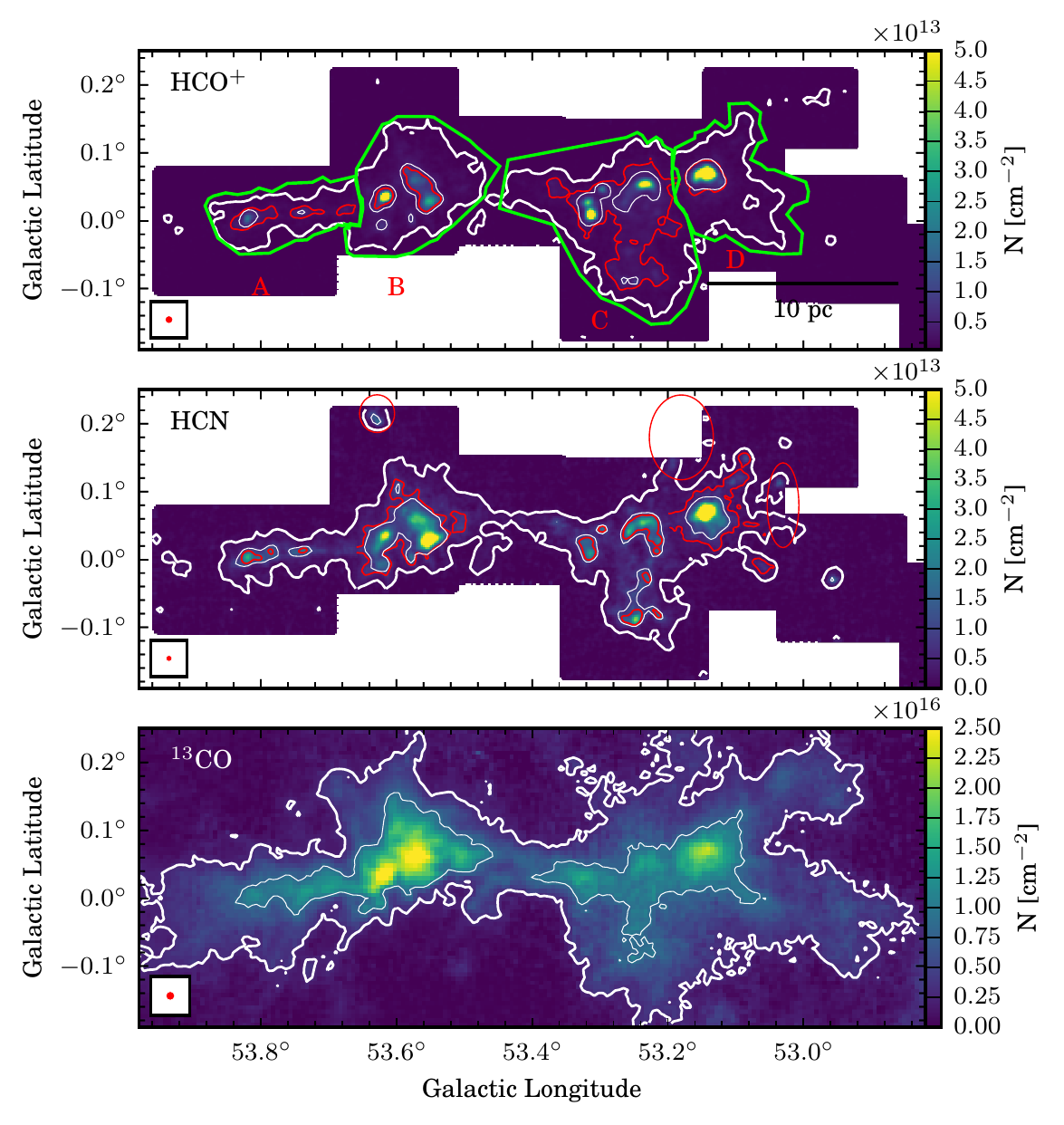}  %for double col mode
\includegraphics[height=10.2cm]{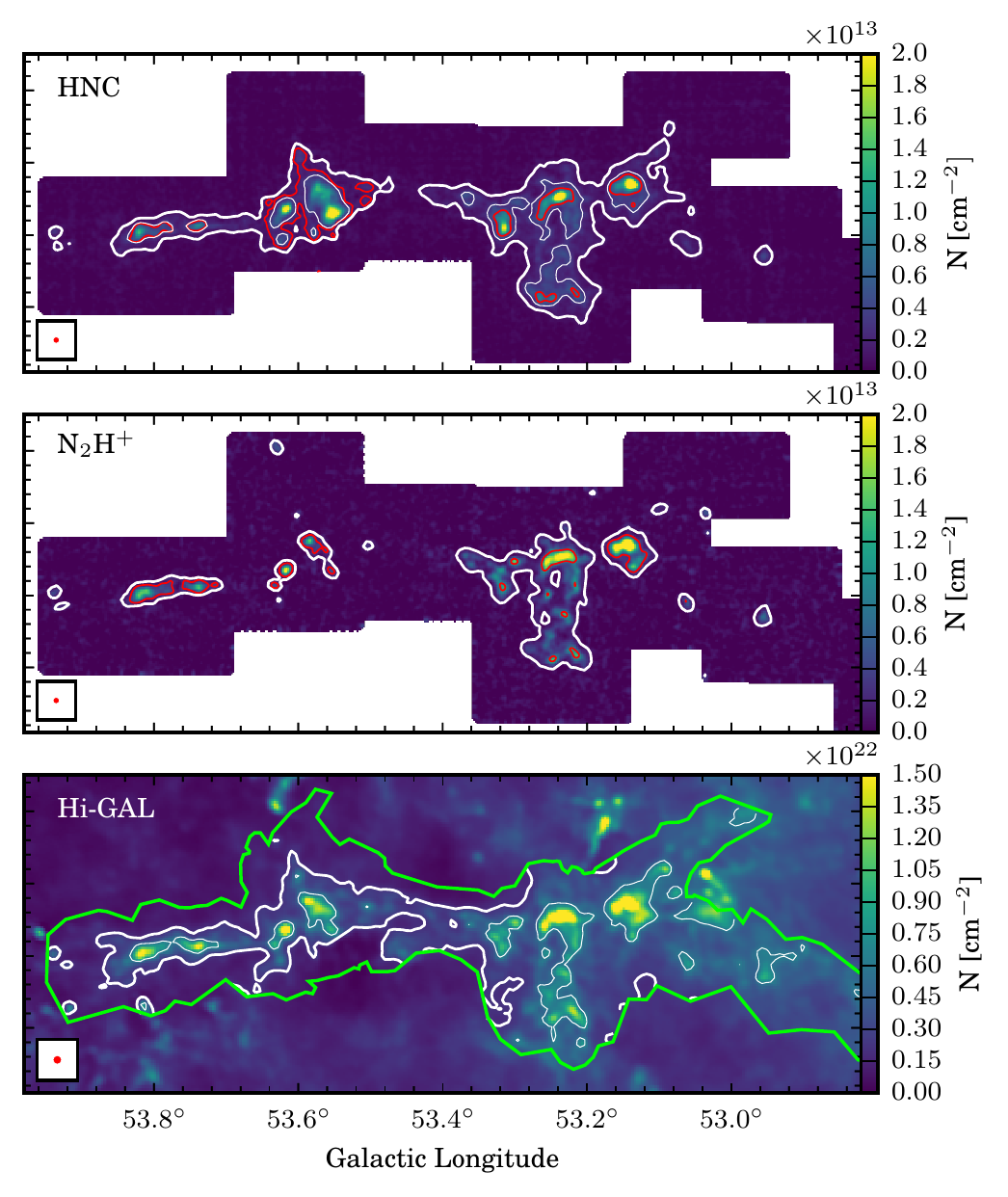}
\caption{Column density maps of HCO$^+$, HCN, $^{13}$CO, HNC, and N$_2$H$^+$, and molecular hydrogen column density map derived from Hi-GAL dust continuum emission. The thick white contours in each panel outline the column densities we used to construct the N-PDFs, and the thin white contours mark the optimal column density threshold ($N_{\rm min}$ for the optimal power-law fit) for the whole filament. The red contours in the dense gas molecular line panels mark $N_{\rm min}$ for the power-law for the sub-regions (see Sect.~\ref{sect_npdf}). The green polygon in the Hi-GAL panel outlines the area we used to construct the N-PDFs from the dense gas tracers and continuum emission in the following sections. The beam of the column density maps (43\arcsec\ for Hi-GAL, 46\arcsec\ for $^{13}$CO, and 32\arcsec\ for the the other panels) is marked in the bottom left of each panel. The polygons in the HCO$^+$ panel and ellipses in the HCN panel are the same as in Fig.~\ref{fig_higal_lines}.}
\label{fig_col_map}
\end{figure*}

As shown in Figs.~\ref{fig_higal_lines} and \ref{fig_ratio_map}, the four sub-regions are: Region A, infrared dark filament dominated; Region B, infrared bright \ion{H}{ii} regions with PDRs \citep{Urquhart2011, Anderson2014}; Region C, infrared dark filament dominated; Region D, high-mass star forming region with a hypercompact (HC) \ion{H}{ii} region, methanol maser, and water maser \citep{Pandian2007, Pandian2009, Sanchez-Monge2011, Urquhart2011, Anderson2014}. Therefore, we propose an evolutionary sequence among these four sub-regions, that Region A and Region C are the youngest, Region D is more evolved, Region B with PDRs is the most evolved region. The mass of the four subregions derived from the Hi-GAL column density map (Fig.~\ref{fig_col_map}) are 1.6, 2.6, 6.1, and 4.4$\times10^3~M_\odot$, respectively. Considering there are only a few dense cores in Regions A and C, with such a large mass reservoir, it is likely that Regions A and C will form massive stars. Regions A and C were also classified as massive star forming regions by a previous study \citep{Dunham2011}.

As shown in Fig.~\ref{fig_ratio_map} where we overlaid the HCN integrated intensity contours on line ratio maps and Hi-GAL dust temperature map \citep{Marsh2017}, compared to HCO$^+$ and HNC, the relative intensity of HCN increases dramatically in regions with high temperature (i.e., Regions B and D). For instance, the HCN/HNC ratio remains relative constant ($\lesssim2$) in Regions A and C, where the dust temperature is $T_{\rm dust}\lesssim18$~K, while in Regions B and D the dust temperature rises to $T_{\rm dust}\gtrsim20$~K and the HCN/HNC ratio also increases to $\gtrsim3$. Similar trends can also be seen in the HCN/HCO$^+$ ratio map, where the HCN/HCO$^+$ ratio is approximately a factor of 3 higher in Region B than in Region C.

\subsection{Column density}
\label{sect_coldens}
Following the standard method described in other works (e.g., Eq.~15.37 in \citealt{wilson2013}, appendix in \citealt{Caselli2002} and \citealt{Feng2016}),  we estimated the column density of different molecules in this section. 

GMF54, as basically all GMFs and giant molecular clouds (GMCs), are active sites of star-formation with warm, IR-bright areas with embedded sources, and cool, IR-dark areas (Sect.~\ref{sect_gas}). This obviously implies a complex temperature structure which has an influence on the determination of the column density. In addition, abundance variations can arise from chemistry and from the radiation field. If the local thermal equilibrium (LTE) assumption is valid and the density is larger than the critical/effective density of the transition, the gas kinetic temperature ($T_{\rm kin}$) corresponds to the excitation temperature ($T_{\rm ex}$) for the molecular lines (which is assumed to be equal for all species). If gas and dust are well mixed, this temperature should also correspond to the dust temperature ($T_{\rm dust}$). However, LTE conditions are not present and the density varies from low values in the diffuse gas phase, mainly seen in $^{12}$CO, to high values in dense clumps in the molecular clouds, traced by dust and the lines of HCO$^+$, HCN, HNC (and their isotopologues), and N$_2$H$^+$. The temperature also varies from a few K inside cold, dense clumps to a few tens of K in UV-heated PDRs. In the following, we determine the excitation temperature and optical depth in different ways, using our observational data sets or literature values.

Assuming that the $^{12}$CO(1--0) is optically thick, \citet{Zhang2019} estimated the excitation temperature $T_{\rm ex}$ for 13 GMFs from $T_{\rm mb}$($^{12}$CO), and they obtained the mean value of $\sim10$~K with a standard deviation of 2.5~K. Since we do not have $^{12}$CO(1--0) observation with compatible angular resolution with GRS for GMF54, we assume $^{13}$CO emission to be optically thin and adopt the mean excitation temperature $T_{\rm ex}$=10~K from \citet{Zhang2019}. We discuss the uncertainty introduced by the possible optical depth of $^{13}$CO(1--0) in Sect.~\ref{sect_uncert}.

For HCN(1--0) and N$_2$H$^+$(1--0), we applied a hyperfine structure (HFS) fit pixel-by-pixel using {\it PySpecKit} \citep{Ginsburg2011} to derive the optical depth $\tau$ and the excitation temperature $T_{\rm ex}$. When fitting the HFS, we set a lower boundary for $T_{\rm ex}$ as 4~K. As the HFS fit only produce $\tau \times T_{\rm ex}$, a lower boundary smaller than 4~K for $T_{\rm ex}$ does not change the results in high S/N ratio regions but produces large $\tau$ (>1) in low S/N ratio regions (along the 3$\sigma$ contour in Fig.~\ref{fig_higal_lines}), which is not physical. Since our maps are over sampled (32\arcsec FWHM beam and 8\arcsec pixel size), we further convolve the $T_{\rm ex}$ map and the $\tau$ map into a gaussian FWHM beam of 32\arcsec (gaussian kernel $\sim$13.59\arcsec). The derived final $T_{\rm ex}$ and $\tau$ maps including a few selected spectra to show the fit results for HCN(1--0) and N$_2$H$^+$(1--0) are shown in Figs.~\ref{fig_tex_tau} and \ref{fig_hfsfit}. The excitation temperatures of HCN(1--0) are between 4 and 74~K with the highest value at the PDR in Region B as we expected. The optical depths of HCN(1--0) are between 0.1 and 3.5 with the highest value in Region C and D. For N$_2$H$^+$(1--0), the derived $T_{\rm ex}$ is between 4 and 29~K and $\tau$ is between 0.1 and 2.4.

For HCO$^+$(1--0) and HNC(1--0), since we only observed one transition for the whole filament, we can not determine the excitation temperatures from our observations directly. Therefore, we assume HCO$^+$(1--0) and HNC(1--0) share the same excitation temperature as HCN(1--0). We discuss the uncertainty of this in Sect.~\ref{sect_uncert}. Our observations also cover the H$^{13}$CO$^+$(1--0) and HN$^{13}$C(1--0) lines, we assume these lines to be optically thin, and share the similar excitation condition as their $^{12}$C counterpart lines. This allows us to estimate the optical depth for HCO$^+$(1--0) and HNC(1--0). To do this, we estimated the $^{12}$C to $^{13}$C abundance ratio following the relation $[^{12}{\rm C}]/[^{13}{\rm C}]=6.2D_{\rm GC}[{\rm kpc}] +9.0$ \citep{Giannetti2014}. For a Galactocentric distance of $D_{\rm GC} =7.3$~kpc \citep{Ragan2014}, $[^{12}{\rm C}]/[^{13}{\rm C}]\sim54$. For regions where the $^{13}$C line integrated intensity is higher than 5$\sigma$,  we calculated $\tau$ for the $^{12}$C lines following (Eq.~6 in \citealt{Giannetti2014}):
\begin{equation}
R_{12/13}=\frac{\int T_{\rm{MB, 12}}\ dv}{\int T_{\rm{MB, 13}}\ dv}=\frac{1-e^{-\tau}}{1-e^{-\tau/\varphi}},
\label{eq:tex}
\end{equation}
, where $\varphi$ is the $^{12}$C to $^{13}$C relative abundance. We assume the remaining regions where the $^{13}$C line emission is below 5$\sigma$ are optically thin. Since the derived $\tau$ maps are calculated pixel-by-pixel, due to limited signal-to-noise ratio of the data, the $\tau$ value could vary significantly between two neighboring pixels, which is not physically plausible. To avoid this pixelization problem, we further convolved the $\tau$ map into a gaussian FWHM beam of 32\arcsec (gaussian kernel $\sim$13.59\arcsec), which is the beam of our 3~mm data. 

We took into account the optical depth and estimated the column densities of HCO$^+$, HCN, HNC, and N$_2$H$^+$ with the derived $T_{\rm ex}$. For HCO$^+$ and HNC, limited by the S/N ratio of the $^{13}$C line emission (Fig.~\ref{fig_13c}), only a small percentage (8\% and 2\%, respectively) of the column density maps were corrected for optical depth with the factor $\tau/(1-{\rm exp}(-\tau))>1.01$. Since the abundances of these molecules have large uncertainties \citep[e.g., ][]{Gerner2014}, we do not convert the column density of the molecules to molecular hydrogen. Instead, we discuss directly the column density distribution of different molecules, and physical properties of molecular hydrogen traced by these tracers. The column density distributions of these tracers represents the distribution of the molecular cloud at the densities traced by these tracers.

\subsection{Column density probability density function}
\label{sect_npdf}
\begin{figure*}
   \centering
   \includegraphics[width=0.33\hsize]{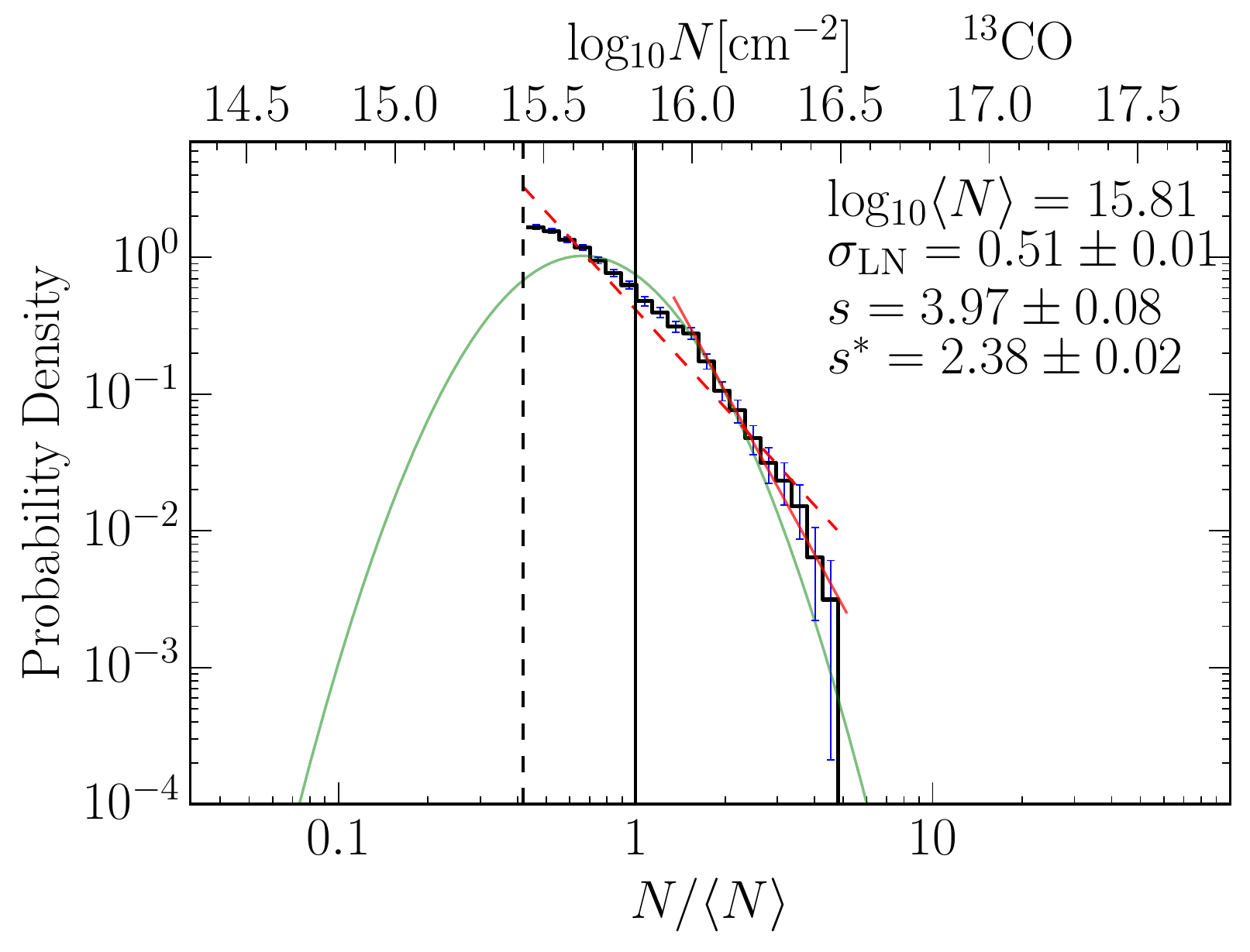}
   \includegraphics[width=0.33\hsize]{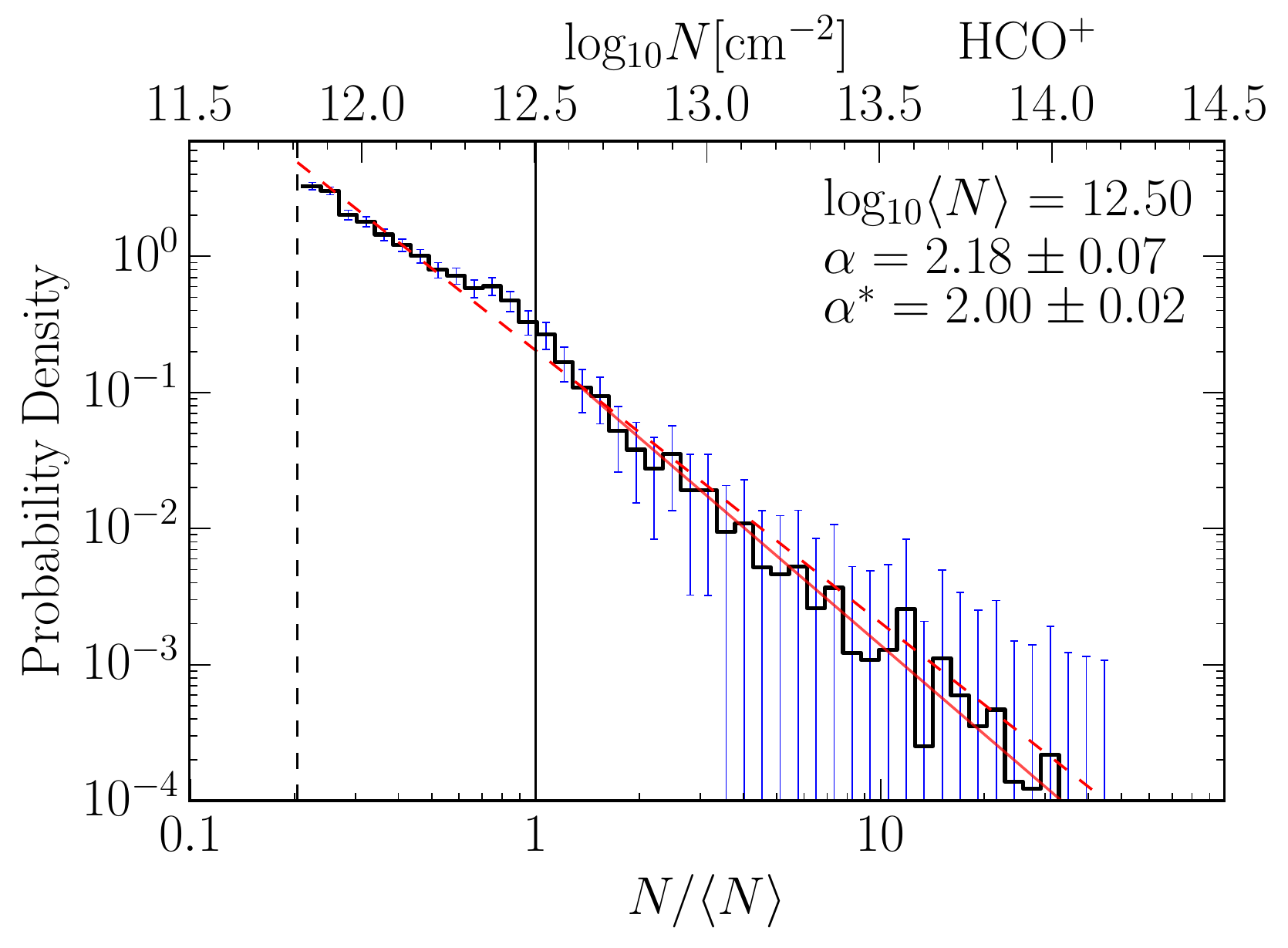}
   \includegraphics[width=0.33\hsize]{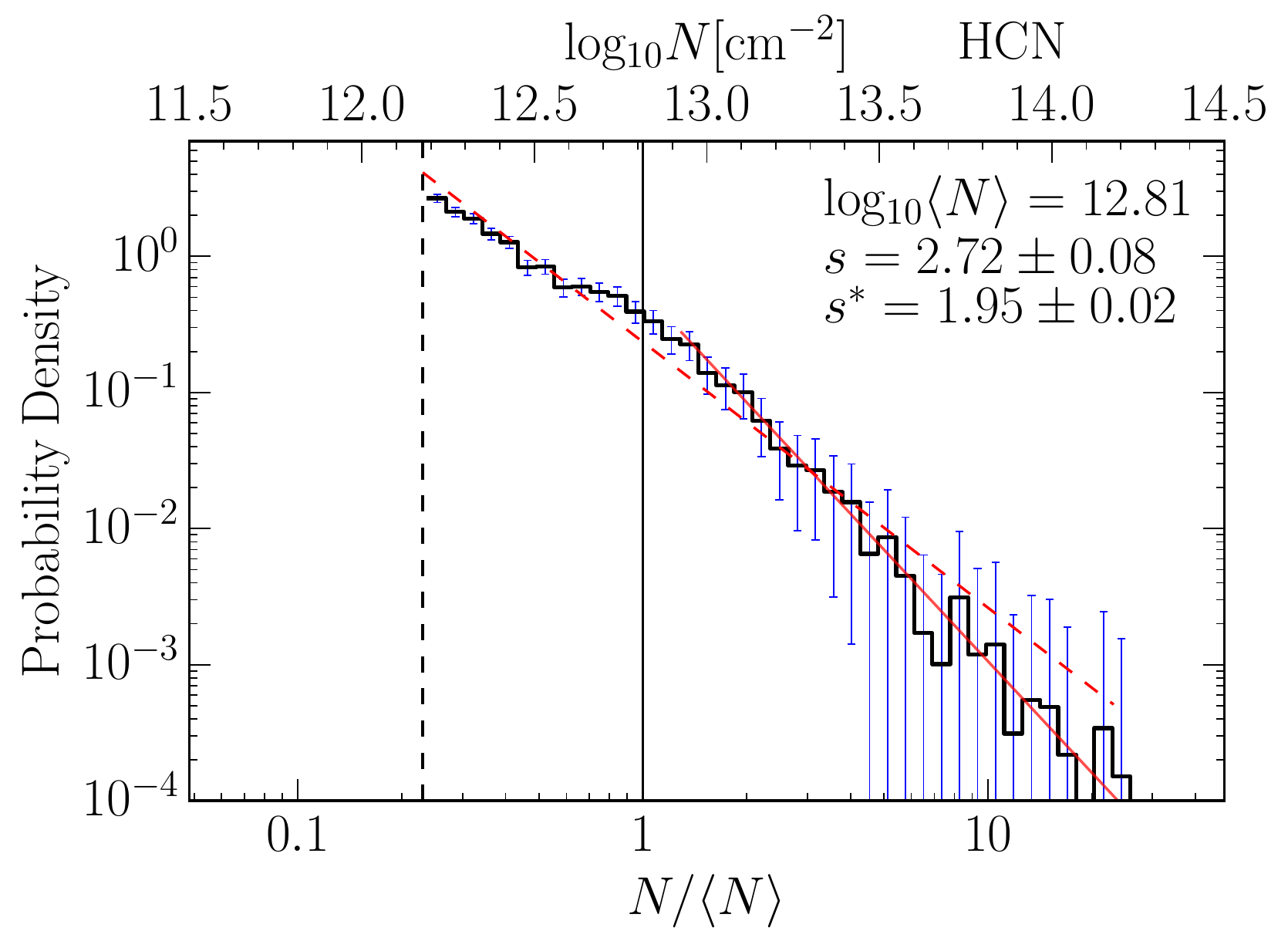}
   \includegraphics[width=0.33\hsize]{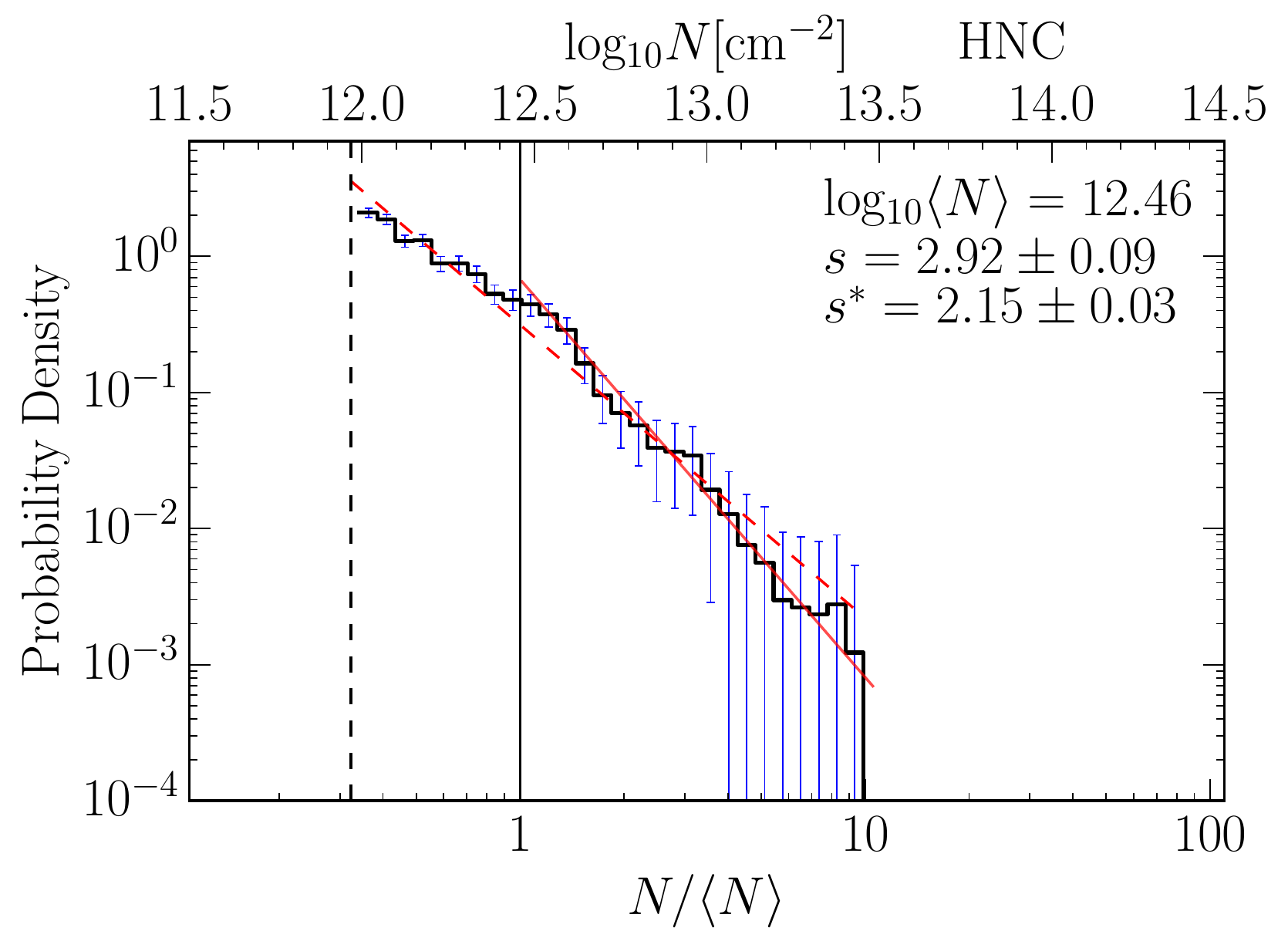}
   \includegraphics[width=0.33\hsize]{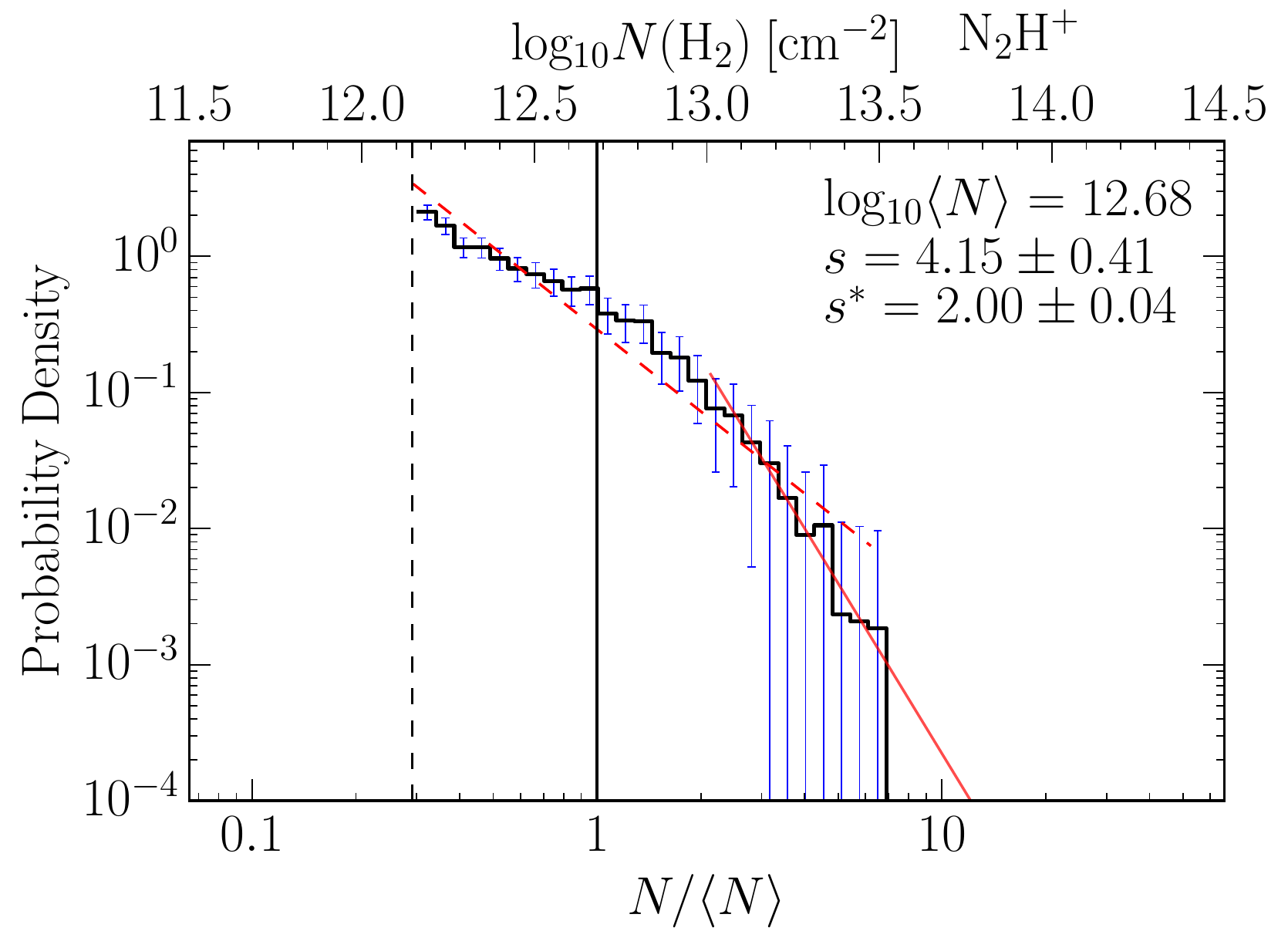}
      \caption{Probability density functions of the $^{13}$CO, HCO$^{+}$, HCN, HNC, and N$_2$H$^+$ column density for the whole filament. The dashed vertical lines in each panel mark the column density threshold (5$\sigma$ level), and the solid vertical lines mark the mean column densities. We list in each panel the mean column density $\langle N\rangle$, the power-law index $s$ (the optimized fit, solid line) and $s^*$ (the fit to all data, dashed line). For the $^{13}$CO N-PDF, we also list the log-normal width $\sigma_{\rm LN}$. Error bars are calculated from Poisson statistics.}
         \label{fig_pdf}
\end{figure*}

We regridded all the column density maps of the dense gas tracers (Fig.~\ref{fig_col_map}) to the same angular resolution of the $^{13}$CO map (FWHM beam size of 46\arcsec and pixel size of 22\arcsec) and constructed the column density probability density functions (N-PDFs) of the filament. Figure~\ref{fig_pdf} presents the N-PDFs for the entire filament traced by different molecular lines. All column density values lower than 5$\sigma$ ($N$ Threshold in Table~\ref{tab_pdf} corresponding to the thick closed contours in Fig.~\ref{fig_col_map}) were ignored when calculating the N-PDFs. We also normalized all N-PDFs by the mean column density of each tracer ($\langle N\rangle$ in Table~\ref{tab_pdf}). We also constructed the N-PDFs of the tracers without the optically depth correction shown in Fig.~\ref{fig_pdf_thin}.

As shown in Fig.~\ref{fig_pdf}, qualitatively the $^{13}$CO N-PDF resembles a log-normal shape, while the N-PDFs of HCO$^+$, HCN, HNC and N$_2$H$^+$ are closer to a broken power-law shape. We fit the N-PDFs with different functions to quantify their shapes. We fit two power-laws ($p(x)\propto x^{-s}$) to all N-PDFs. We first fit the N-PDFs with a optimal power-law to the high column density tail from an optimal column density threshold. The fitting procedure creates power-law fits starting from each unique value in the data, and the optimal column density $N_{\rm min}$ is the one that results in the minimal Kolmogorov-Smirnov (K-S) distance between the data and the fit. The fitting was performed with the python package {\it Powerlaw}\footnote{\url{https://github.com/jeffalstott/powerlaw}} \citep{Alstott2014}. The obtained $N_{\rm min}$ (marked with the thin white contours in Fig.~\ref{fig_col_map}) and power-law index $s$ are listed in Table~\ref{tab_pdf}. Among different tracers, the N-PDF of HCO$^+$ has the lowest power-law index of 2.18, and N$_2$H$^+$ has the highest index of 4.15. We further fit a power-law function which includes all data points to each N-PDF, and all the derived power-law indices are quite similar to each other (between 1.95 and 2.15), which are lower than the ones using the optimal column density threshold $N_{\rm min}$.

We also fit a single log-normal function\footnote{The $lognorm$ function in the Python package $SciPy$ was used to perform the fit (\url{https://docs.scipy.org/doc/scipy/reference/generated/scipy.stats.lognorm.html}).} to $^{13}$CO N-PDF and derive a width of 0.51. The error of the log-normal fitting are estimated with bootstrap method, that we ran the fitting procedure 100 times while removing 10\% of the data points randomly in each run, and we take the difference between the maximum and minimum width as the error of the fit. We cannot identify a clear peak in N-PDF, thus the width we derived could have large uncertainties. A simple model comparison performed with the python package {\it Powerlaw} shows the $^{13}$CO N-PDF can be better described by a log-normal function than a power-law with a loglikelihood ratio of 22 (log-normal over power-law) and $p$-value$<0.001$.

\begin{table*}
\caption[]{Summary of the fitting results of the N-PDFs.}
\label{tab_pdf}
\centering 
\begin{tabular}{l c c c c r}
\hline\hline 
Tracers & $N$ Threshold\tablefootmark{a} & $\langle N\rangle$\tablefootmark{b} & optimal power-law &single power-law&$N_{\rm min}$\tablefootmark{c}\\
    & $\times10^{12}$~cm$^{-2}$& $\times10^{12}$~cm$^{-2}$ & index $s$& index $s^*$ &$\times10^{12}$~cm$^{-2}$\\
\hline                  
\multicolumn{6}{l}{The whole filament}\\
\hline                  
$^{13}$CO & 2700 & 6500  & 3.97$\pm$0.08& 2.36$\pm$0.02& 8694\\
 HCO$^+$ & 0.65 & 3.2 & 2.18$\pm$0.07& 2.00$\pm$0.02& 4.4 \\
 HCN     & 1.5  &6.5 & 2.72$\pm$0.08& 1.95$\pm$0.03& 8.4 \\
 HNC     & 0.93 & 2.9 &2.92$\pm$0.09& 2.15$\pm$0.03& 2.9  \\
 N$_2$H$^+$&1.4& 4.8 & 4.15$\pm$0.41& 2.00$\pm$0.04& 10.3\\
\hline                  
\multicolumn{6}{l}{Region A}\\
\hline                  
 HCO$^+$ &1.3 & 3.0& 3.12$\pm$0.10& --& 2.3 \\
 HCN     &3.2  &8.4 &5.15$\pm$0.39& --& 1.4 \\
 HNC     &1.6 & 3.5 & 3.54$\pm$0.16& --& 3.4  \\
 N$_2$H$^+$&2.8& 6.2& 2.52$\pm$0.09& --& 2.8 \\
\hline                  
\multicolumn{6}{l}{Region B}\\
\hline                  
 HCO$^+$ &1.3 &4.8& 2.61$\pm$0.08& --& 7.0 \\
 HCN     &3.2  &11.0 & 2.42$\pm$0.03& --&6.2 \\
 HNC     &1.6 & 4.8 & 2.19$\pm$0.03& --& 1.6 \\
 N$_2$H$^+$&2.8& 6.2& 2.62$\pm$0.12& --&2.8\\
\hline                  
\multicolumn{6}{l}{Region C}\\
\hline                  
 HCO$^+$ &1.3 & 4.3 & 2.68$\pm$0.03& --& 2.1 \\
 HCN     &3.2  &8.0 &3.76$\pm$0.12& --& 1.2 \\
 HNC     &1.6 & 3.8 & 3.24$\pm$0.11& --& 5.6  \\
 N$_2$H$^+$&2.8& 6.6&3.89$\pm$0.17& --& 9.4\\
\hline                  
\multicolumn{6}{l}{Region D}\\
\hline                  
 HCO$^+$ &1.3 & 8.8 & 1.76$\pm$0.04& --& 3.4 \\
 HCN     &3.2  &1.5 & 2.05$\pm$0.03& --& 3.2 \\
 HNC     &1.6 & 4.6 & 2.41$\pm$0.10& --& 4.4  \\
 N$_2$H$^+$&2.8&9.9 &2.23$\pm$0.07& --& 4.4\\
\hline                                  
\end{tabular}
\tablefoot{
\tablefoottext{a}{Five $\sigma$ threshold for the N-PDFs and the thick white contours in Fig.~\ref{fig_col_map}.}
\tablefoottext{b}{The mean column density above 5$\sigma$ and used to normalize the column density for the N-PDFs.}
\tablefoottext{c}{The optimal column density determined by the program $Powerlaw$.}
} 
\end{table*}

\begin{figure*}
   \centering
   \includegraphics[height=3.4cm]{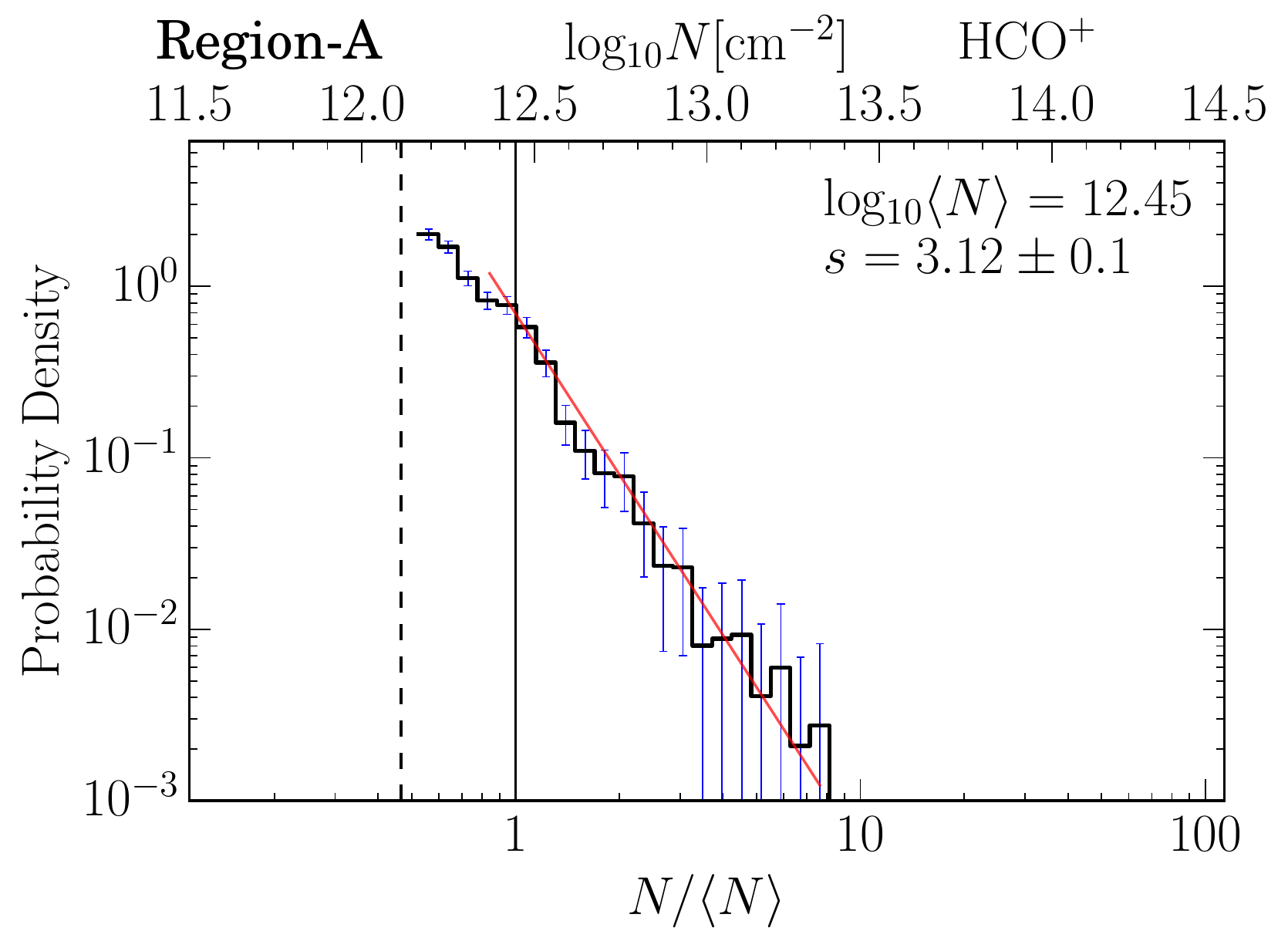}
   \includegraphics[height=3.4cm]{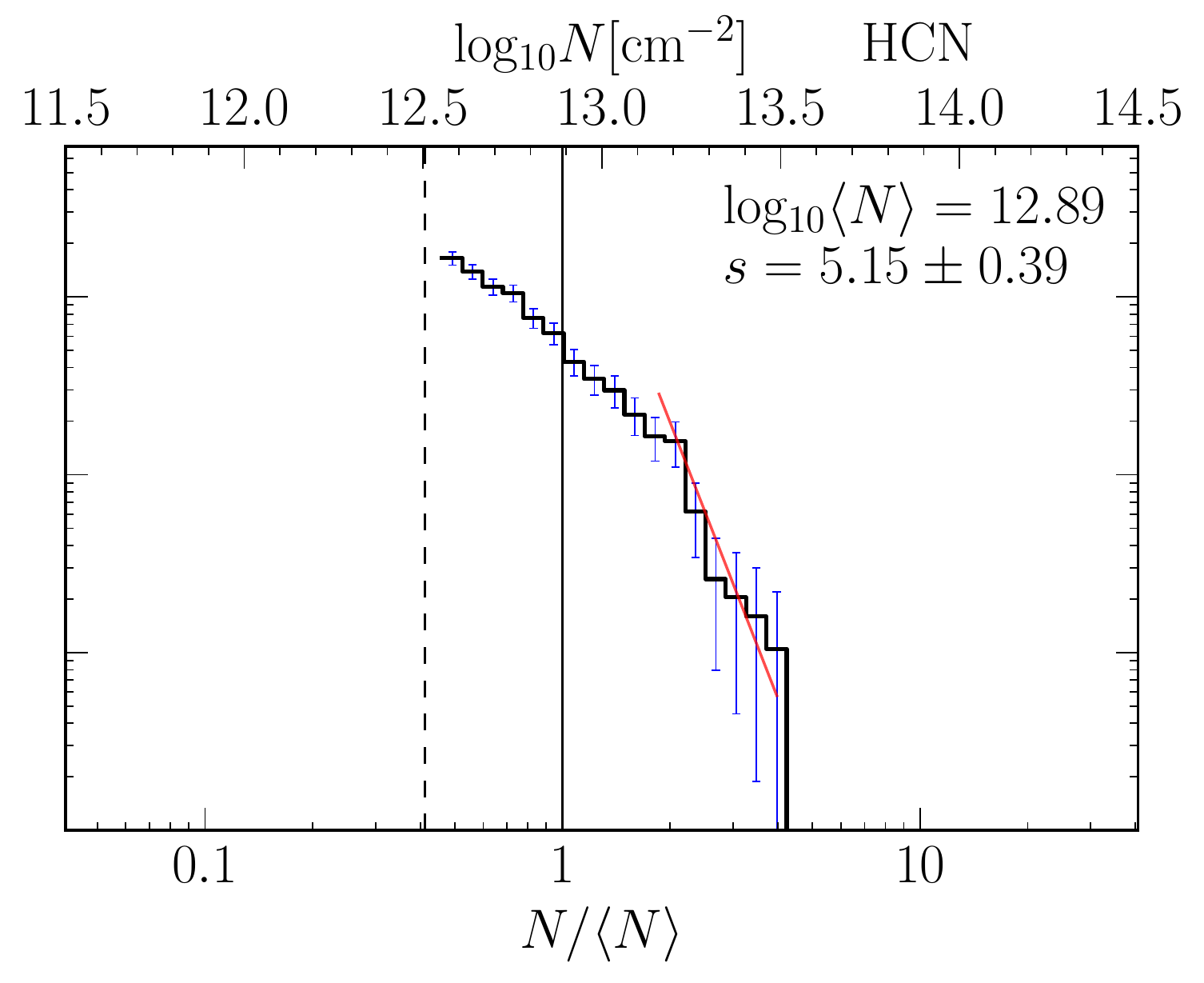}
   \includegraphics[height=3.4cm]{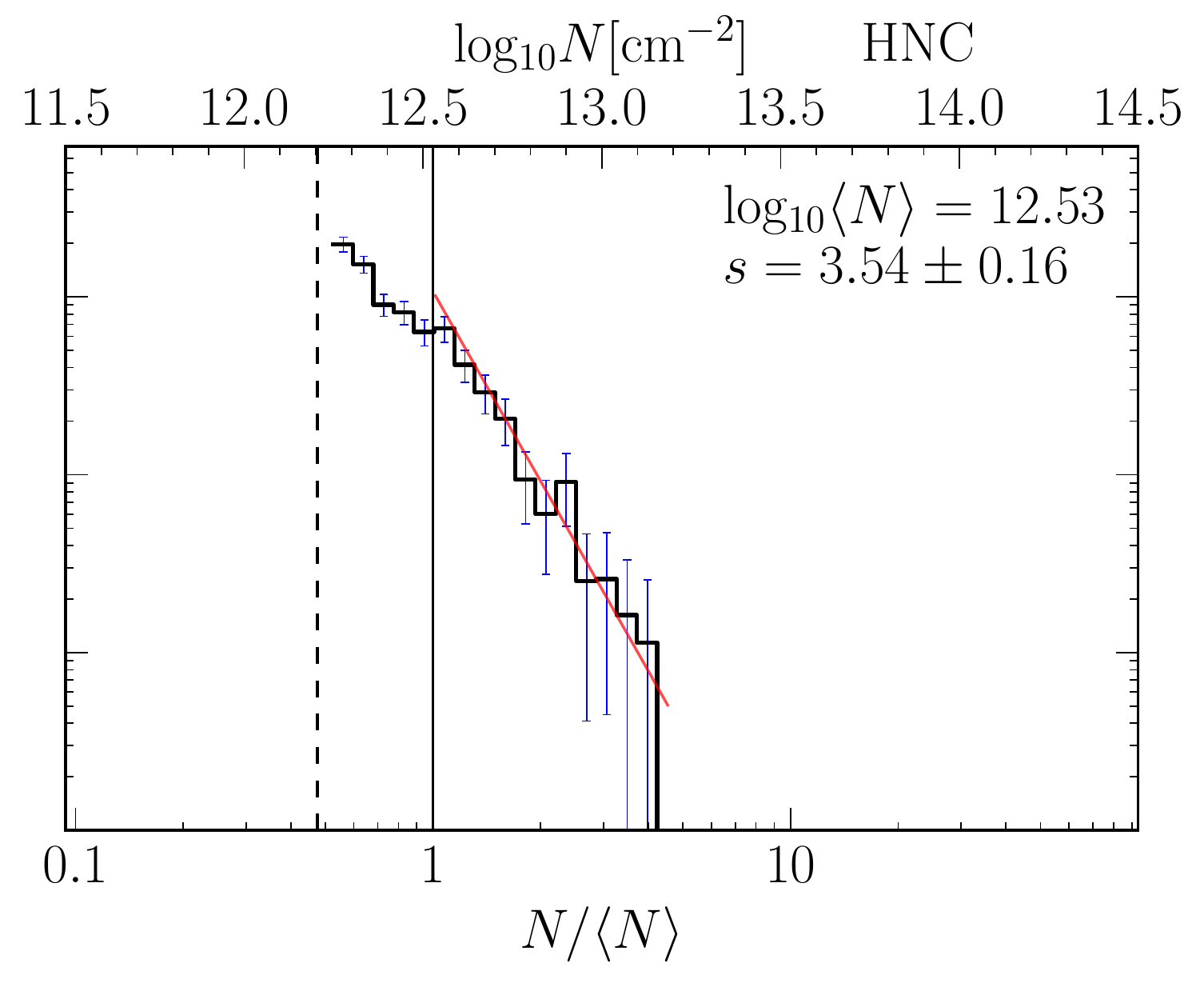}
   \includegraphics[height=3.4cm]{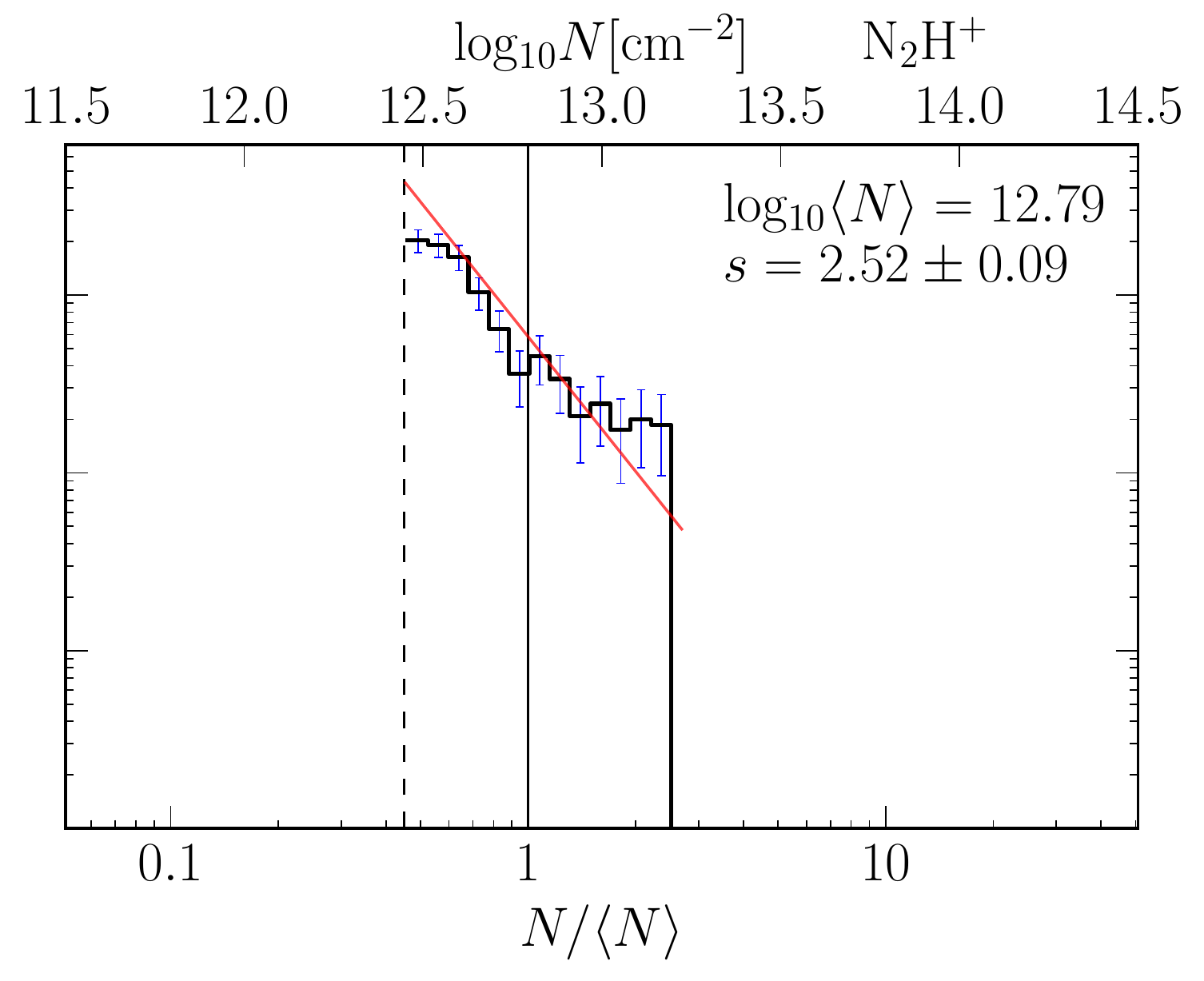}
   \includegraphics[height=3.4cm]{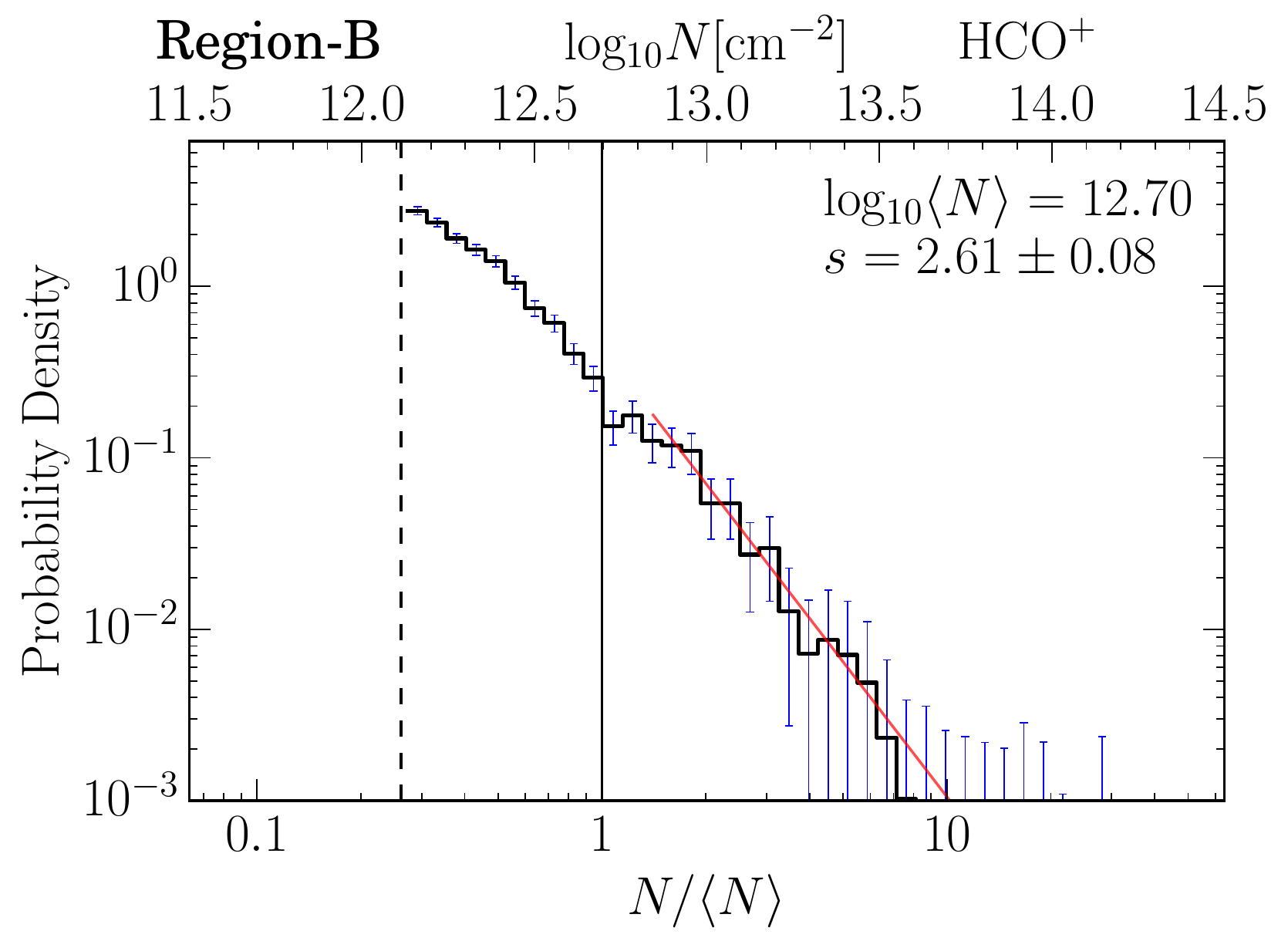}
   \includegraphics[height=3.4cm]{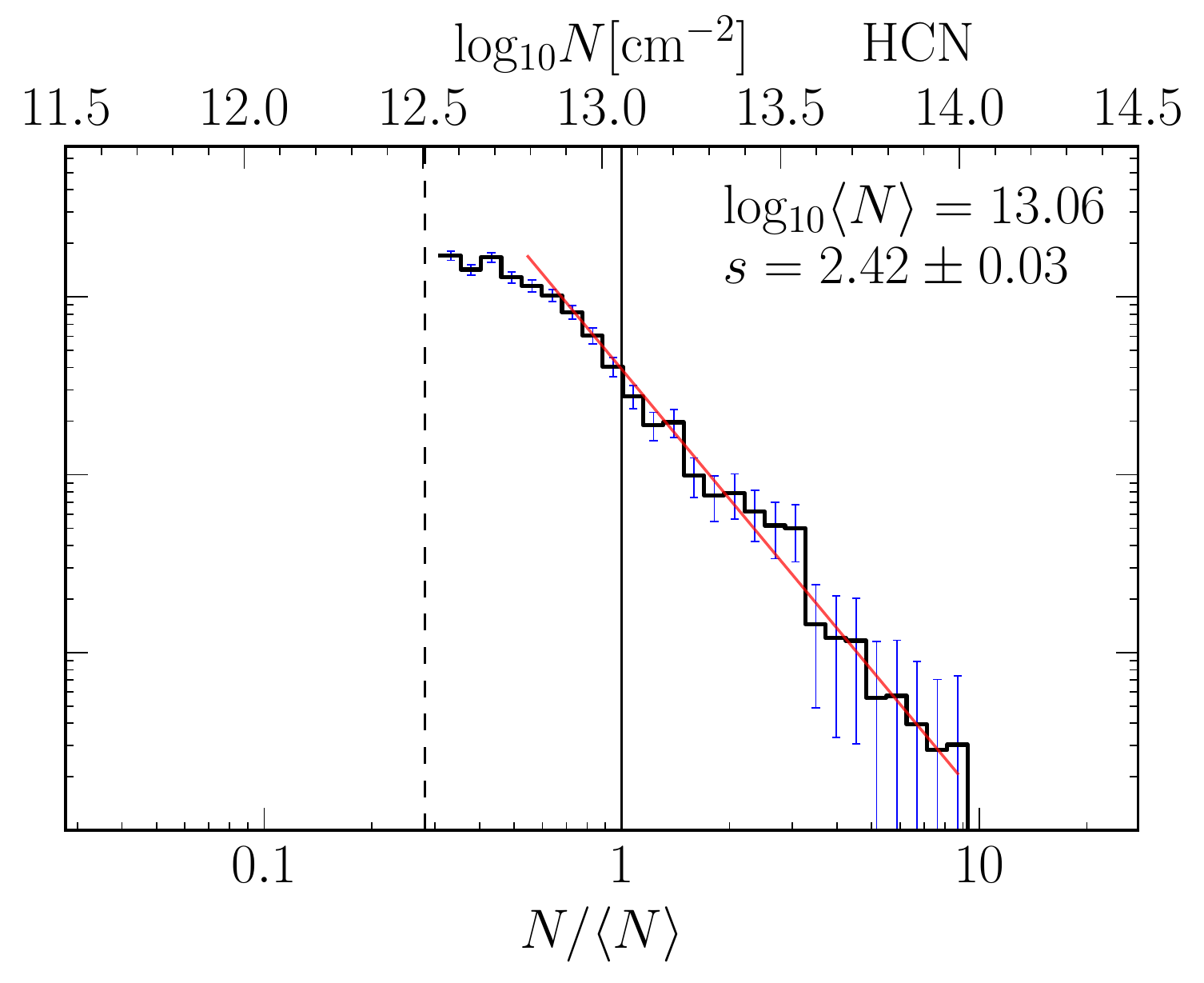}
   \includegraphics[height=3.4cm]{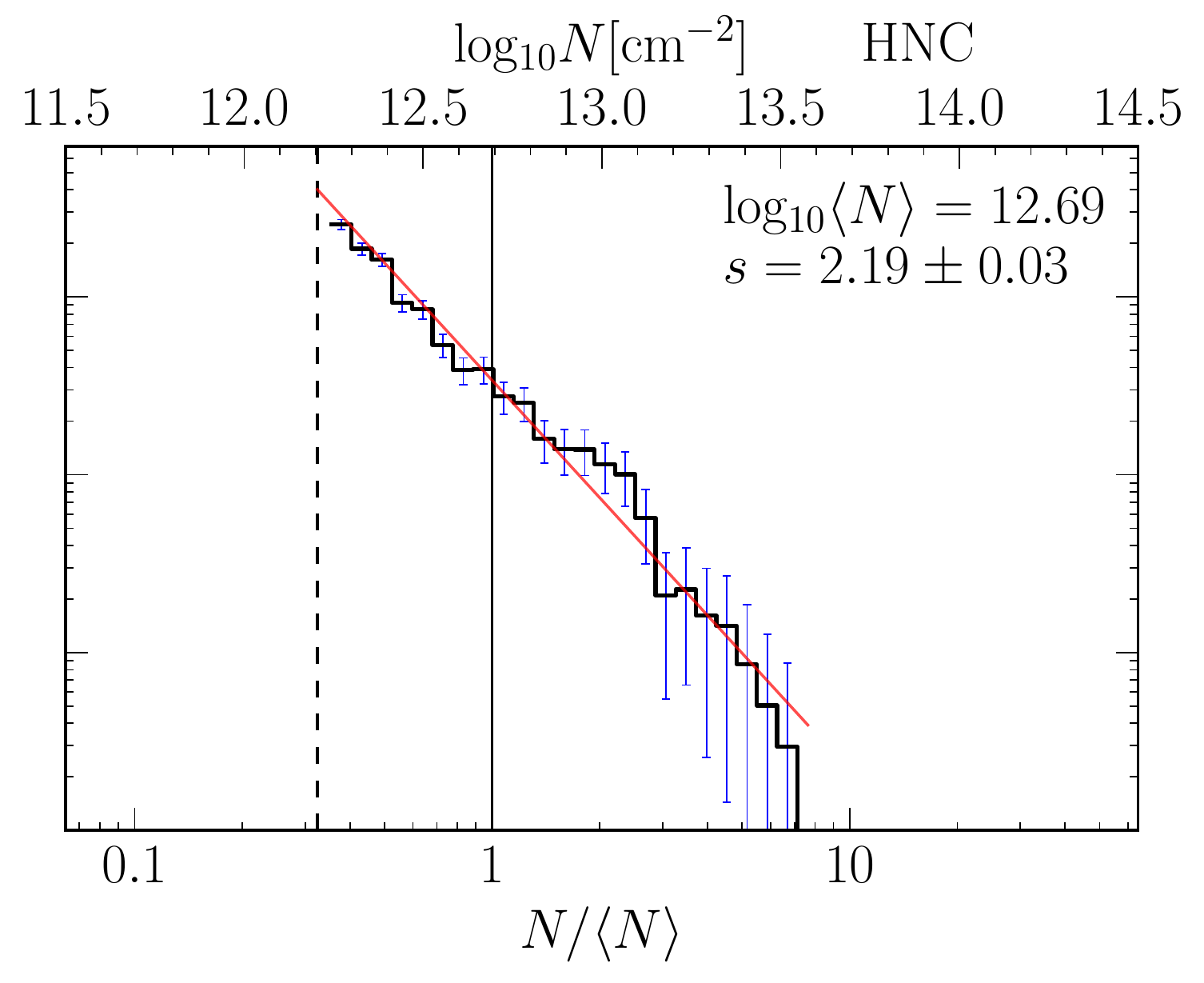}
   \includegraphics[height=3.4cm]{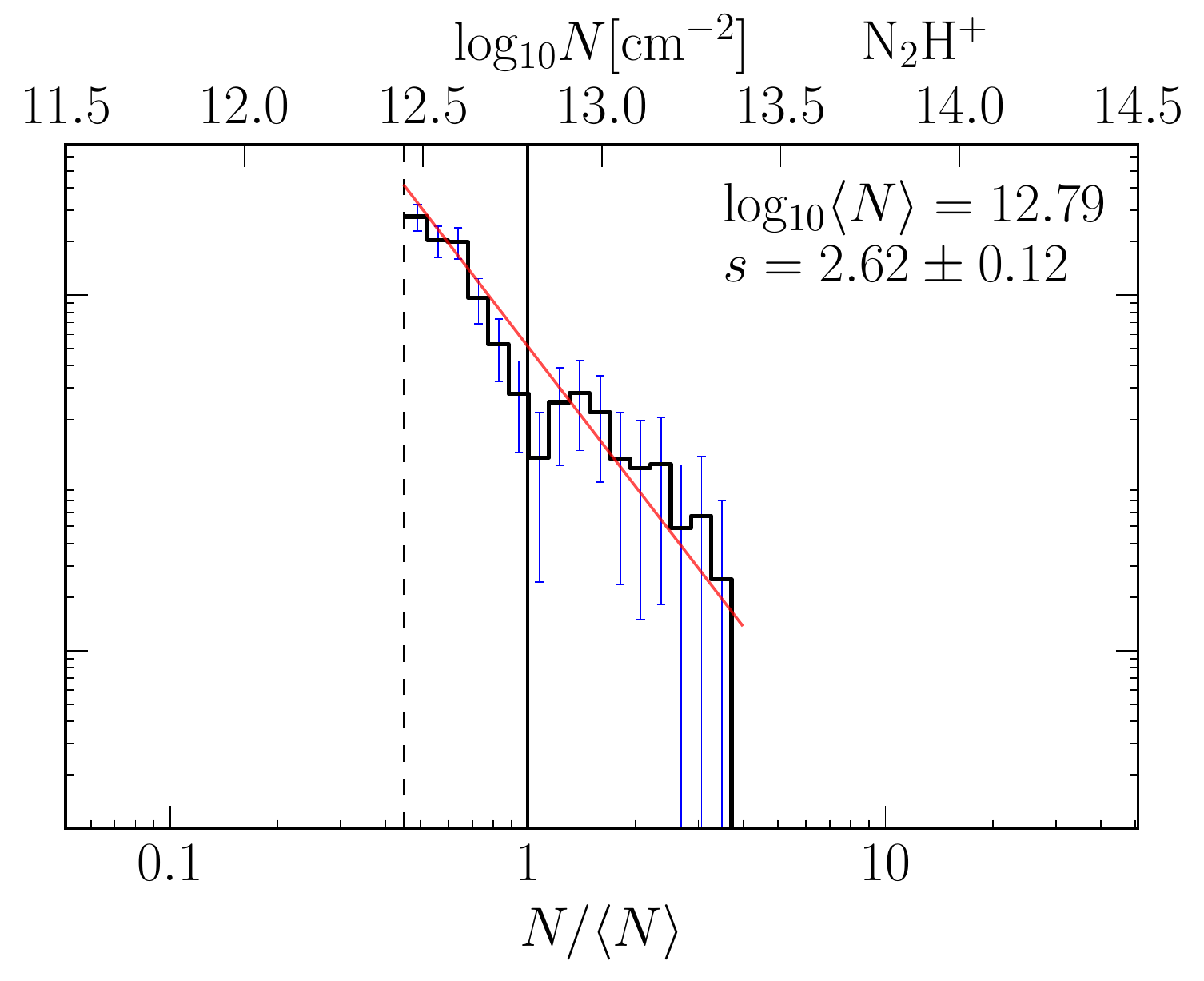}
   \includegraphics[height=3.4cm]{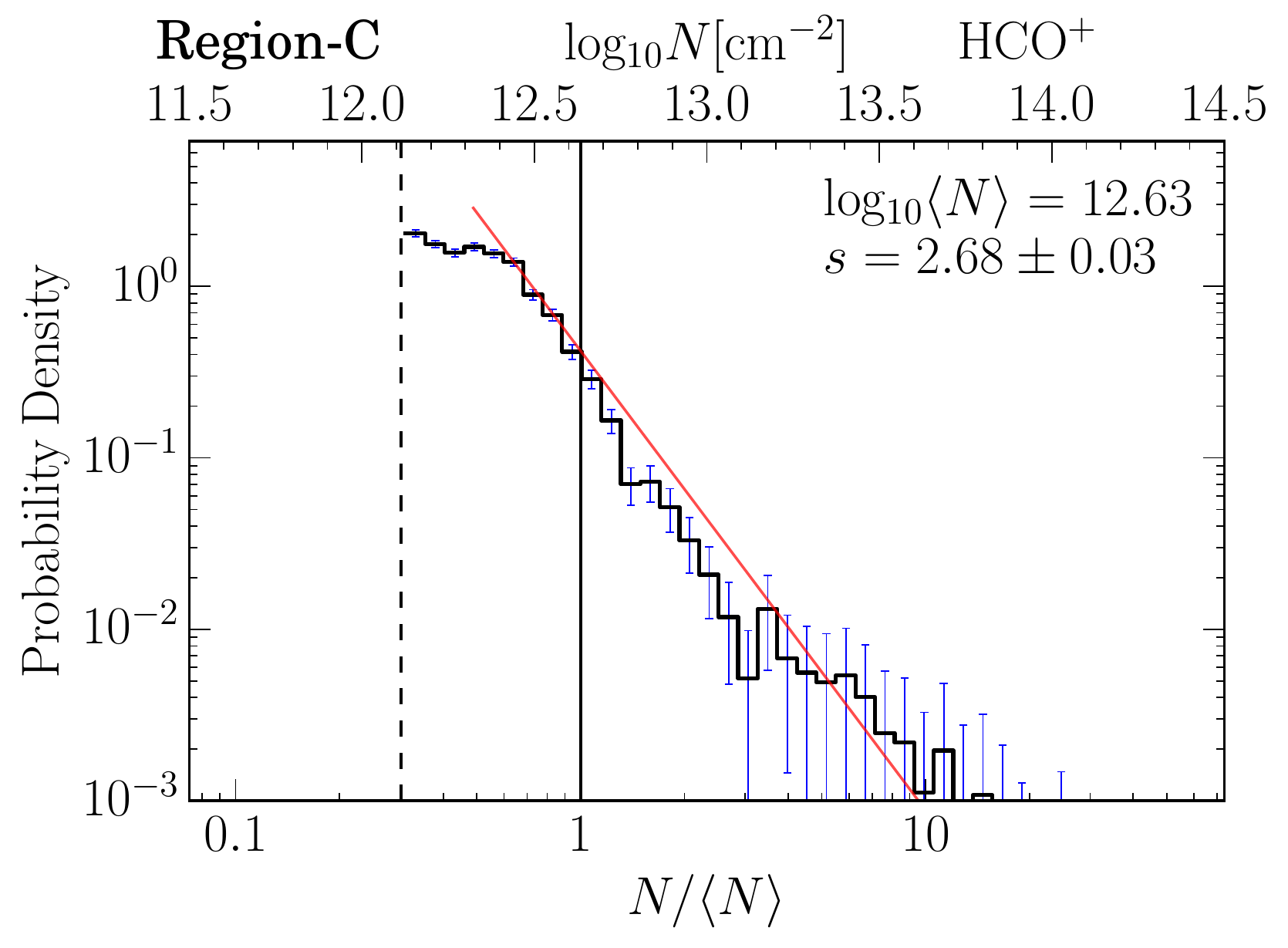}
   \includegraphics[height=3.4cm]{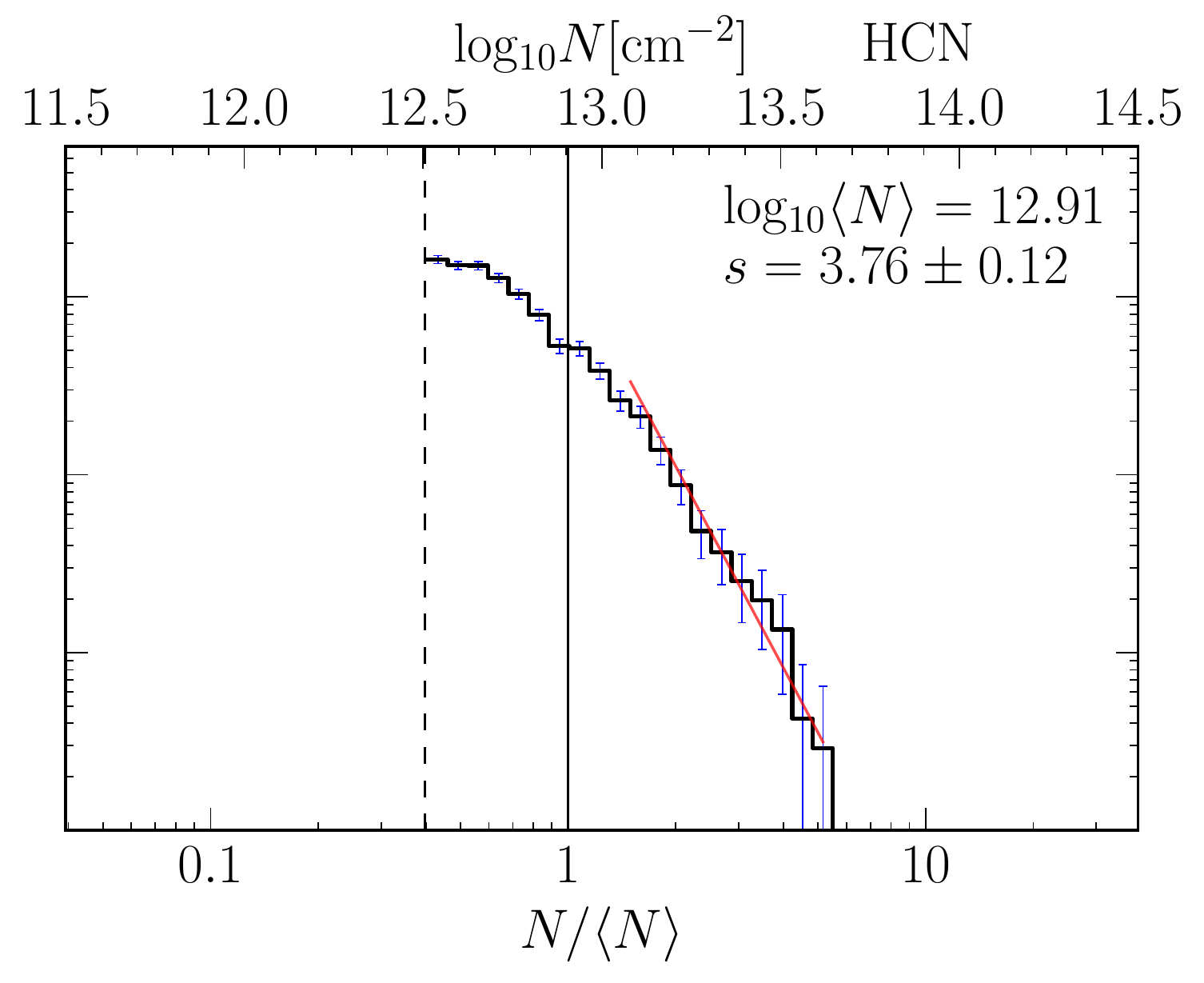}
   \includegraphics[height=3.4cm]{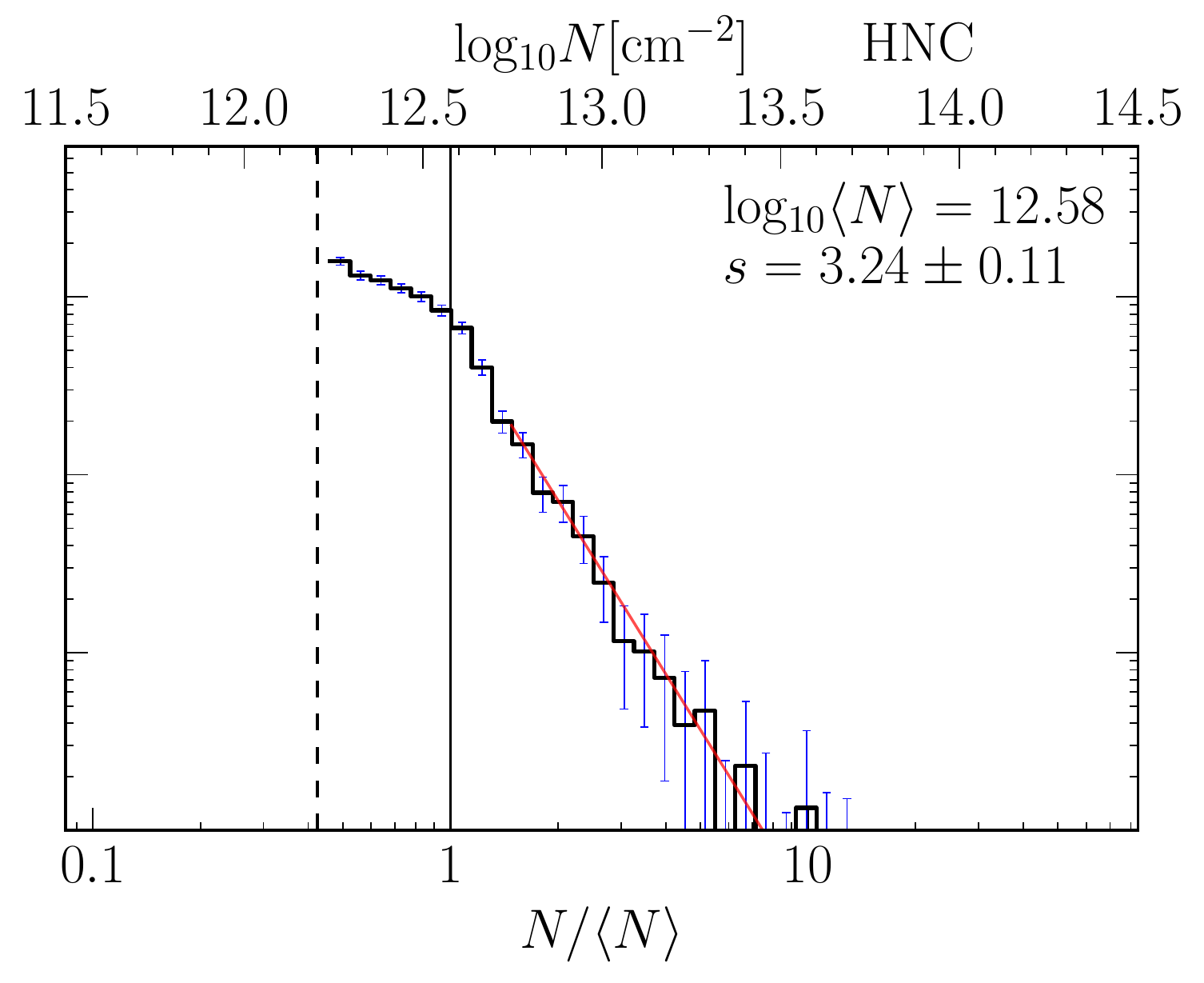}
   \includegraphics[height=3.4cm]{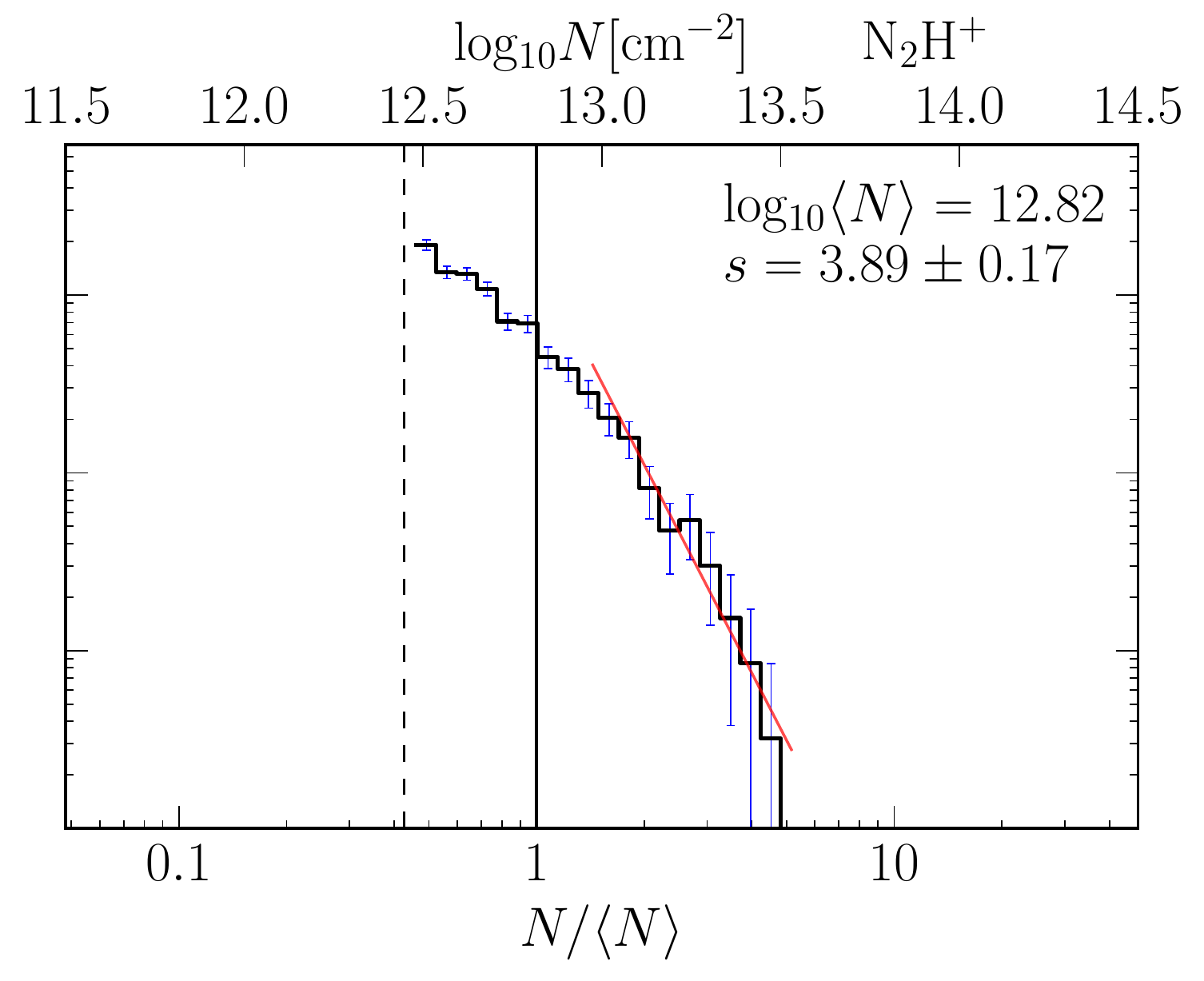}
   \includegraphics[height=3.4cm]{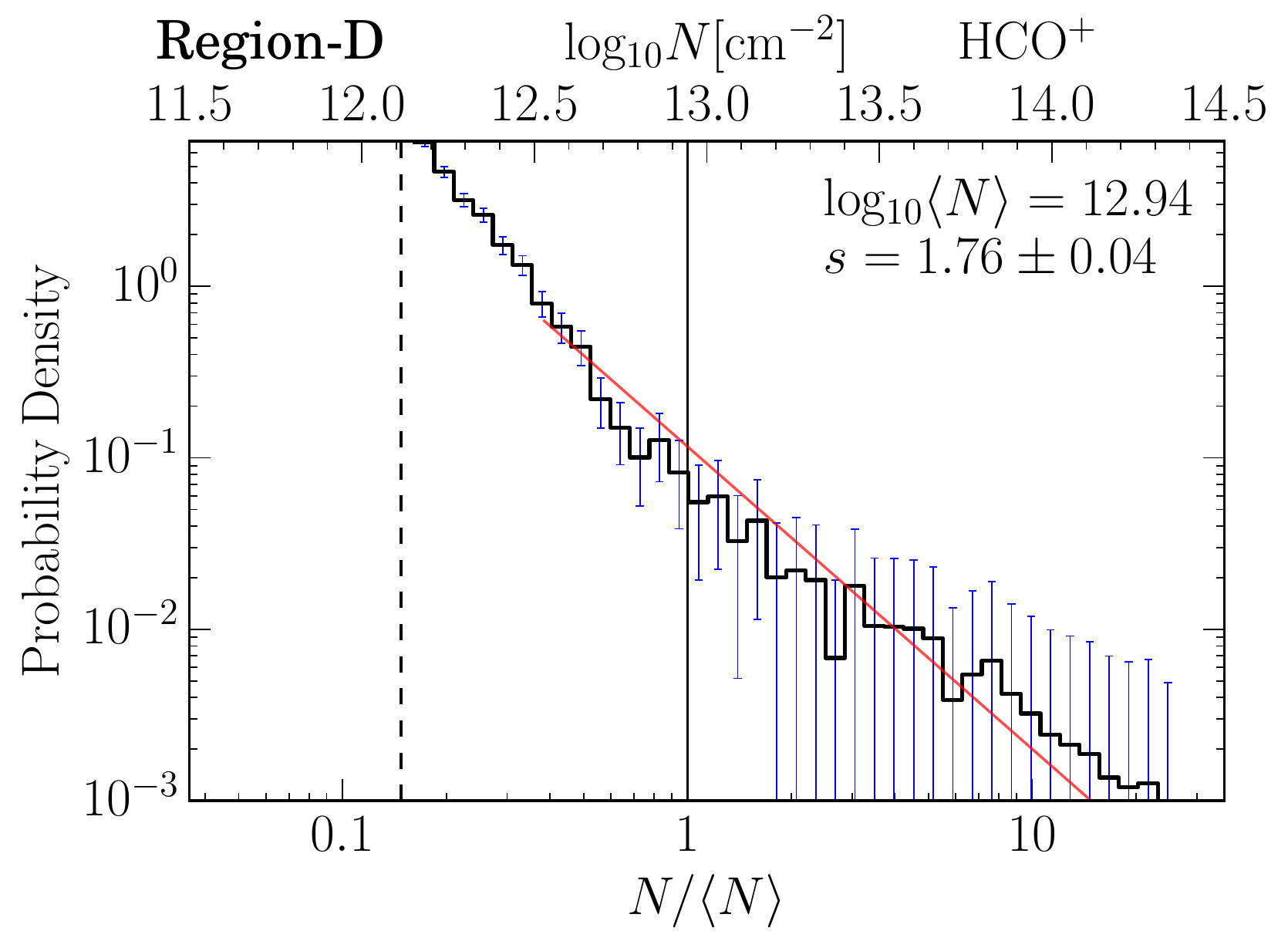}
   \includegraphics[height=3.4cm]{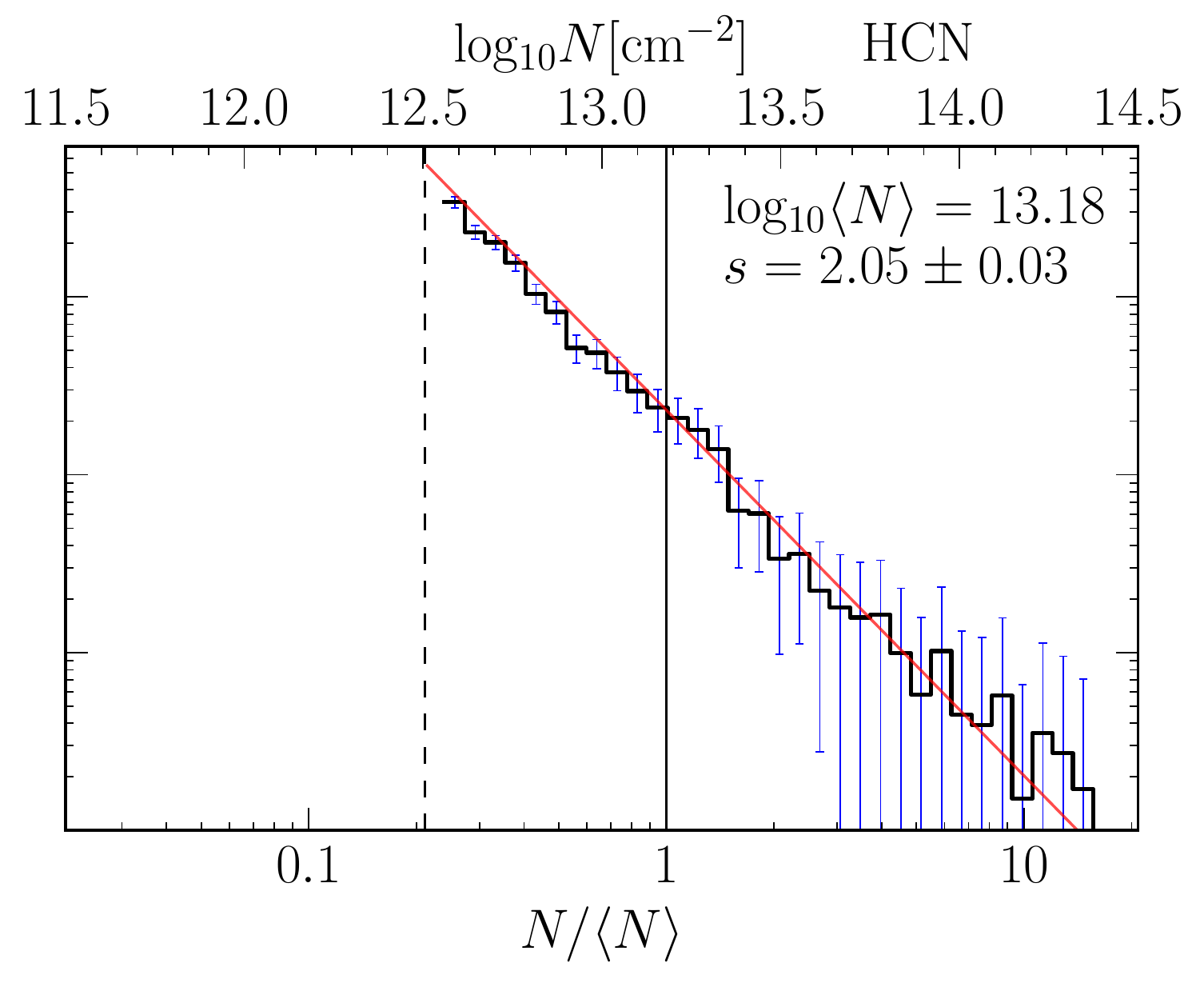}
   \includegraphics[height=3.4cm]{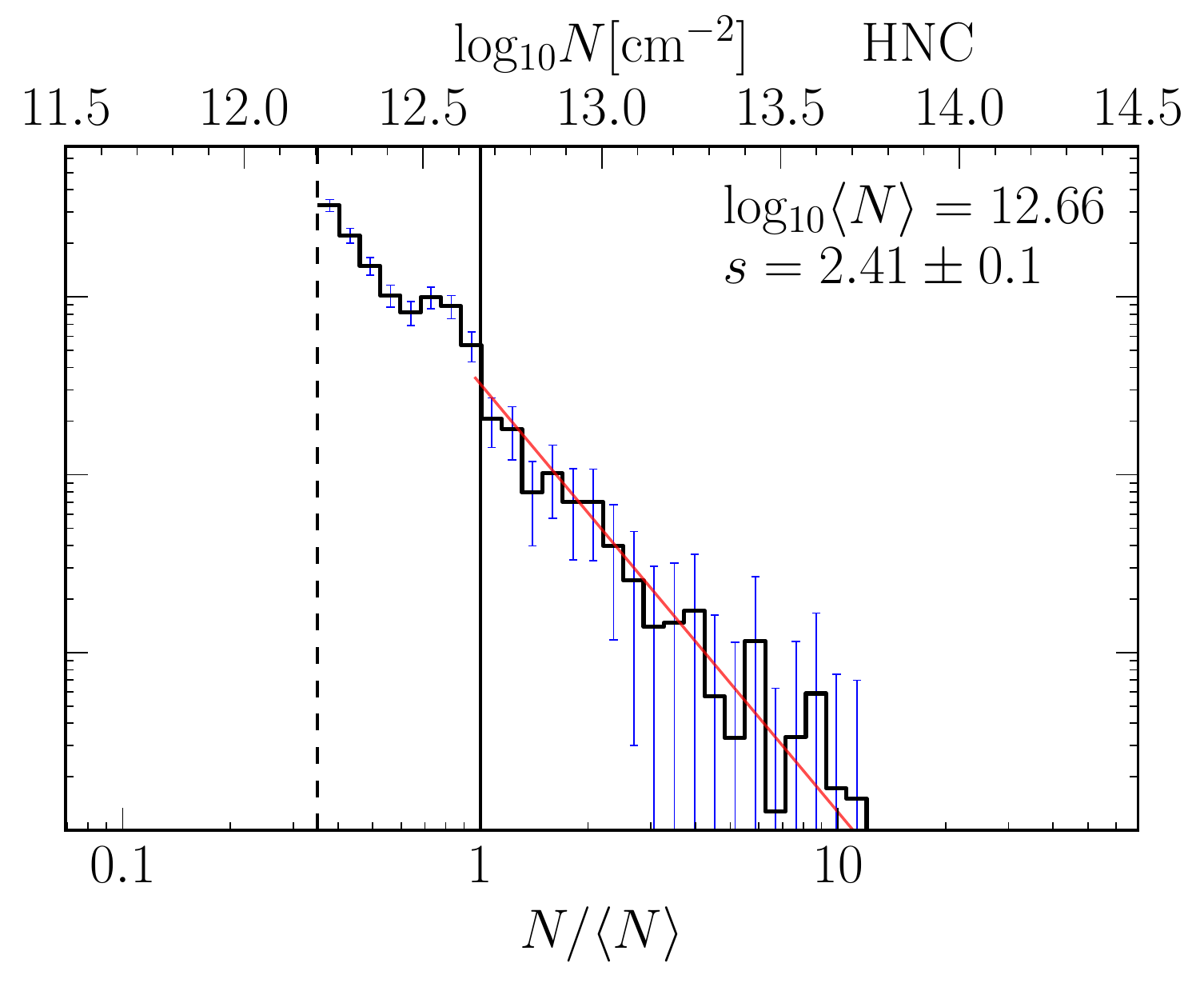}
   \includegraphics[height=3.4cm]{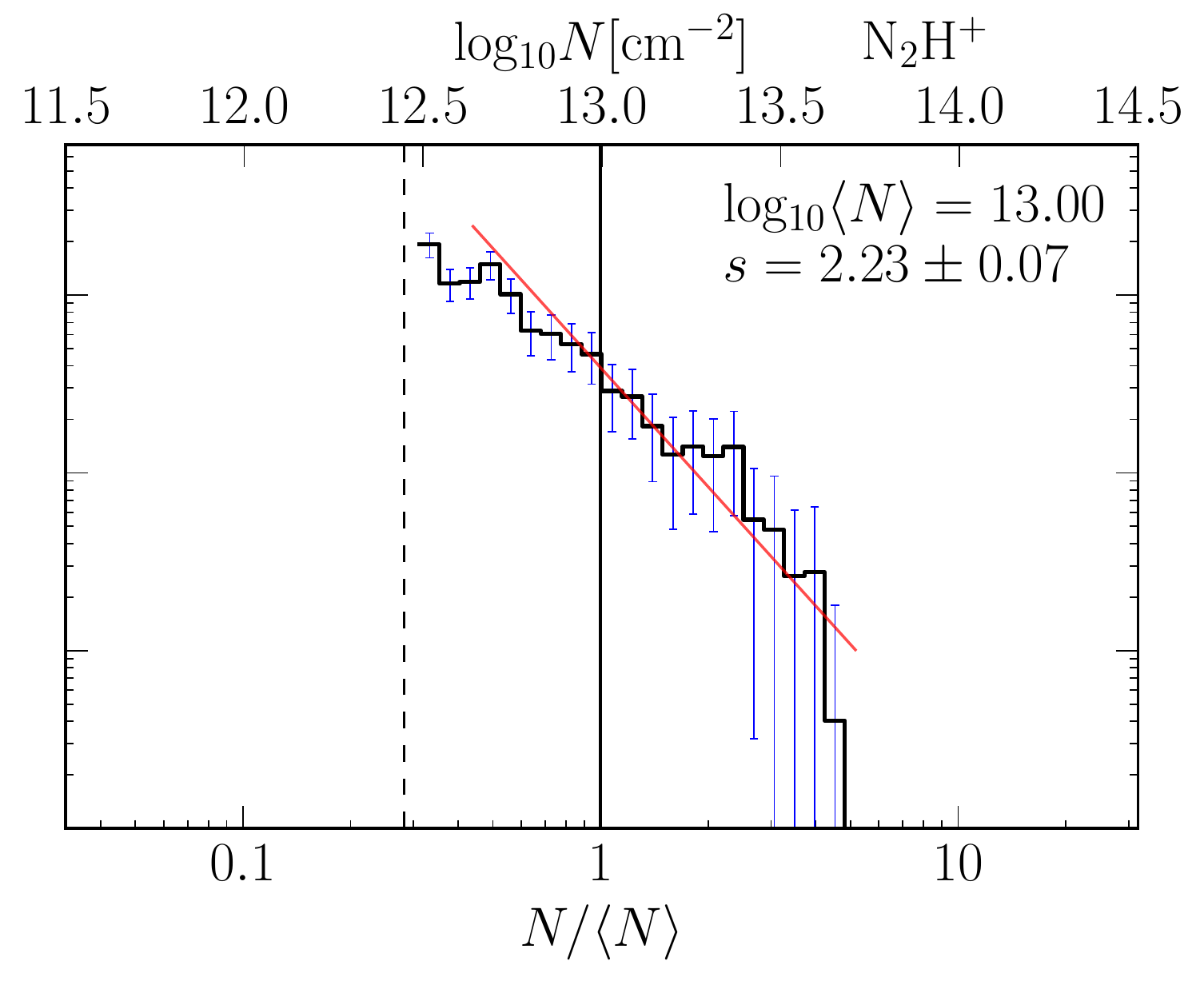}
  \caption{Probability density functions of the $^{13}$CO, HCO$^{+}$, HCN, HNC, and N$_2$H$^+$ column density for Regions A, B, C, and D. The dashed vertical lines in each panel mark the column density threshold, and the solid vertical lines mark the mean column densities. We list in each panel the mean column density $\langle N\rangle$, the power-law index $s$ (the optimal fit, solid line). Error bars are calculated from Poisson statistics.}
         \label{fig_pdf_regs}
\end{figure*}

To study the dense gas properties in different evolutionary stages, we constructed the dense gas N-PDFs for each sub-region (Fig.~\ref{fig_higal_lines}). Since we only compare between the dense gas tracers, the N-PDFs were constructed with the column density maps at the original resolution (8\arcsec pixel size and 32\arcsec beam size). The derived N-PDFs of the dense gas tracers for each sub-region are shown in Fig.~\ref{fig_pdf_regs}.

\begin{figure}
   \centering
   \includegraphics[width=1\hsize]{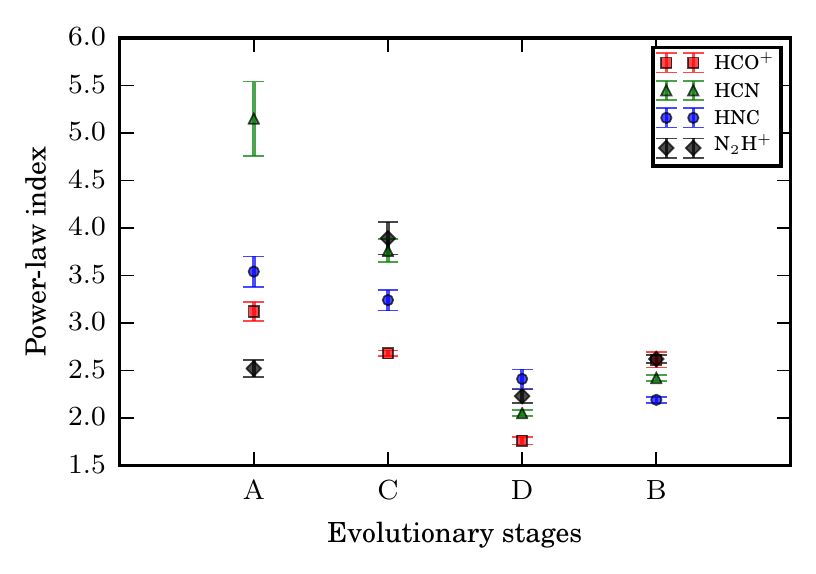}
   \includegraphics[width=1\hsize]{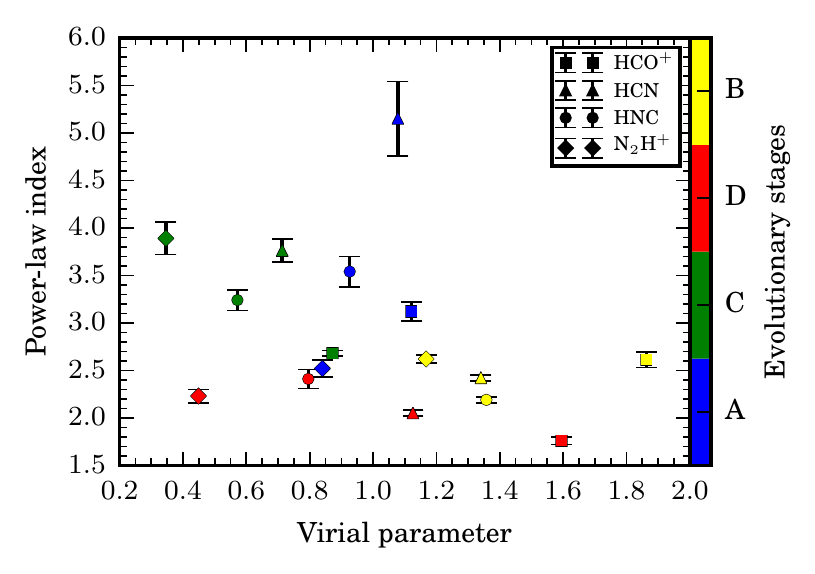}
      \caption{Power-law indices of the N-PDFs plotted as a function of the evolutionary stages (top panel) and of the virial parameter (bottom panel) the sub-regions. The data points in the bottom panel are colored according to the subregions and evolutionary stages, that Regions A and C are in relative early and quiescent stage, Region D is more evolved with an HC\ion{H}{ii} region, and Region B is the most evolved region with PDRs.}
         \label{fig_virial}
\end{figure}

Qualitatively, the N-PDFs of each dense gas tracer show clear differences from region to region. As we proposed in Sect.~\ref{sect_gas}, Region A and Region C are in earlier relative quiescent stage, Region D is more evolved with HC\ion{H}{ii} region, Region B with PDRs is the most evolved region. An evolutionary trend can be seen in some molecular lines, for instance, the HCN and HNC N-PDFs {\bf resemble more of a break power-law shape in the younger Regions C and A to an almost straight single power-law shape in the more evolved Region D. To quantify this evolutionary trend}, we fit an optimal power-law function to each N-PDF with the python package {\it Powerlaw} \citep{Alstott2014}. The obtained optimal column density $N_{\rm min}$ for each subregion (marked with the thin red contours in Fig.~\ref{fig_col_map}) and power-law indices $s$ are listed in Table~\ref{tab_pdf}. As shown in the top panel in Fig.~\ref{fig_virial}, from quiescent regions to HC\ion{H}{ii} region, the power-law slope becomes flatter systematically, then from HC\ion{H}{ii} region to more evolved PDRs the power-law slope steepens again. The lowest contour in HCO$^+$, HCN and HNC column density maps for these subregions are not closed. However, when we fit the N-PDFs with the optimal power-law, the optimal column density threshold (N$_{\rm min}$) as marked with red contours in Fig.~\ref{fig_col_map} are closed contours except in HCO$^+$ for Region C and HCN for Region D. If we fit the power-law from the close contour value (HCO$^+$ 2.2$\times 10^{12}$~cm$^{-2}$, HCN 3.71$\times 10^{12}$~cm$^{-2}$) for these two region, the index for HCO$+$ (Region C) increases by 0.03 and and decreases by 0.01 for HCN (Region D).

To investigate the evolution of the N-PDFs, we estimated the virial parameter $\alpha_{\rm virial}$ of each sub-region following the method described in \citet{Bertoldi1992}. To eliminate the uncertainty brought in by the abundances of the dense gas tracers, we estimated the mass of each sub-region with the Hi-GAL column density map (Fig.~\ref{fig_col_map}). For HCO$^+$ and HNC, we fit the spectra with a Gaussian profile to derive the velocity dispersion. For HCN and N$_2$H$^+$ we fit the HFS to derive the velocity dispersion. All fittings were performed with {\it PySpecKit} \citep{Ginsburg2011}. Assuming a circular shape, we estimated the effective radius from the projected pixel area of each sub-region. As shown in Fig.~\ref{fig_virial} where the power-law indices are plotted as a function of the virial parameters, the virial parameters of all regions and tracers are around or smaller than two.

There is no clear correlation between the virial parameters and the power-law slope (Fig.~\ref{fig_virial}). Among all four molecular lines, N$_2$H$^+$ always has the lowest virial parameter, which makes sense, since N$_2$H$^+$ traces the dense cores that are gravitational bond. Virial parameters of Regions A and C measured with all four tracers are all $\lesssim1$, indicating the gas traced by these four tracers in these two relative quiescent regions are all at gravitational virial equilibrium and could be undergoing gravitational collapsing \citep[e.g.,][]{Bertoldi1992, McKee1992}. For the more evolved Regions B and D, while the gas traced by N$_2$H$^+$ is still gravitational bond ($\alpha_{\rm virial}\lesssim1$), the gas traced by HCO$^+$ show much larger $\alpha_{\rm virial}$ and could be confined by pressure \citep{Bertoldi1992}.

\section{Discussion}
\subsection{N-PDFs of the whole filament}
As mentioned in Sect.~\ref{sect_intro}, there is only one prior dense gas N-PDF study. \citet{Schneider2016} studied the N-PDFs of Cygnus X with {\it Herschel} dust continuum, $^{12}$CO(1--0), $^{13}$CO(1--0), C$^{18}$O(1--0) and the dense gas tracers N$_2$H$^+$(1--0) and CS(2--1). They found that compared to the dust N-PDF, the N-PDFs derived from CO observations can recover the log-normal shape in the lower column density regime, whereas, the N-PDFs derived from dense gas tracers only recover the power-law tail in the high column density regime. Although the absolute column density depends on the abundance and excitation temperature adopted, the power-law index of the dense gas is robust (--1.41 for N$_2$H$^+$, and --1.56 for CS). Compared to \citet{Schneider2016}, our observation is deeper; for instance, the 3$\sigma$ level of the N$_2$H$^+$(1--0) observation from \citet{Schneider2016} corresponds to an $N({\rm H_2})$ of $\sim1\times10^{22}$~cm$^{-2}$, while applying the same abundance, the 3$\sigma$ level of our N$_2$H$^+$(1--0) observation corresponds to an $N({\rm H_2})$ of $3.6\times10^{21}$~cm$^{-2}$. The $^{13}$CO N-PDF can be better described with a log-normal function, but we still could not recover the log-normal peak of the N-PDFs.

\citet{wang2020} studied the N-PDFs of the atomic and molecular gas in the giant molecular filament GMF38a, and found that the N-PDF of \ion{H}{i} emission has the smallest width, followed by the cold neutral media traced by \ion{H}{i} self absorption (HISA) and the $^{13}$CO N-PDF has the largest width. Furthermore, the log-normal widths of the dense gas for the whole filament are all larger than the $^{13}$CO N-PDF as listed in Table~\ref{tab_pdf}. It seems that the width of the N-PDF is correlated with the density it traces. 

The power-law indices derived by fitting all data points in the N-PDFs are quite similar for all molecular lines, between 1.95 and 2.15 ($s^*$ in Table~\ref{tab_pdf}). On the other hand, the optimal power-law indices ($s$ in Table~\ref{tab_pdf}) are different among different lines. A power-law tail in the N-PDF requires a a power-law distribution in the density structure. Though such a density configuration can be achieved by a hydrostatic configuration, where the power-law arises from a balance of gravitational forces and pressure gradients, it is more likely that in a massive GMF self-gravity is the dominant process. Large-scale collapse of filaments and clumps as well as small-scale collapse of high-density cores give then rise to the power-law slopes, seen in the different tracers. Following the method described in Appendix D of \citet{Schneider2016}, we correlate the slope of the power-law tail $s$ and the exponent $\alpha$ of a spherical (for clumps and cores, $\rho(r)\propto r^{-\alpha_{\rm c}}$) and a cylindrical (for filaments, $\rho(r)\propto r^{-\alpha_{\rm f}}$) density distribution via $\alpha_c = 1+2/s$ and $\alpha_f = 1+1/s$, respectively (see also Appendx~\ref{app_cy}). Similar calculations are also discussed by \citet{Federrath2013}, \citet{Fischera2014}, and \citet{Myers2015}. The power-law indices $s^*$ for the dense gas are all around 2, thus the radial density profile $\alpha_c$ is around 2 and $\alpha_f$ around 1.5. These values are fully consistent with the collapse of an isothermal sphere \citep[e.g.,][]{Shu1977} and with a self-gravitating filament model \citep[e.g.,][]{Myers2015}.

The N-PDFs of molecular clouds in the nearby galaxy M33 derived from $^{12}$CO observations \citep{Druard2014, Corbelli2018} peak at lower column density, but the shape and width is similar to the $^{13}$CO one in Fig.~\ref{fig_pdf}. Our study provides the foundation to interpret future high angular resolution dense gas tracer observations towards nearby galaxies with ALMA.

\subsection{N-PDFs and evolutionary stages}
Previous observations show that clouds without star formation have log-normal shape N-PDF with little or no excess in the high column density tail, while active star-forming clouds present prominent power-law-like wings \citep[e.g.,][]{Kainulainen2009, Schneider2013}. Furthermore, the slope index of the power-law tail is correlated with star formation activities, as clumps with flatter power-law tails are more efficient at forming Class 0 protostars \citep{Sadavoy2014, Stutz2015}. \citet{Louvet2014} demonstrated that the star formation rate per free fall time steeply decreases with the virial parameter. The dense gas N-PDFs for the sub-regions further confirm this in a way that more quiescent regions have flatter N-PDFs (Table~\ref{tab_pdf}). Although with large scatter, we also find a correlation between the evolutionary stages of subregions and N-PDF power-law indices (Fig.~\ref{fig_virial}). Sources at early quiescent stages have a lower fraction of high density gas which would result in a steeper power-law slope, while in high-mass star forming region with HC\ion{H}{ii} region the fraction of high density gas is higher which results in a flatter power-law slope. When the source evolves into the PDR stage, the feedback processes from high-mass stars could reduce the dense gas fraction, and the power-law slope of the N-PDF is also steeper again.

\subsection{Comparison to the dust continuum N-PDF}
\begin{figure}
   \centering
   \includegraphics[width=\hsize]{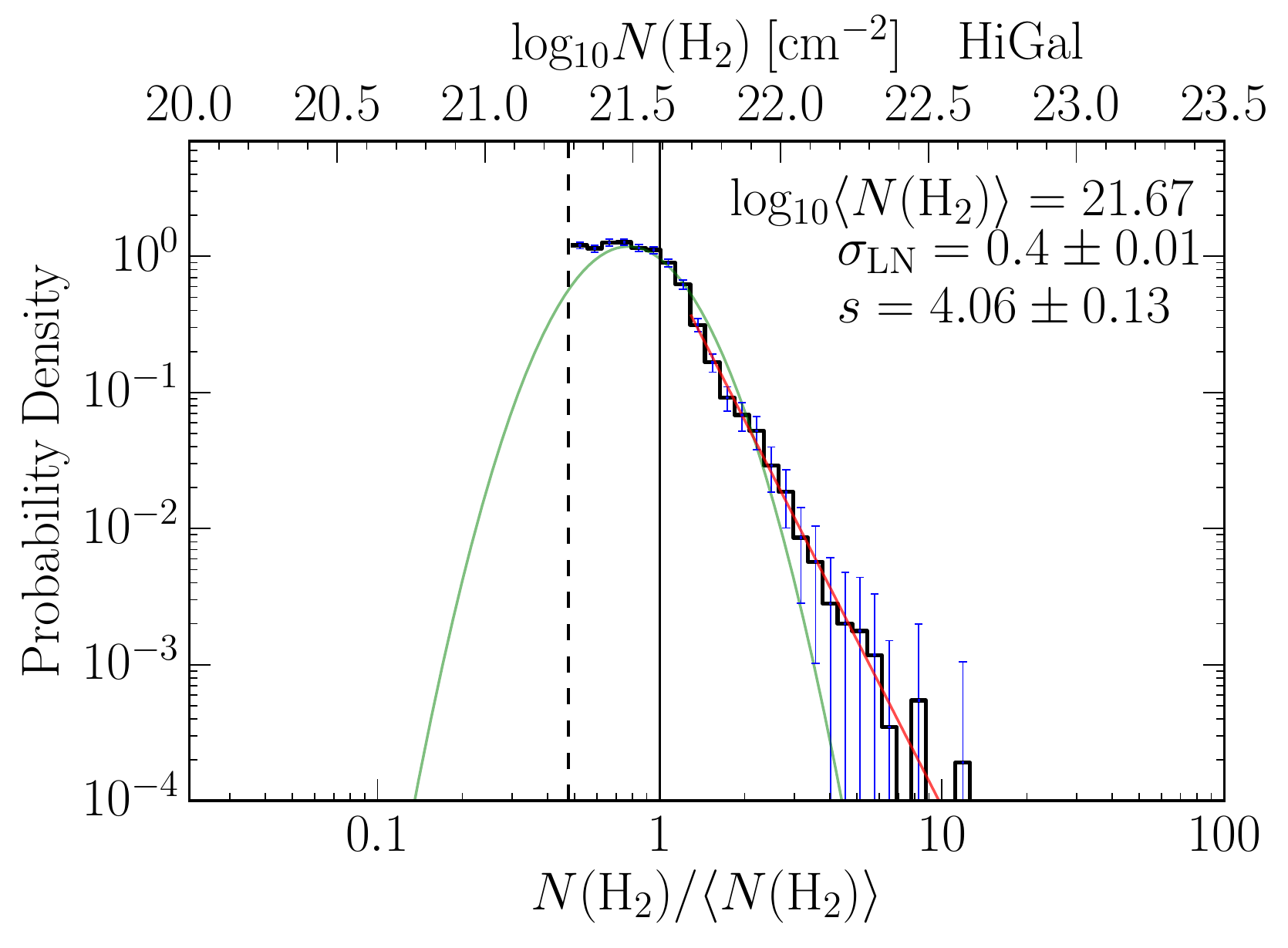}
      \caption{Probability density functions of the molecular hydrogen column density derived from Hi-GAL (Fig.~\ref{fig_higal_lines}). The dashed vertical line marks the column density threshold, and the solid vertical line marks the mean column density. We list in the figure the mean column density $\langle N({\rm H_2})\rangle$, the log-normal (green curve) width $\sigma_{\rm LN}$, the power-law index $s$ (solid line). Error bars are calculated from Poisson statistics.}
         \label{fig_dust}
   \end{figure}

To compare with the N-PDFs we derived from dense gas tracers, we also constructed the N-PDF from the {\it Herschel} dust continuum column density map. We convolved and regridded the dust column density map to the same angular resolution and pixel size as the GRS $^{13}$CO(1--0) map, and constructed the N-PDF shown in Fig.~\ref{fig_dust}. We also fitted an optimal power-law function and a log-normal function (with a closed contour $N$(H$_2$)=6.0$\times10^{21}$~cm$^{-2}$) to the dust N-PDF. Since there is no closed contour in the low column density regime of the dust N-PDF, and the log-normal width is not reliable \citep{Alves2017}, we only compare the power-law slope and the column density range to the ones traced by molecular lines.

Compared to N-PDFs from dense gas tracers, the dust N-PDF has a similar power-law slope to N$_2$H$^+$. Indeed, Figs.~\ref{fig_higal_lines} and \ref{fig_col_map} also show that the N$_2$H$^+$ emission follows the dust column density very well, and traces the dense cores (see also \citealt{Kauffmann2017}). Similar to \citet{Schneider2016}, we can also shift the  N$_2$H$^+$ N-PDF  to``calibrate'' the N$_2$H$^+$ abundance with the dust N-PDF, and the ``calibrated'' abundance of N$_2$H$^+$ is 1.6$\times 10^{-9}$. This ``calibrated'' abundance is similar to the abundance \citet{Gerner2014} estimated for high-mass protostellar objects. 

\subsection{Line ratios}
Integrated intensity ratios of molecular lines are often used to trace molecular gas properties, which can be linked to star formation activities. Numerous surveys have been carried out towards nearby galaxies in the 3~mm band \citep{Watanabe2014, Aladro2015, Meier2015, Nishimura2016a, Nishimura2016b, Jimenez-Donaire2019}. We would like to compare our observations of GMF54 with the observed line ratios in such nearby galaxies. Figure~\ref{fig_ratio_map} shows the line ratio map of GMF54, and we list the averaged line ratios of GMF54 and in the nearby galaxies in Table~\ref{tab_ratio}. Most of the observations of dense gas in the nearby galaxies in Table~\ref{tab_ratio} were carried out with single-dish telescopes, where the spatial resolution varies from $\sim10$~pc towards the Large Magellanic Cloud (LMC) to $\sim$24~kpc towards some active galactic nucleus (AGNs). In comparison to this, GMF54 has a size of 68~pc and our observations have a spatial resolution of 0.32~pc. Despite the huge spatial resolution differences, the line ratios we derived in GMF54 are in general comparable to those seen in the nearby galaxies.

\begin{table*}
\caption[]{Integrated intensity ratios.}
\label{tab_ratio}
\centering 
\begin{tabular}{l c c c c c r r}
\hline\hline 
Sources & HCN/HCO$^+$ & HCN/HNC & HCN/$^{13}$CO& HCO$^+$/$^{13}$CO & N$_2$H$^+$/$^{13}$CO &Telescope &Ref.  \\
GMF54 & 1.09 & 1.95 & 0.15 & 0.15 & 0.10 & IRAM 30~m&1\\
W51 arm& 1.29& 2.78 & 0.35 & 0.27 &0.06 & IRAM 30~m&2 \\
Starbursts & 0.70--1.18 & 1.93--2.47 &  0.59--0.84 & 0.54--1.19 & 0.06--0.13& IRAM 30~m&3\\
AGNs & 0.82--2.02 & 2.19--3.03 & 0.81--1.80& 0.40--1.24& 0.11--0.15& IRAM 30~m& 3\\
ULIRGs& 1.54--2.39&1.42--3.03&2.29--5.02& 0.96--3.27& 0.49 & IRAM 30~m&3\\
LMC & 0.40--1.36&1.13--3.36&0.05--0.60& 0.13--0.89 &0.01--0.11 & Mopra 22~m&4\\
Dwarf galaxy IC10 &0.04&3.15&0.22 &0.56 & 0.07 &Nobeyama 45~m&5\\
Barred disk& 1.30 & 1.75 &  0.29 & 0.25& --& IRAM 30~m&6\\
Unbarred disk& 1.29 & 2.52 &  0.25 & 0.27 & --& IRAM 30~m&6\\
\hline                                  
\end{tabular}
\tablefoot{References: 1. This work; 2. \citet{Watanabe2014}; 3, \citet{Aladro2015}; 4, \citet{Nishimura2016b}; 5, \citet{Nishimura2016a}; 6, EMPIRE project, \citet{Jimenez-Donaire2019}}
\end{table*}

The HCN/CO ratio is used to trace the dense gas fraction, which is directly related to star formation activity \citep{Gao2004, Usero2015, Bigiel2016}. Since we do not have CO observations, we use the HCN/$^{13}$CO ratio. Compared to nearby galaxies, the HCN/$^{13}$CO ratio in GMF54 is much lower than in nearby galaxies, while ULIRGs present the highest HCN/$^{13}$CO ratio. As discussed by \citet{Shirley2015, Kauffmann2017, Pety2017}, HCN traces much lower density than its critical density $n_{\rm crit}$, and typically traces densities around $\sim10^2-10^3$~cm$^{-3}$. The HCN flux is also influenced by far ultra-violet (UV) radiation \citep{Pety2017}. Similarly, HCO$^+$(1--0) also traces only moderate dense gas, and the flux can be affected by far-UV radiation \citep{Shirley2015, Pety2017}. Thus ULIRGs and AGNs have the highest HCO$^+$/$^{13}$CO ratio, while GMF54 has the lowest ratio. 

On the other hand, N$_2$H$^+$ traces the real dense gas fraction with an effective density $n_{\rm eff}~10^4$~cm$^{-3}$($T_{\rm k}$=10~K). The N$_2$H$^+$/$^{13}$CO ratio, which traces the high density gas, shows similar values in GMF54 and nearby galaxies, except in ULIRGs which have a much higher N$_2$H$^+$/$^{13}$CO ratio. Considering ULIRGs are forming many stars, it is reasonable that they have a higher fraction of high density gas. It is surprising that GMF54 has a similar or even higher N$_2$H$^+$/$^{13}$CO ratio than the starbursts in Table~\ref{tab_ratio}. Considering the filling factor of N$_2$H$^+$ is $\lesssim10$\% (with respect to $^{13}$CO), it is likely that the extragalactic studies are underestimating the true N$_2$H$^+$/$^{13}$CO ratio.

\begin{figure}
   \centering
   \includegraphics[width=1\hsize]{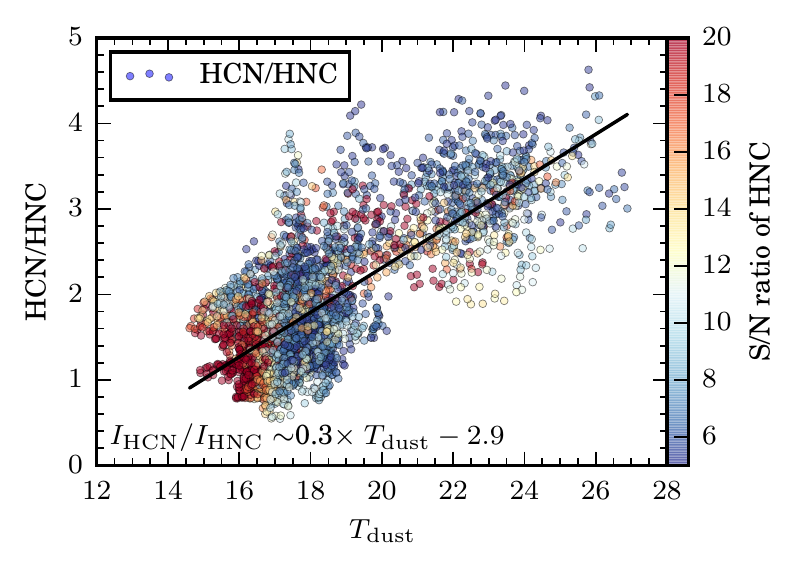}
      \caption{Integrated intensity ratio of HCN/HNC plotted as a function of the dust temperature for the whole filament pixel by pixel. The points are colored according to the S/N ration of the HNC integrated intensity. Pixels below 5$\sigma$ were masked out. A linear fit result is plotted and labeled in the figure.}
\label{fig_hcn_hnc}
\end{figure}

Another useful line ratio is HCN/HNC, which is proposed as a tracer of evolutionary stages of the molecular cloud because at temperature $\gtrsim30$~K HNC starts to convert to HCN \citep{Schilke1992, Herbst2000, Graninger2014, Hacar2019}. Therefore, the HCN/HNC ratio would increase as temperature increases. As we demonstrate in Fig.~\ref{fig_ratio_map}, the HCN/HNC ratio remains relative constant $\lesssim2$ in infrared dark regions, and increases to $\gtrsim3$ in the PDRs and around infrared bright HC\ion{H}{ii} region. To further demonstrate this, we plotted the HCN/HNC ratio as a function of the dust temperature derived from Hi-GAL \citep{Zucker2018}. A clear correlation between the HCN/HNC ratio and $T_{\rm dust}$ can be seen in Fig.~\ref{fig_hcn_hnc}, with a Pearson correlation coefficient of 0.77 and $p$-value $<0.001$. The mean HCN/HNC ratio across the whole GMF54 is also comparable to nearby galaxies. 

\subsection{Uncertainties}
\label{sect_uncert}
Besides the rms noise of the data, the main factors that introduce uncertainties to the column densities and N-PDFs are excitation temperature and optical depth. For N$_2$H$+$ and HCN we derive T$_{\rm ex}$ and $\tau$ from the HFS fitting. The uncertainties of T$_{\rm ex}$ and $\tau$ of the fitting for N$_2$H$+$ have a median value of 2~K and 0.5, respectively. Therefore, the column density uncertainty of the N$_2$H$+$ column density brought in by HFS fitting is $\sim50$\%. The uncertainties of T$_{\rm ex}$ and $\tau$ for HCN have a median value of 1~K and 0.1, respectively, which brings in $\sim10$\% uncertainty to the HCN column density estimation. 

Since we cannot derive the excitation temperature for HCO$^+$ and HNC directly from our observations, we assume they share the same excitation temperature as HCN. If we assume the excitation temperatures of HCO$^+$(1--0) and HNC(1--0) are systematically 50$\%$ higher or lower than HCN(1--0), the average uncertainty introduced to the column densities of HCO$^+$ and HNC is $\sim$7$\%$. The uncertainties of the N-PDF power-law indices for the whole filament brought in by the excitation temperature are between 0.1 and 0.5. For the subregions, this uncertainty of power-law indices of HCO$^+$ is $<0.6$. For HNC, the changes in T$_{\rm ex}$ also change the derived $N_{\rm min}$ for the optimal fit, and further increase the power-law indices for Regions C and D by $<0.4$ and Regions A and B by $<2$. However, the evolutionary trend of the power-law index shown in Fig.~\ref{fig_virial} does not change, i.e., the quiescent Regions A and C show the steepest power-law index, the HC\ion{H}{ii} region D has the flattest power-law index, and Region B with PDRs has an intermediate power-law index.

To estimate the $^{13}$CO column density we adopted a uniform excitation temperature of 10~K, which was derived by \citet{Zhang2019} from estimating the $T_{\rm ex}$ for 13 GMFs using $T_{\rm mb}$($^{12}$CO). They derived a mean excitation temperature of $\sim10$~K with a standard deviation of 2.5~K, and the $^{13}$CO column density uncertainty brought in is $\sim19$\%. Another uncertainty source for the $^{13}$CO column density is optical depth. Since we do not have $^{12}$CO or C$^{18}$O observation with compatible angular resolution, we cannot estimate the $^{13}$CO(1--0) optical depth directly. With a similar angular resolution, \citet{Schneider2016} estimated the $^{13}$CO(1--0) optical depth for the Cygnus X North and found that the optical depth is negligible and does not influence the N-PDF significantly. \citet{Riener2020} estimated the $^{13}$CO(1--0) optical depth for GMF24, and found a median $\tau$ value of $\sim0.5$. \citet{wang2020} found that the GRS $^{13}$CO(1--0) optical depth for another filament GMF38a ($l\sim33\degr-37\degr$) is mostly smaller than 1 with an median value of $\sim$0.4. If we apply this 0.4 to GMF54, the column density increases by $\sim$21\%. In total the $^{13}$CO column density has a uncertainty of $\sim40$\%.

\section{Conclusions}
We studied the dense gas properties of one giant molecular filament, GMF54, with typical dense gas tracers at 3~mm, HCO$^+$(1--0), HCN(1--0), HNC(1--0), and N$_2$H$^+$(1--0). The main results are summarized as follows:
\begin{enumerate}
\item All dense gas molecular lines trace the extended structure of the filament, the area filling factors (with respect to $^{13}$CO(1--0) of HCO$^+$, HCN, HNC, and N$_2$H$^+$ are, 0.28, 0.21, 0.17, and 0.06, respectively. If we convolve and regrid the dense gas maps to the same angular resolution and pixel size as the $^{13}$CO map, the filling factors increase by $\sim70\%$.
\item We constructed the N-PDFs of the $^{13}$CO column density and the optical depth corrected HCO$^+$, HCN, HNC and N$_2$H$^+$column densities. While the $^{13}$CO N-PDF can be better fitted with a log-normal function, the others are closer to a power-law shape. We fitted power-law functions to the N-PDFs. The HCO$^+$ N-PDF has the flattest power-law slope. The radial density profile of the filament is estimated to be $\sim$1.5, which is consistent with the self-gravitating filament model.
\item We studied the N-PDFs for sub-regions of GMF54, and found an evolutionary trend in the N-PDFs, in the sense that more evolved regions have a flatter power-law index.
\item The integrated intensity ratios of the observed molecular lines in GMF54 are comparable to those in nearby galaxies. The N$_2$H$^+$/$^{13}$CO ratio, which traces the dense gas fraction, has similar values in GMF54 and all nearby galaxies except ULIRGs. A clear correlation between the HCN/HNC ratio and the dust temperature was detected with a Pearson correlation coefficient of 0.74 and $p$-value $<0.001$.
\end{enumerate}

As the largest coherent cold gas structures in our Milky Way, GMFs are excellent candidates to connect studies of star formation on Galactic and extragalactic scales.

\begin{acknowledgements}
Y.W., H.B., S.B., and J.D.S. acknowledge support from the European Research Council under the Horizon 2020 Framework Program via the ERC Consolidator Grant CSF-648505. H.B. also acknowledges support from the Deutsche Forschungsgemeinschaft in the Collaborative Research Center (SFB 881) “The Milky Way System” (subproject B1). F.B. acknowledges funding from the European Research Council (ERC) under the European Union’s Horizon 2020 research and innovation programme (grant agreement No. 726384/``EMPIRE''). N.S. acknowledges support from the ANR/France and DFG/Germany project `GENESIS' (ANR-16-CE92-0035-01/DFG1591/2-1) and from the DFG project SFB 956.C.B. gratefully acknowledges support from the National Science Foundation under Award No. 1816715. This research made use of Astropy and affiliated packages, a community-developed core Python package for Astronomy \citep{astropy2018}, Python package {\it SciPy}\footnote{\url{https://www.scipy.org}}, APLpy, an open-source plotting package for Python \citep{robitaille2012}.
\end{acknowledgements}

\bibliographystyle{aa}

\bibliography{filament_wang.bib}

\begin{appendix}
\section{Excitation temperature and optical depth maps}
\begin{figure*}
   \centering
   \includegraphics[height=6cm]{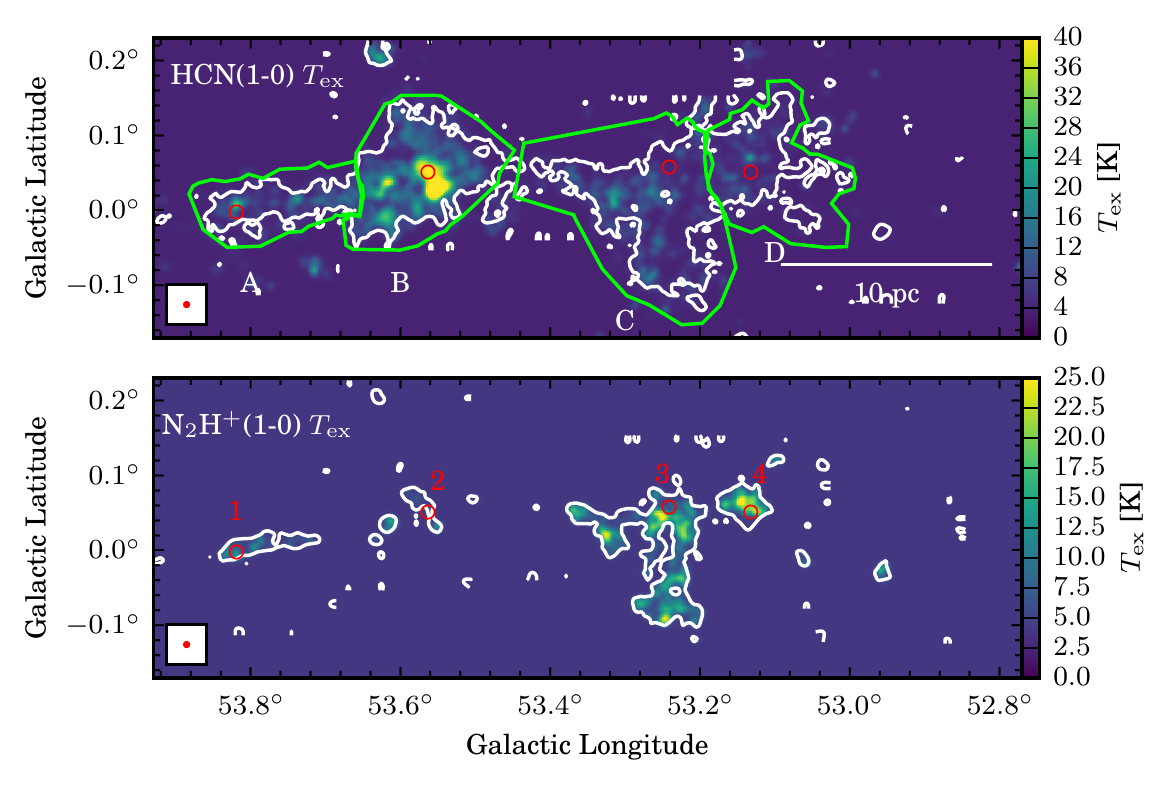}
   \includegraphics[height=6cm]{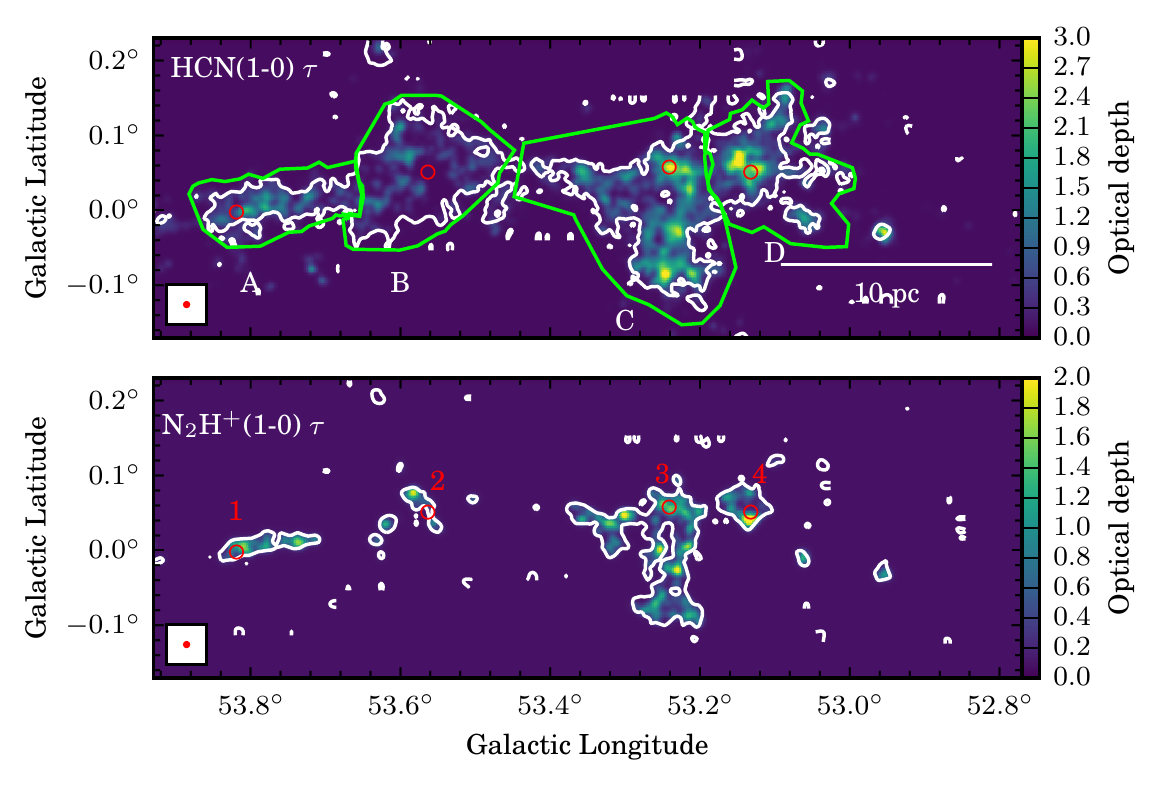}
      \caption{The excitation temperature ($T_{\rm ex}$, left) and optical depth ($\tau$, right) maps derived from the HFS fitting after convolved with a gaussian FWHM beam of 32\arcsec\ (gaussian kernel $\sim$13.59\arcsec). The thick white contour in each panel traces the 3$\sigma$ level the respect $^{12}$C isotopologue line integrated intensity shown in Fig.~\ref{fig_higal_lines}. The red circles mark the locations whose spectra are shown in Fig.~\ref{fig_hfsfit}. The beams and polygons are the same as in Fig.~\ref{fig_higal_lines}.}
 \label{fig_tex_tau}
\end{figure*}

\begin{figure*}
   \centering
   \includegraphics[width=0.24\hsize]{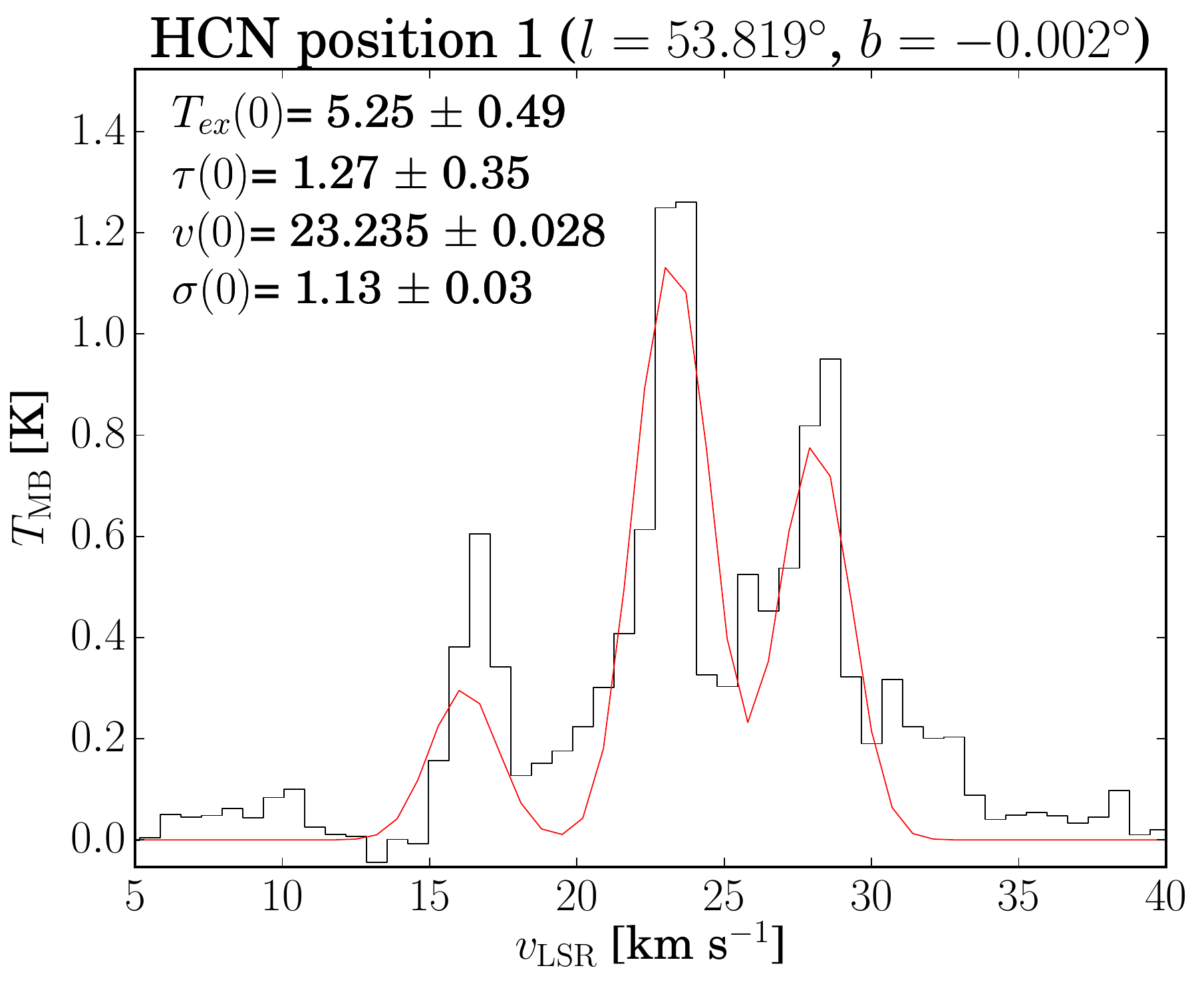}
   \includegraphics[width=0.24\hsize]{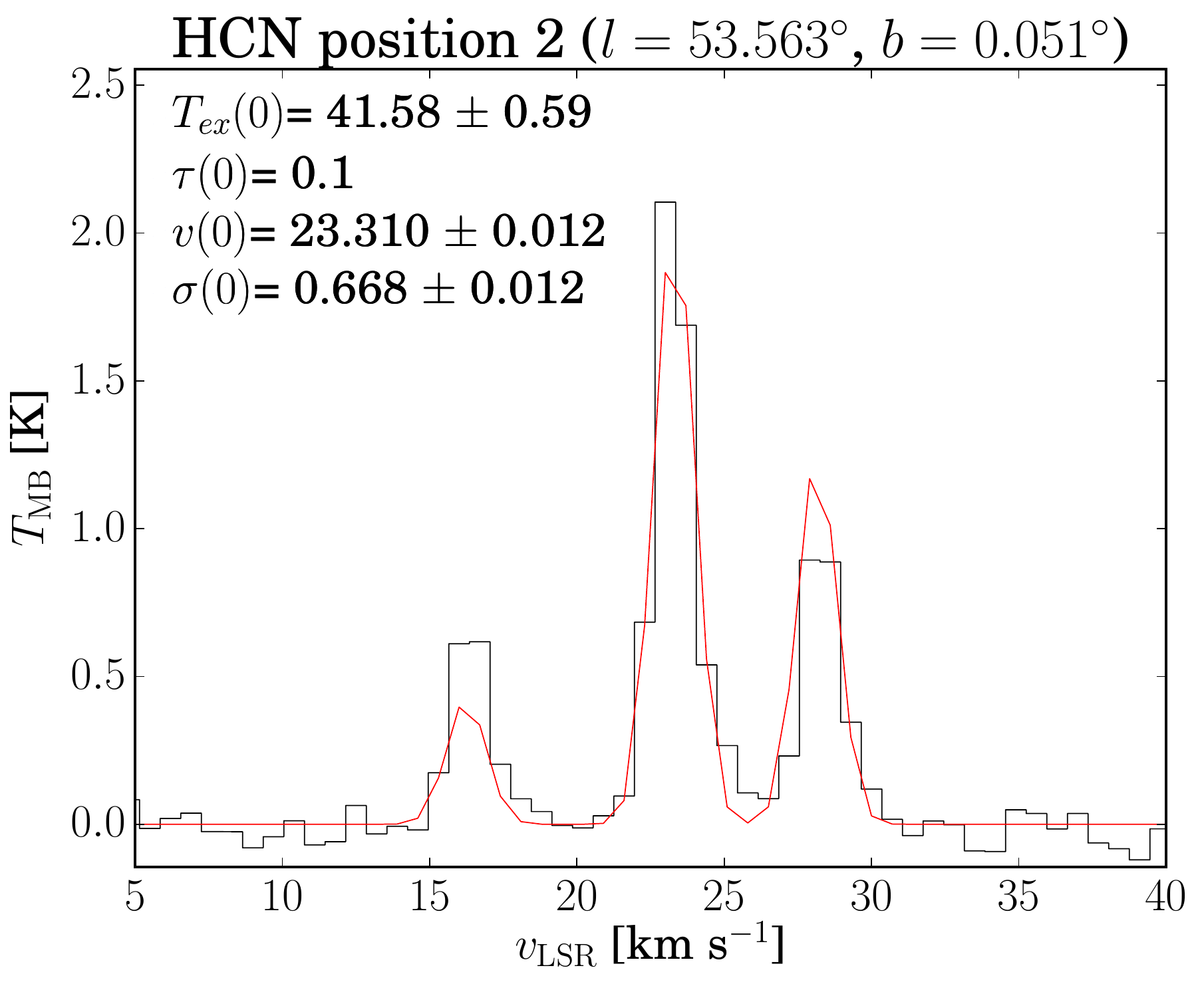}
   \includegraphics[width=0.24\hsize]{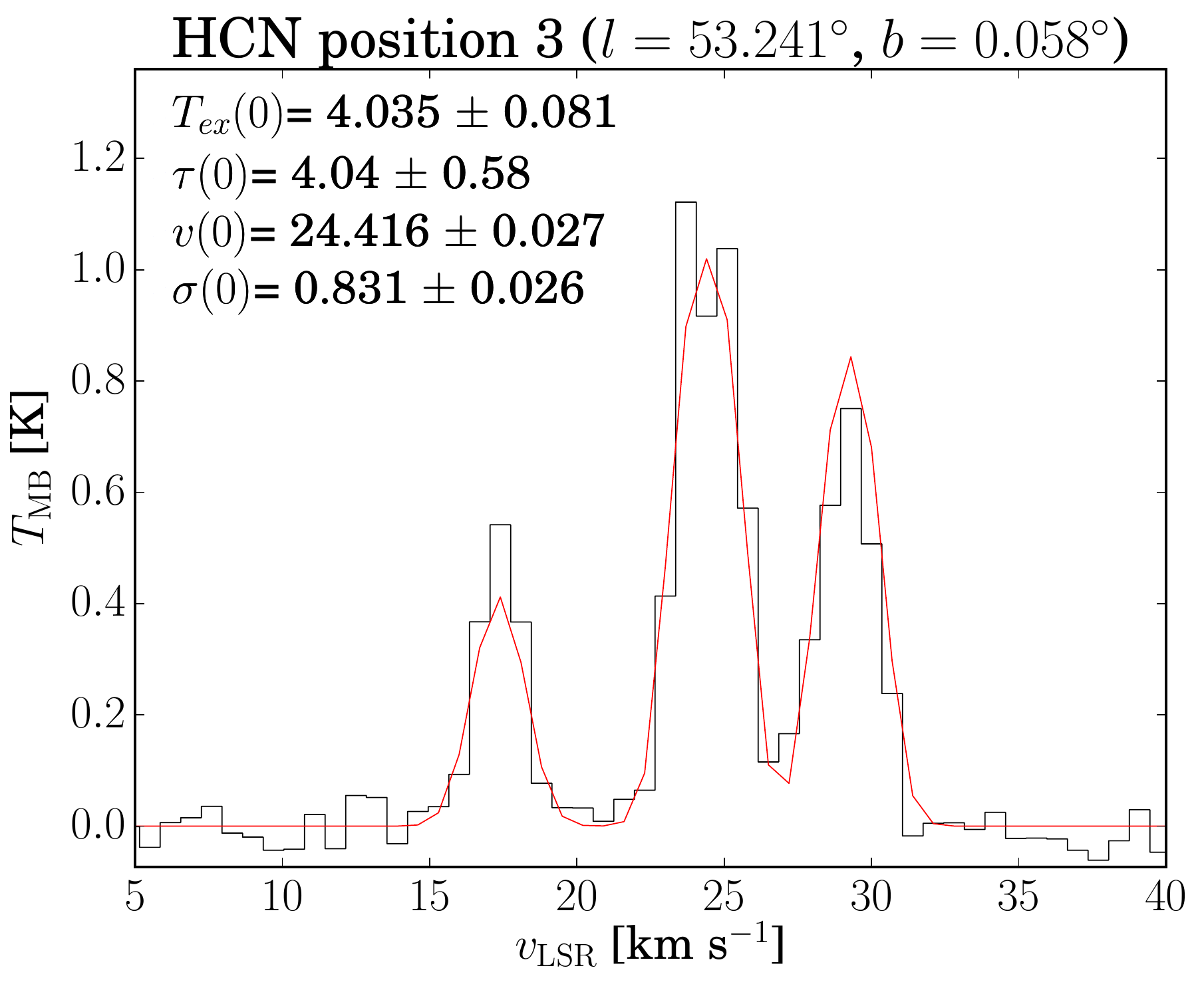}
   \includegraphics[width=0.24\hsize]{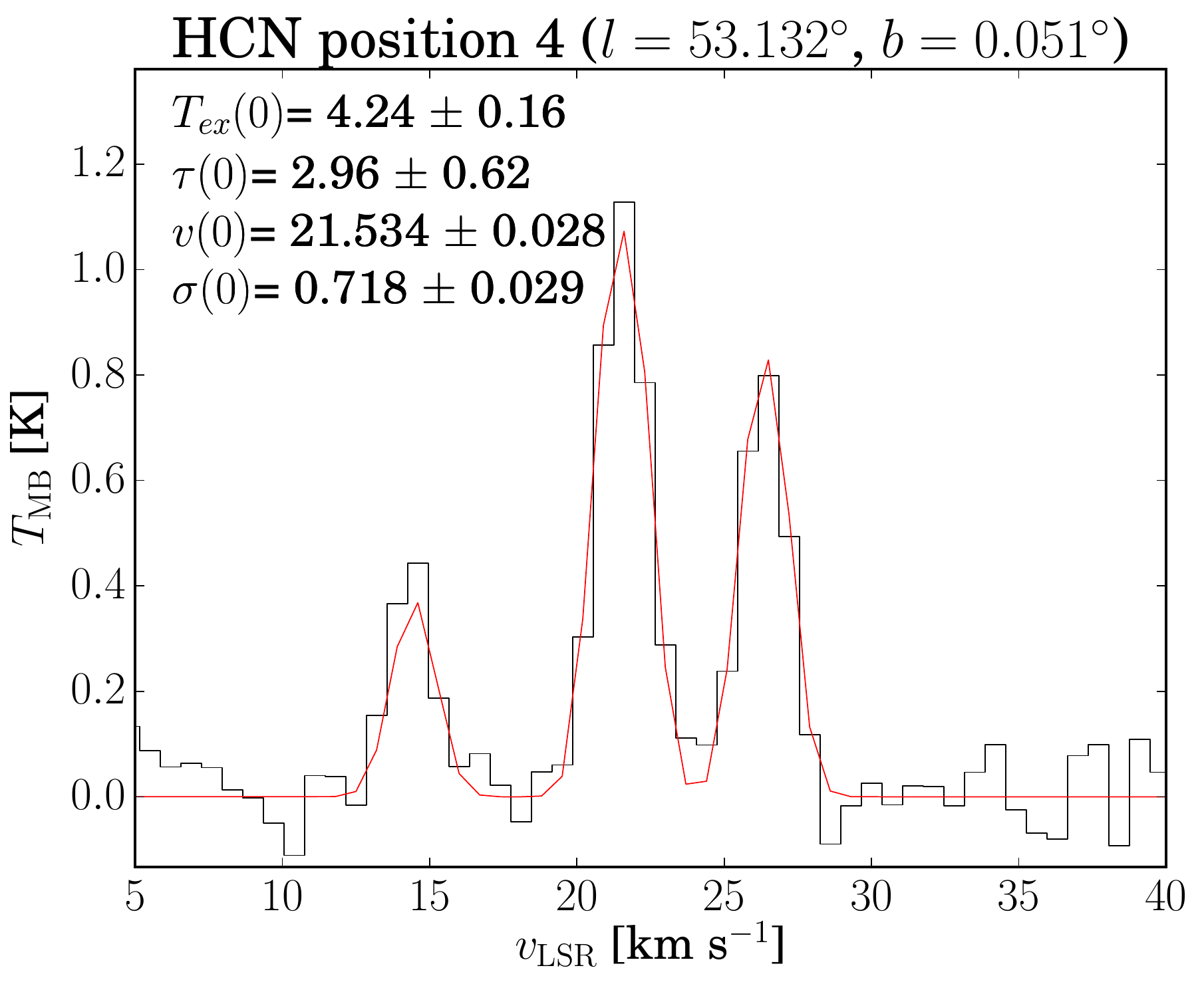}
   \includegraphics[width=0.24\hsize]{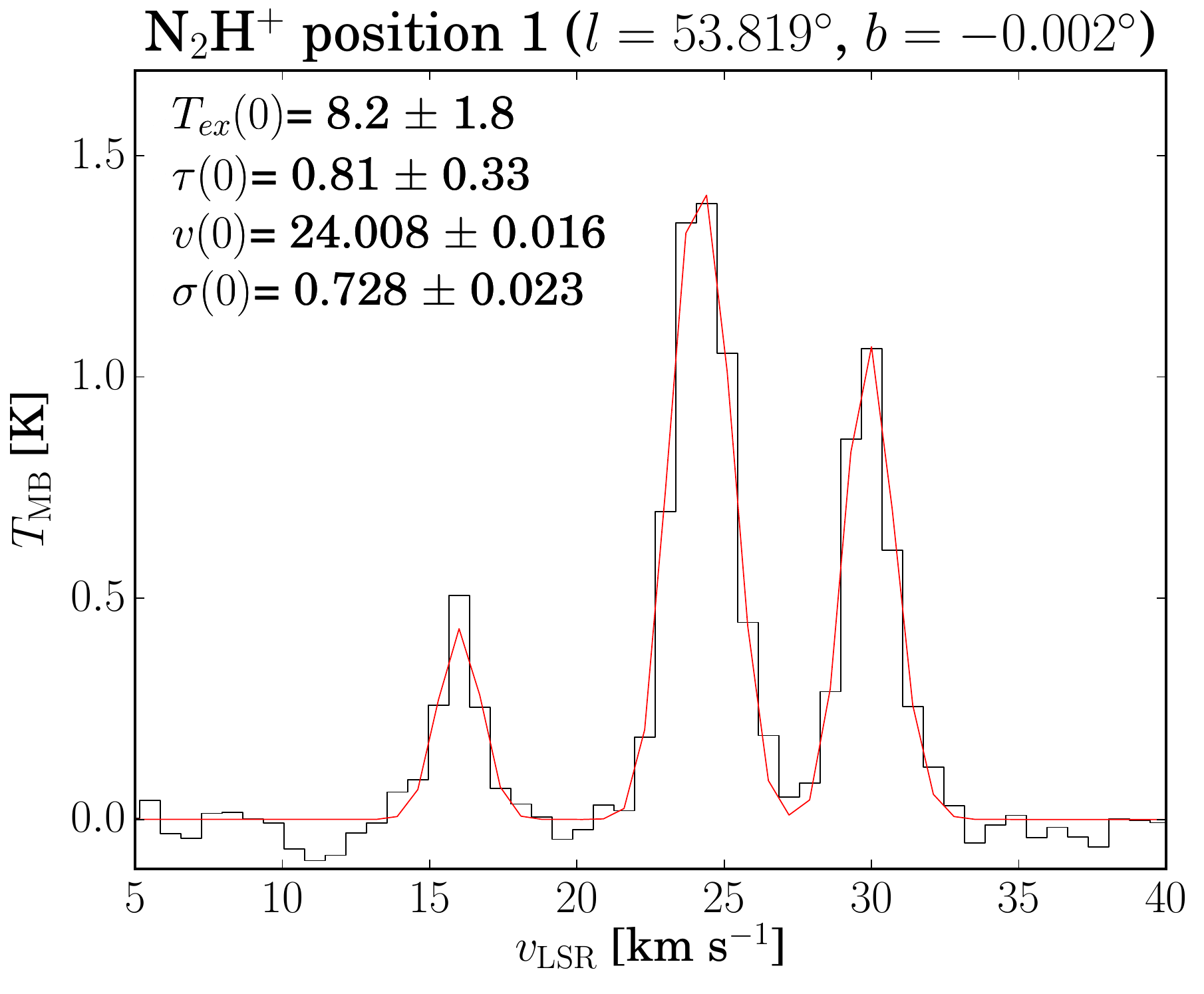}
   \includegraphics[width=0.24\hsize]{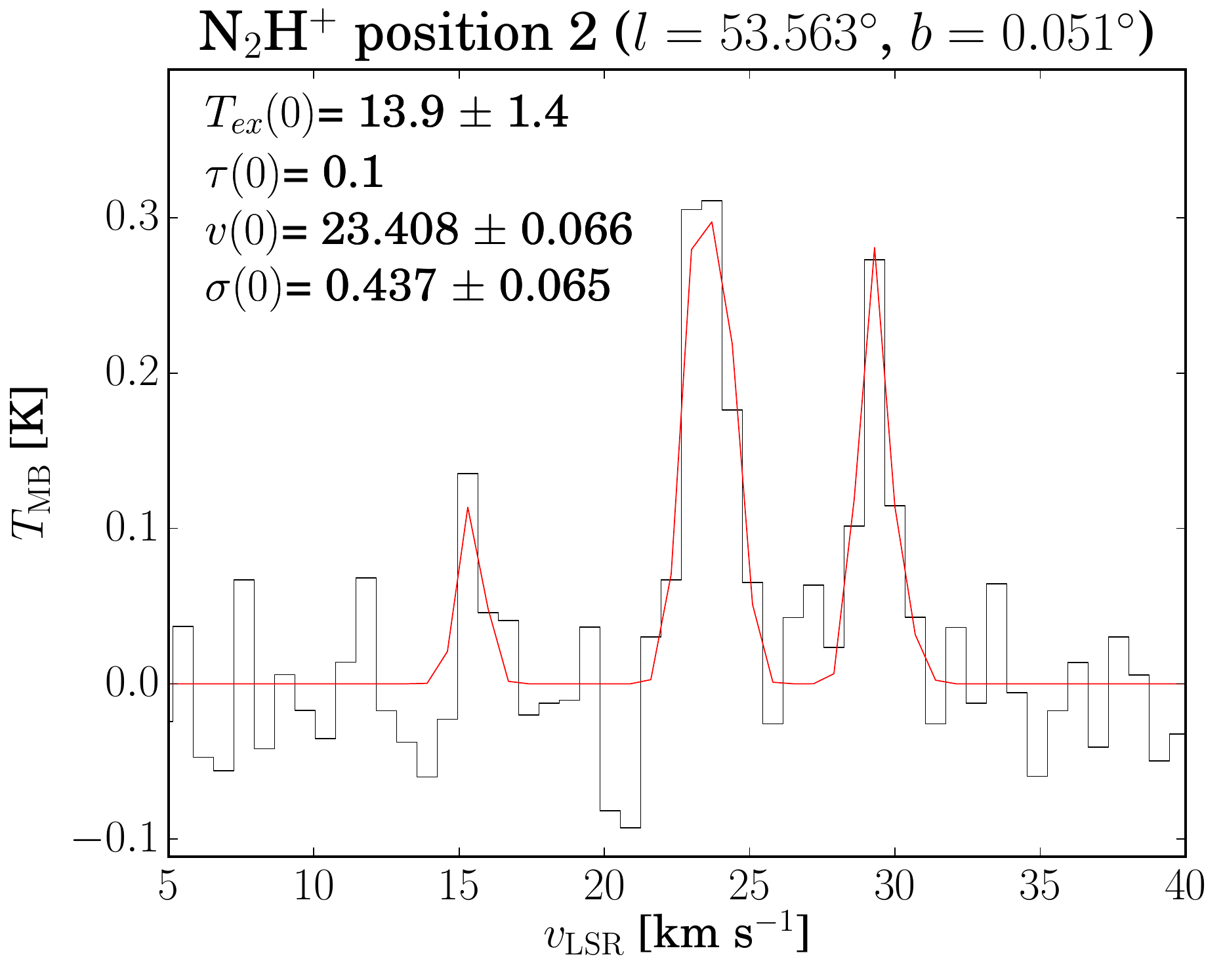}
   \includegraphics[width=0.24\hsize]{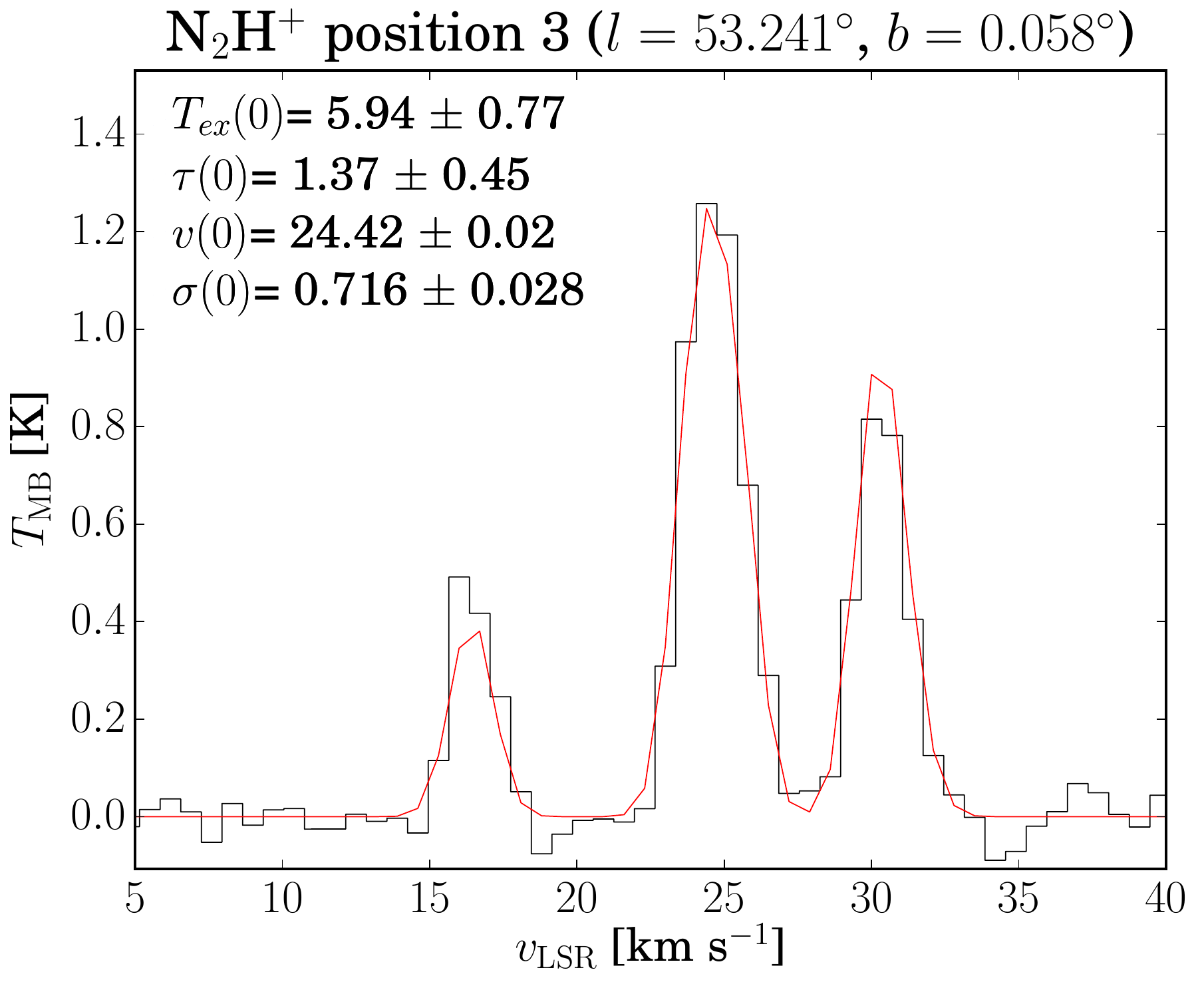}
   \includegraphics[width=0.24\hsize]{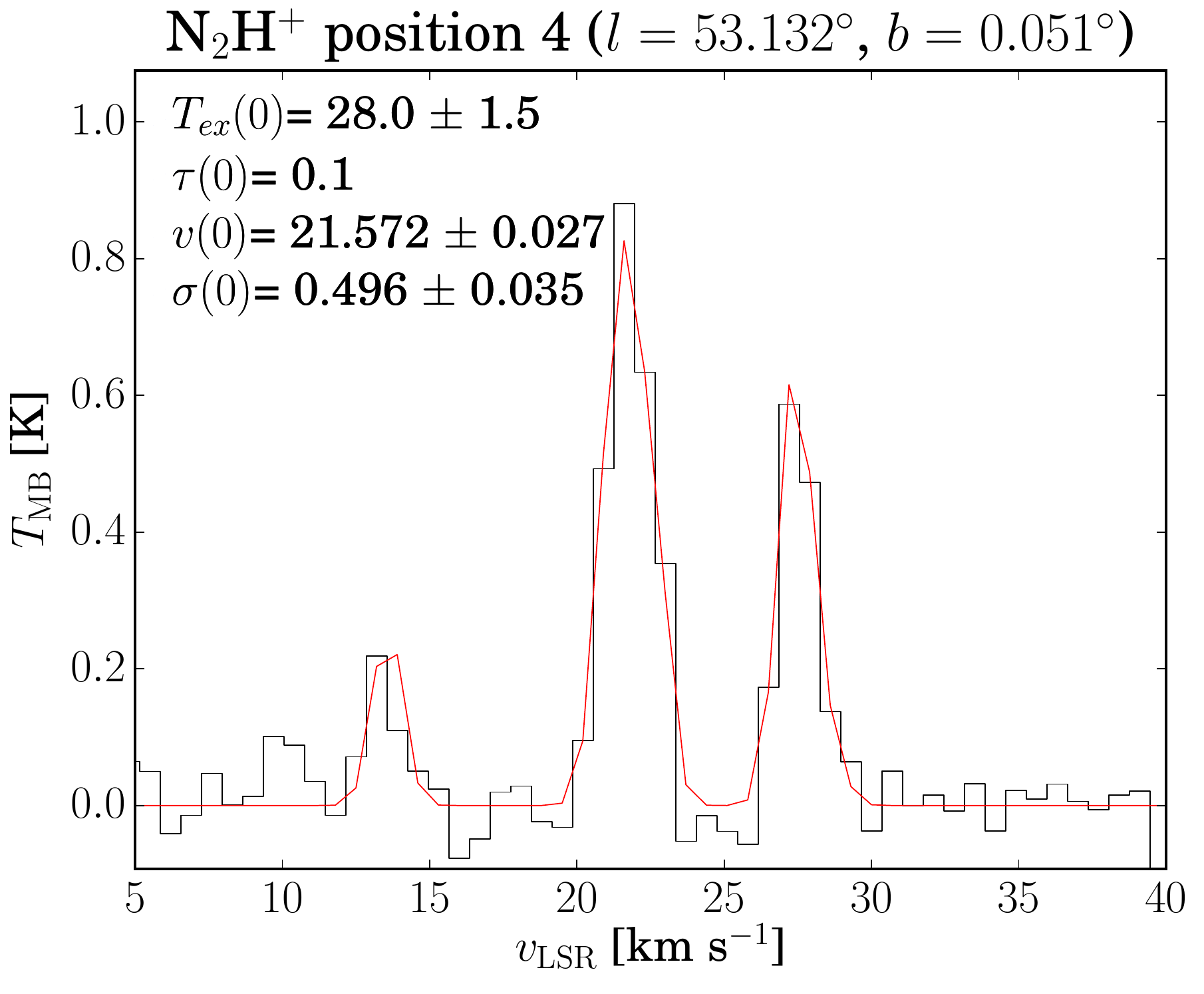}
      \caption{Selected HCN(1--0) (top) and N$_2$H$^+$(1--0) (bottom) spectra from pixeles corresponding the positions indicated in Fig.~\ref{fig_tex_tau}. The HFS fit are shown in red. The fitting results, excitation temperature ($T_{\rm ex}$), optical depth ($\tau$), line center ($v$), and line width ($\sigma$, velocity dispersion) are listed in each panel.}
 \label{fig_hfsfit}
\end{figure*}

\begin{figure}
   \centering
   \includegraphics[width=1\hsize]{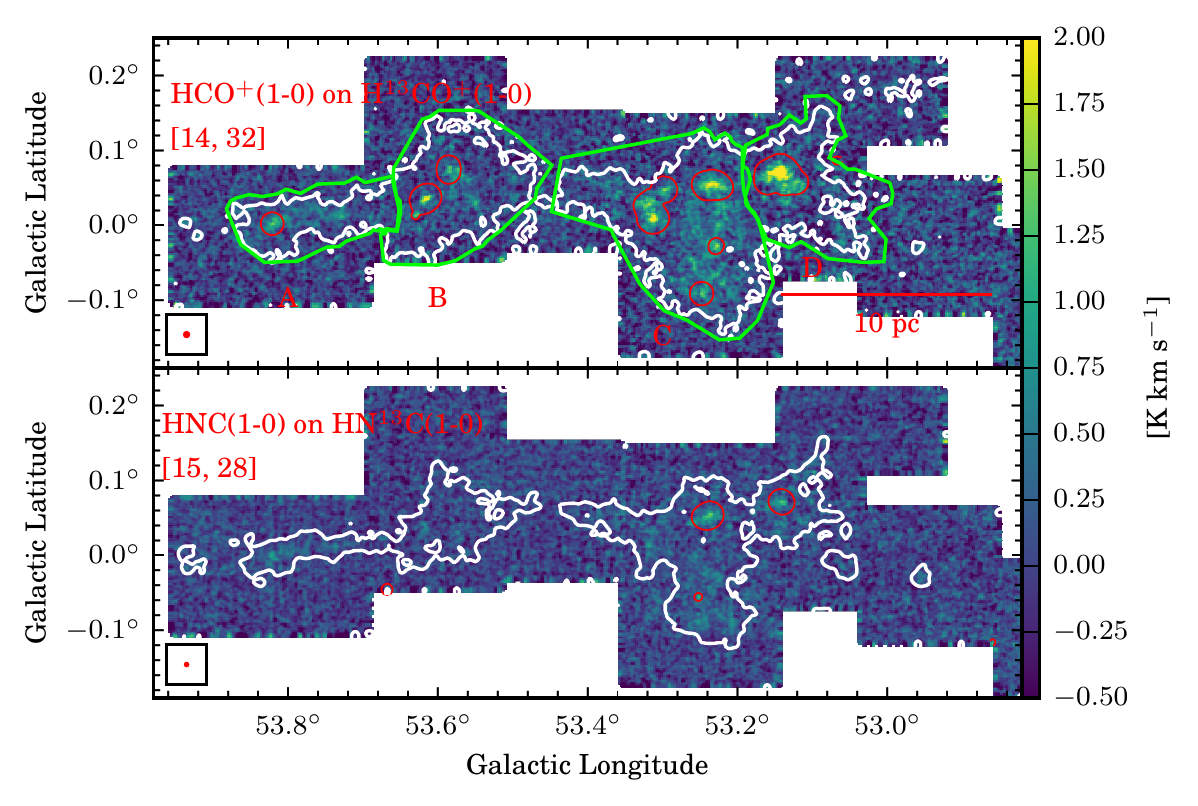}
      \caption{H$^{13}$CO$^+$(1--0) and HN$^{13}$C(1--0) integrated intensity maps (color scale and red contours). The red contours outline the area where the $^{13}$C lines have good signal-to-noise ratio and the optical depth correction was applied for the column density calculation. The thick white contour in each panel traces the 3$\sigma$ level the respect $^{12}$C isotopologue line integrated intensity shown in Fig.~\ref{fig_higal_lines}. The beams, ellipses, and polygons are the same as in Fig.~\ref{fig_higal_lines}.}
 \label{fig_13c}
\end{figure}

In this section we show the excitation temperature and optical depth maps derived from the HFS fitting for HCN(1--0) and N$_2$H$^+$(1--0). All four maps are convolved with a gaussian FWHM beam of 32\arcsec\ (gaussian kernel $\sim$13.59\arcsec) in Fig.~\ref{fig_tex_tau} and four selected spectra of each transition to show the HFS fitting results (Fig.~\ref{fig_hfsfit}). We also show the H$^{13}$CO$^+$(1--0) and HN$^{13}$C(1--0) integrated intensity maps in Fig.~\ref{fig_13c}.

\section{N-PDFs with optically thin assumption}
\begin{figure*}
   \centering
   \includegraphics[width=0.24\hsize]{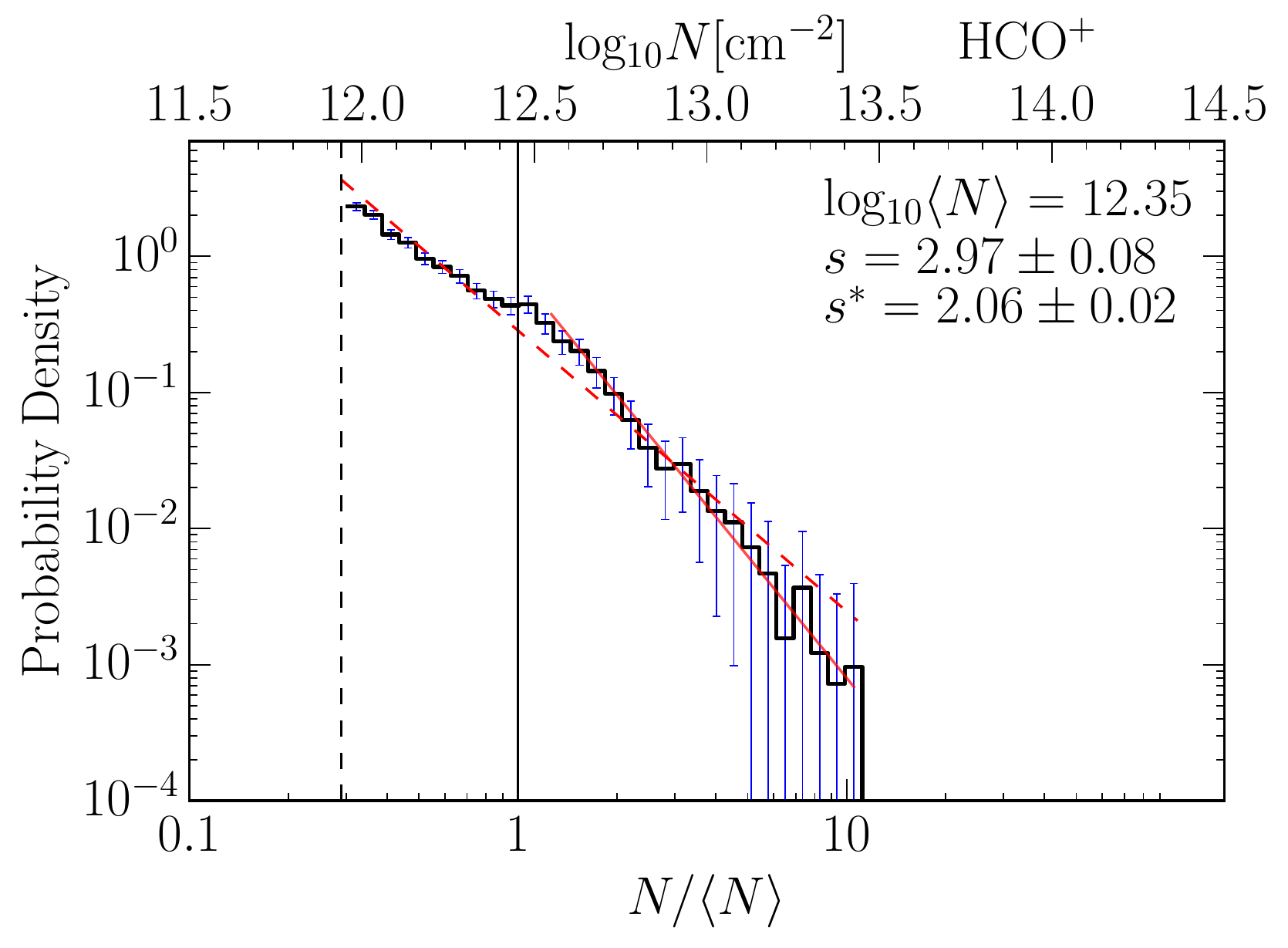}
   \includegraphics[width=0.24\hsize]{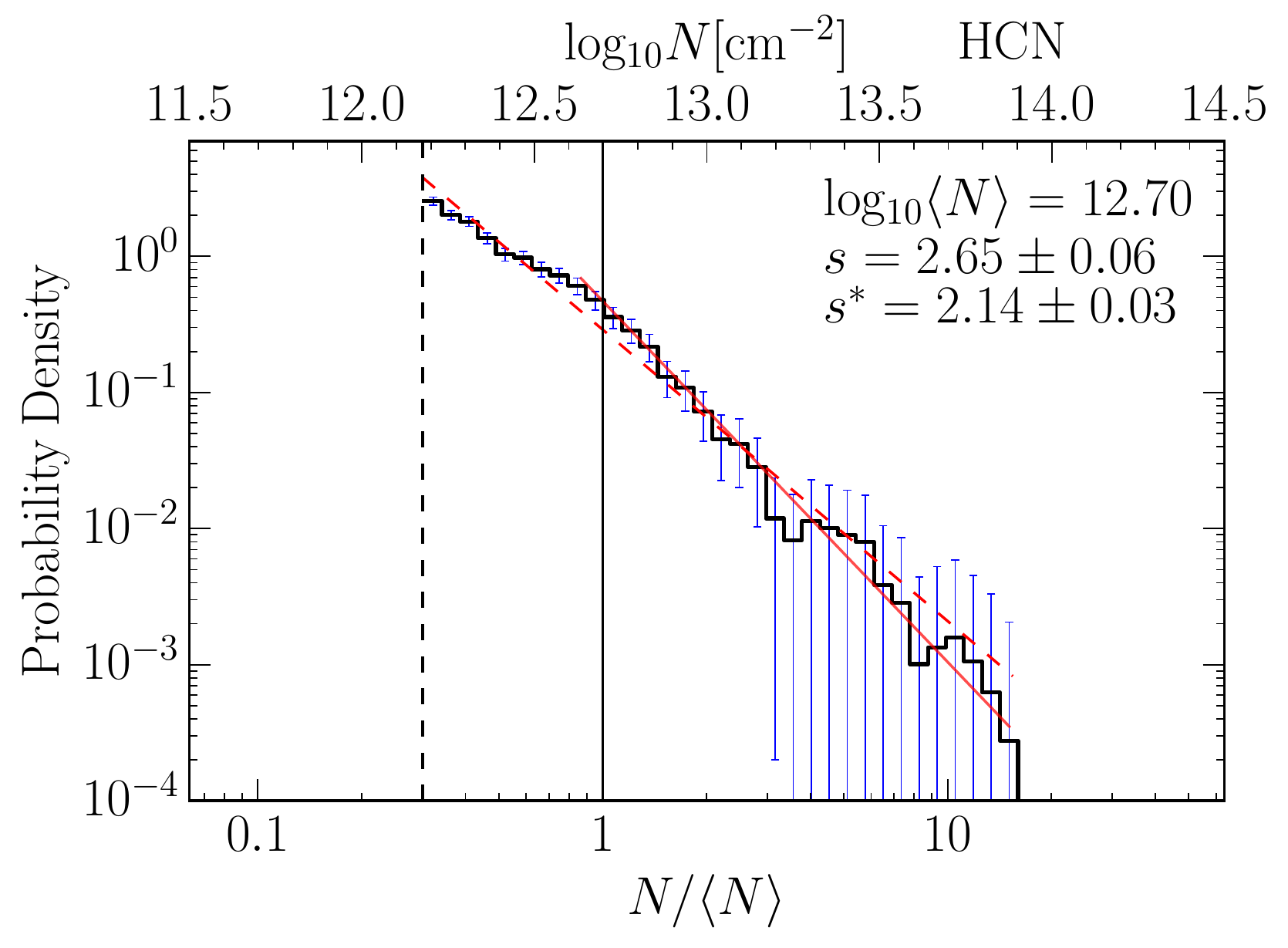}
   \includegraphics[width=0.24\hsize]{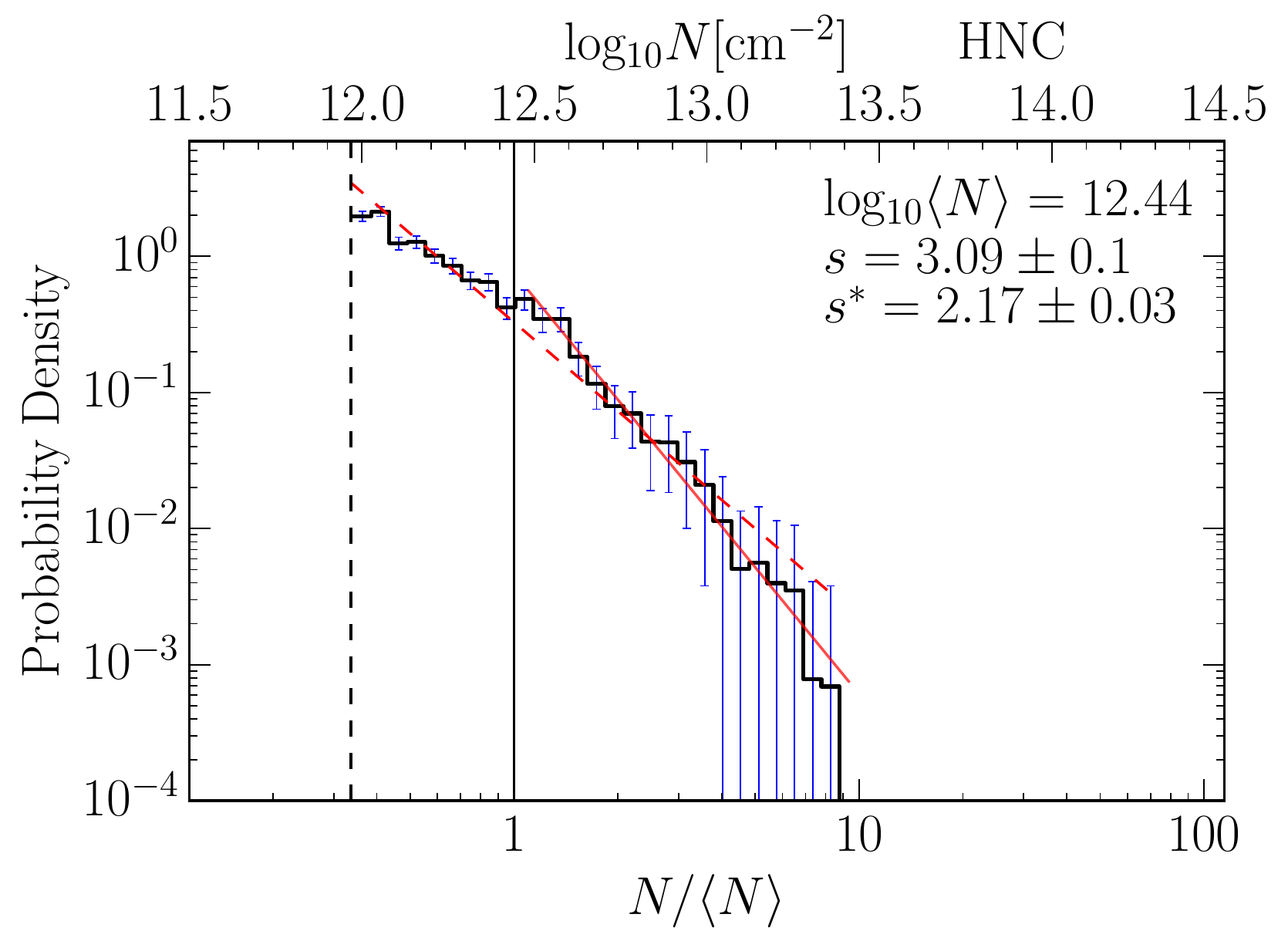}
   \includegraphics[width=0.24\hsize]{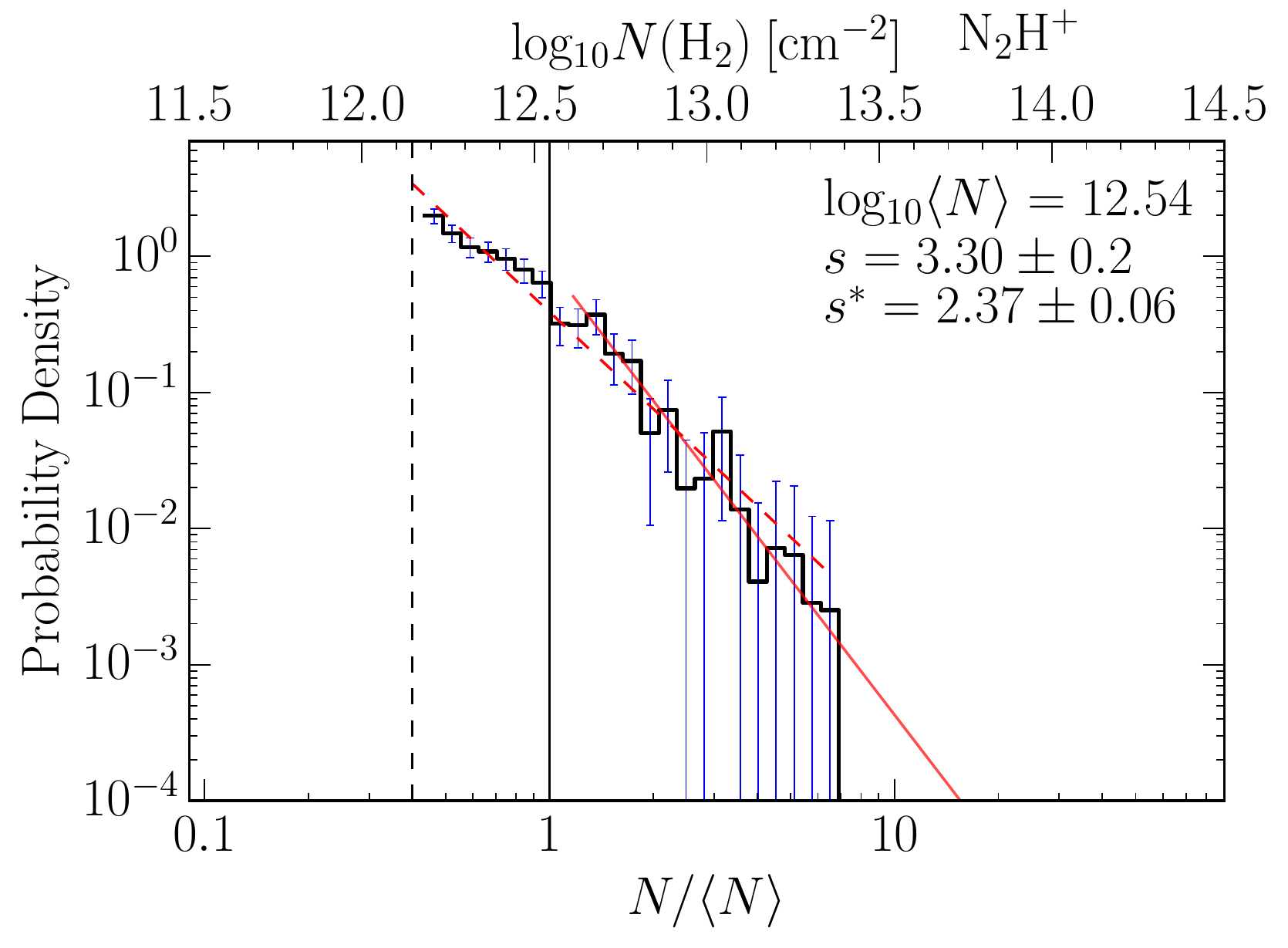}
      \caption{N-PDFs of column density derived from the HCO$^{+}$(1--0), HCN(1--0), HNC(1--0), and N$_2$H$^+$(1--0) assuming optically thin emission for the whole filament. We listed in each panel the mean column density $\langle N\rangle$, the power-law index $s$ (the optimized fit, solid line).}
 \label{fig_pdf_thin}
\end{figure*}

\section{The density profile of a singular polytropic cylinder}
\label{app_cy}
This section is adopted from the appendix in \citet{Schneider2016}.

The N-PDF is defined as the area fraction of gas $A$ with column density $N$, $p_N\propto {\rm d}A/{\rm d}N$. In this paper, we calculate the probability distribution $p$ of the normalized column density log$_{10}$($N/\langle N\rangle$), therefore $p \propto {\rm d}A/{\rm d(log}_{10}(N/\langle N\rangle)) \propto Np_N$. For a singular polytropic cylinder, the density profile from the axis of the cylinder filament is $\rho(r)\propto r^{-\alpha_{\rm f}}$. The column density N is an integral of the density, hence $N\propto \rho r \propto r^{-\alpha_{\rm f}+1}$. The area fraction $A$ is proportional to the cylinder length $z$ and radius $r$, $A\propto zr\propto r$, therefore
\begin{gather}
p_N \propto dA/dN \propto 1/r^{-\alpha_{\rm f}} \propto N^{\alpha_{\rm f}/(1-\alpha_{\rm f})}\\
p \propto Np_N \propto N^{1/(1-\alpha_{\rm f})}.
\end{gather} 

The power-law index of the N-PDF is $s$ ($p\propto (N/\langle N\rangle)^{-s}$), hence $\alpha_{\rm f}=1+1/s$. 

Similarly for a spherical core or clump with a density profile of $\rho(r)\propto r^{-\alpha_{\rm c}}$, column density $N\propto \rho r \propto r^{-\alpha_{\rm c}+1}$. In this case, the are $A \propto r^2$. Therefor,
\begin{gather}
p_N \propto dA/dN \propto r/r^{-\alpha_{\rm c}} \propto N^{1+\alpha_{\rm c}/(1-\alpha_{\rm c})}\\
p \propto Np_N \propto N^{2/(1-\alpha_{\rm c})},
\end{gather} 
and $\alpha_{\rm c} =1+2/s$.

Similar calculations are also discussed by \citet{Federrath2013}, \citet{Fischera2014}, and \citet{Myers2015}.

\end{appendix}

\end{document}